\documentclass[prd,notitlepage,nofootinbib,superscriptaddress]{revtex4-1}

\usepackage{amsmath}
\usepackage{amsfonts}
\usepackage[utf8]{inputenc}
\usepackage{graphicx}
\usepackage{hyperref}
\usepackage{slashed}
\usepackage{dcolumn}
\hypersetup{colorlinks, linkcolor = [rgb]{0,0.0,0.75}, citecolor = [rgb]{0,0.0,0.75}, urlcolor = [rgb]{0,0.0,0.75}}

\DeclareMathOperator{\Tr}{Tr}

\makeatletter
\g@addto@macro\bfseries{\boldmath}
\makeatother

\begin{document}

\title{Up, down, and strange nucleon axial form factors from lattice QCD}

\author{Jeremy~Green}
\email{jeremy.green@desy.de}
  \affiliation{NIC, Deutsches Elektronen-Synchrotron, 15738 Zeuthen, Germany}

\author{Nesreen~Hasan}
\email{n.hasan@fz-juelich.de}
  \affiliation{Bergische Universität Wuppertal, 42119 Wuppertal, Germany}
  \affiliation{IAS, Jülich Supercomputing Centre, Forschungszentrum Jülich, 52425 Jülich, Germany}

\author{Stefan~Meinel}
\email{smeinel@email.arizona.edu}
  \affiliation{Department of Physics, University of Arizona, Tucson, AZ 85721, USA}
  \affiliation{RIKEN BNL Research Center, Brookhaven National Laboratory, Upton, NY 11973, USA}

\author{Michael~Engelhardt}
  \affiliation{Department of Physics, New Mexico State University, Las Cruces, NM 88003-8001, USA}

\author{Stefan~Krieg}
  \affiliation{Bergische Universität Wuppertal, 42119 Wuppertal, Germany}
  \affiliation{IAS, Jülich Supercomputing Centre, Forschungszentrum Jülich, 52425 Jülich, Germany}

\author{Jesse~Laeuchli}
  \affiliation{Department of Computer Science, College of William and Mary, Williamsburg, VA 23187, USA}
  
\author{John~Negele}
  \affiliation{Center for Theoretical Physics, Massachusetts Institute of Technology, Cambridge, MA 02139, USA}

\author{Kostas~Orginos}
  \affiliation{Physics Department, College of William and Mary, Williamsburg, VA 23187, USA}
  \affiliation{Thomas Jefferson National Accelerator Facility, Newport News, VA 23606, USA}

\author{Andrew~Pochinsky}
  \affiliation{Center for Theoretical Physics, Massachusetts Institute of Technology, Cambridge, MA 02139, USA}

\author{Sergey~Syritsyn} 
  \affiliation{Thomas Jefferson National Accelerator Facility, Newport News, VA 23606, USA}

\date{\today}

\begin{abstract}

  We report a calculation of the nucleon axial form factors
  $G_A^q(Q^2)$ and $G_P^q(Q^2)$ for all three light quark flavors
  $q\in\{u,d,s\}$ in the range $0\leq Q^2\lesssim 1.2\text{ GeV}^2$
  using lattice QCD. This work was done using a single ensemble with
  pion mass 317~MeV and made use of the \emph{hierarchical probing}
  technique to efficiently evaluate the required disconnected
  loops. We perform nonperturbative renormalization of the axial
  current, including a nonperturbative treatment of the mixing between
  light and strange currents due to the singlet-nonsinglet difference
  caused by the axial anomaly. The form factor shapes are fit using
  the model-independent $z$ expansion. From $G_A^q(Q^2)$, we determine
  the quark contributions to the nucleon spin and axial radii. By
  extrapolating the isovector $G_P^{u-d}(Q^2)$, we obtain the induced
  pseudoscalar coupling relevant for ordinary muon capture and the
  pion-nucleon coupling constant. We find that the disconnected
  contributions to $G_P$ form factors are large, and give an
  interpretation based on the dominant influence of the pseudoscalar
  poles in these form factors.

\end{abstract}

\maketitle

\section{Introduction}

The axial and induced pseudoscalar form factors\footnote{We also
  denote flavor combinations using, e.g., $G_A^{u-d}(Q^2)\equiv
  G_A^u(Q^2)-G_A^d(Q^2)$.}, $G_A^q(Q^2)$ and $G_P^q(Q^2)$,
parameterize matrix elements of the axial current between proton
states:
\begin{equation}\label{eq:GA_GP}
  \langle p',\lambda'|A_\mu^q|p,\lambda\rangle
 = \bar u(p',\lambda') \left[ \gamma_\mu G_A^q(Q^2)
 + \frac{(p'-p)_\mu}{2m_N} G_P^q(Q^2) \right] \gamma_5 u(p,\lambda),
\end{equation}
where $Q^2=-(p'-p)^2$ and $A_\mu^q=\bar q\gamma_\mu\gamma_5 q$. It has
been shown that $G_A^q(Q^2)$ can be interpreted as the two-dimensional
Fourier transform of the difference $q_\uparrow(\mathbf{b}_\perp)+\bar
q_\uparrow(\mathbf{b}_\perp)-q_\downarrow(\mathbf{b}_\perp)-\bar
q_\downarrow(\mathbf{b}_\perp)$ between transverse densities of
helicity aligned and anti-aligned quarks plus antiquarks in a
longitudinally polarized nucleon, in the infinite momentum
frame~\cite{Burkardt:2002hr}.

At $Q^2=0$, the axial form factor gives the fractional contribution
from the spin of quarks $q$ and $\bar q$ to the proton's spin, which
can also be obtained from a moment of polarized parton distribution
functions:
\begin{equation}
  \Delta q \equiv g_A^q \equiv G_A^q(0)
 = \int_0^1 dx \left( \Delta q(x) + \Delta \bar q(x) \right).
\end{equation}
Understanding the constituents of the proton's spin has been of great
interest ever since the European Muon Collaboration found, by
measuring the spin asymmetry in polarized deep inelastic scattering,
that the total contribution from quark spin to the proton's spin is less
than half~\cite{Ashman:1987hv}.

Axial form factors naturally arise in the interactions of nucleons
with $W$ and $Z$ bosons. Assuming isospin symmetry, the $W$ boson is
sensitive to the $u-d$ flavor combination, whereas the $Z$ boson is
also sensitive to strange quarks. Neutron beta decay, mediated by
$W$-boson exchange, is used to determine the ``axial charge''
$g_A\equiv g_A^{u-d}$. Quasielastic neutrino scattering, $\nu n\to
\ell^- p$ or $\bar\nu p\to \ell^+ n$, has been used to measure the
isovector axial form factor $G_A^{u-d}(Q^2)$, whereas elastic neutrino
scattering is also sensitive to $G_A^s(Q^2)$. The shape of the
isovector axial form factor is often assumed to be a dipole,
$G_A^{u-d}(Q^2)=g_A/(1+Q^2/m_A^2)^2$; rather than assume a dipole, we
will use a more general fit and characterize the shape using the
squared axial radii $(r_A^2)^q$. These are defined from the slope of the
form factors at zero momentum transfer\footnote{In contrast with the
  strange magnetic radius $(r_M^2)^s\equiv
  -6\frac{d}{dQ^2}G_M^s(Q^2)|_{Q^2=0}$, we choose to normalize the
  strange axial radius relative to the value of the form factor at
  $Q^2=0$, the same as for all the axial radii. Note that this means
  the flavor combinations satisfy, e.g., $g_A^{u-d}(r_A^2)^{u-d}=
  g_A^u(r_A^2)^u-g_A^d(r_A^2)^d$.}:
\begin{equation}
  G_A^q(Q^2) = g_A^q\left(1-\frac{1}{6}(r_A^2)^qQ^2 + O(Q^4)\right).
\end{equation}
The ordinary ``axial radius'' is the isovector one, $r_A\equiv
\sqrt{(r_A^2)^{u-d}}$; in the dipole model, $r_A^2=12/m_A^2$. It can
also be determined from pion electroproduction, using chiral
perturbation theory~\cite{Bernard:2001rs}.

In addition to the valence up and down quarks, quantum fluctuations
cause other quarks to play a role in the structure of nucleons; the
strange quark is the next lightest, and is expected to be the next
most important. In this paper, we report a calculation of the nucleon
axial form factors using a single lattice QCD ensemble. This
calculation includes both quark-connected and disconnected diagrams,
which allows us to determine the up, down, and strange form
factors. Using the same dataset, we previously reported a
high-precision calculation of the strange nucleon electromagnetic form
factors~\cite{Green:2015wqa}.

A lattice QCD study of the axial form factors of the nucleon is
timely not least in view of experimental efforts underway using
the MicroBooNE liquid Argon time-projection chamber, which, in
particular, will be able to map out the strange axial form factor
of the nucleon to momentum transfers as low as
$Q^2 = 0.08\text{ GeV}^2$~\cite{Papavassiliou:2009zz}. This is
achieved by combining neutrino-proton neutral and charged current
scattering cross section measurements with available polarized
electron-proton/deuterium cross section data, and is expected to
reduce the experimental uncertainty of the extrapolated value at
$Q^2 =0$, i.e., the strange quark spin contribution $\Delta s$,
by an order of magnitude. Such an extraction is complementary to
polarized DIS determinations that access the strange quark helicity
distribution function, but suffer from lack of coverage at low and
high momentum fraction $x$ when evaluating the first $x$-moment.
The $Q^2 $ range explored by the MicroBooNE experiment, between
$Q^2 = 0.08\text{ GeV}^2 $ and about $Q^2 = 1\text{ GeV}^2 $, matches the range
covered by the present lattice calculation well, enabling a future
comparison of the $Q^2 $-dependence obtained for the strange axial
form factor.

This paper is organized as follows. Section~\ref{sec:methodology}
describes our methodology: the approach used to isolate the nucleon
ground state and determine the form factors, the methods used to
determine the numerically-challenging disconnected diagrams, the
details of the lattice ensemble, and the fits to the $Q^2$-dependence
of the form factors using the $z$ expansion. The unwanted
contributions from excited states to the different observables are
examined in detail, and the estimation of systematic uncertainty is
described. Our nonperturbative calculation of the renormalization
factors, including a nonperturbative treatment of the flavor singlet
case, is presented in Sec.~\ref{sec:renormalization}. The main results
are in Sec.~\ref{sec:formfacs}: the axial and induced pseudoscalar
form factors for light and strange quarks, as well as the quark
contributions to the nucleon spin. Finally, we present our conclusions
in Sec.~\ref{sec:conclusions}. In an
\hyperref[app:fit_params]{appendix}, we give the parameters for our
fits to the form factors.

\section{\label{sec:methodology}Lattice methodology}

\subsection{Computation of matrix elements}

To determine nucleon matrix elements, we compute two-point and
three-point functions,
\begin{gather}
  C_\text{2pt}(\vec p,t) = \sum_{\vec x} e^{-i\vec p\cdot\vec x}
  \Tr\left[\Gamma_\text{pol} 
    \langle \chi(\vec x,t)\bar\chi(\vec 0,0) \rangle \right] \\
  C_\text{3pt}^{A_\mu^q}(\vec p,\vec p\,',\tau,T) = \sum_{\vec x,\vec y}
  e^{-i\vec p\,'\cdot\vec x}e^{i(\vec p\,'-\vec p)\cdot y} \Tr\left[\Gamma_\text{pol}
   \langle \chi(\vec x,T) A_\mu^q(\vec y,\tau) \bar\chi(\vec 0,0)\rangle\right],
\end{gather}
where
$\chi=\epsilon^{abc}(\tilde u^T_a C\gamma_5 \frac{1+\gamma_4}{2}
\tilde d_b) \tilde u_c$ is a proton interpolating operator and
$\Gamma_\text{pol}$ is a spin and parity projection matrix. In the
interpolating operator, we use Wuppertal-smeared~\cite{Gusken:1989qx}
quark fields $\tilde q=(\frac{1+\alpha H}{1+6\alpha})^N q$, where $H$
is the nearest-neighbor gauge-covariant hopping matrix constructed
using spatially APE-smeared~\cite{Albanese:1987ds} gauge links.

The proton ground state can be obtained in the limit where all time
separations $t$, $\tau$, and $T-\tau$ are large. In this limit, the
following ratio does not depend on the time separations or on the
interpolating operator:
\begin{equation}\label{eq:ratio}
  \begin{aligned}
    R_\mu^q(\vec p,\vec p\,',\tau,T) &\equiv \frac{C_\text{3pt}^{A_\mu^q}(\vec p,\vec p\,',\tau,T)}{
      \sqrt{C_\text{2pt}(\vec p,T)C_\text{2pt}(\vec p\,',T)}}
    \sqrt{\frac{C_\text{2pt}(\vec p,T-\tau)C_\text{2pt}(\vec p\,',\tau)}{
        C_\text{2pt}(\vec p\,',T-\tau)C_\text{2pt}(\vec p,\tau)}} \\
    &= M_\mu^q(\vec p,\vec p\,') + O(e^{-\Delta E_{10}(\vec p)\tau})
 + O(e^{-\Delta E_{10}(\vec p\,')(T-\tau)}),
  \end{aligned}
\end{equation}
where $M_\mu^q(\vec p,\vec p\,')$ contains the desired nucleon matrix
element $\langle\vec p\,',\lambda'|A_\mu^q|\vec p,\lambda\rangle$
(with spins depending on $\Gamma_\text{pol}$) and some kinematic
factors (see, e.g., \cite{Green:2014xba}), and $\Delta E(\vec p)$ is
the energy gap between the ground and lowest excited state with
momentum $\vec p$.

For each source-sink separation $T$, for the \emph{ratio-plateau
  method}, we take the average of the central two or three points
$R_\mu^q(\vec p,\vec p\,',\tau,T)$ near $\tau=T/2$. This gives an
estimate of $M_\mu^q(\vec p,\vec p\,')$ (and thus the nucleon matrix
element) with a systematic error coming from excited-state
contamination that decays exponentially as
$e^{-\Delta E_\text{min} T/2}$, where
$\Delta E_\text{min}=\min\{\Delta E_{10}(\vec p),\Delta E_{10}(\vec
p\,')\}$. We also use the \emph{summation method}, computing the sums
\begin{equation}
  S^q_\mu(\vec p,\vec p\,',T) \equiv a\sum_{\tau/a=1}^{T/a-1} R^q_\mu(\vec p,\vec p\,',\tau,T) = c + TM_\mu^q(\vec p,\vec p\,') + O(Te^{-\Delta E_\text{min}T}).
\end{equation}
Fitting the slope with respect to $T$ yields an estimate of
$M_\mu^q(\vec p,\vec p\,')$ that has a greater suppression of unwanted
excited-state contributions~\cite{Capitani:2010sg,Bulava:2010ej},
which now decay as $Te^{-\Delta E_\text{min}T}$.

For each $Q^2$, we construct a system of equations parameterizing the
corresponding set of matrix elements of the axial current with
$G_A(Q^2)$ and $G_P(Q^2)$. We combine equivalent matrix elements to
improve the condition number~\cite{Syritsyn:2009mx}, and then solve
the resulting overdetermined system of
equations~\cite{Hagler:2003jd}. This approach makes use of all
available data to minimize the statistical uncertainty. In particular,
for disconnected diagrams, we are able to compute correlators for all
polarizations and all equivalent momenta, maximizing the amount of
averaging.

\subsection{Disconnected diagrams}
\label{disc}
\begin{figure}
  \centering
  \includegraphics[width=0.2\textwidth]{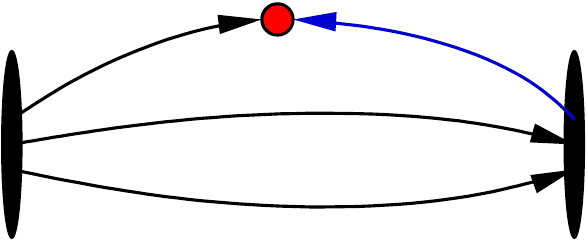}
\hspace{0.05\textwidth}
  \includegraphics[width=0.2\textwidth]{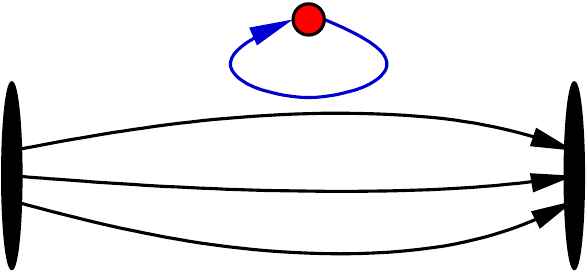}
  \caption{Two classes of quark contractions for $C_\text{3pt}$. Left:
    \emph{connected}, which is evaluated using a sequential propagator
    through the sink (shown in blue). Right: \emph{disconnected},
    where the loop containing the axial current is evaluated
    stochastically.}
  \label{fig:conn_disc}
\end{figure}

There are two kinds of quark contractions that contribute to
$C_\text{3pt}$: \emph{connected} and \emph{disconnected}, shown in
Fig.~\ref{fig:conn_disc}. We evaluate the former exactly for each
source on each gauge configuration, using sequential propagators
through the sink~\cite{Martinelli:1988rr}. For the latter, we perform
a stochastic evaluation of the \emph{disconnected loop},
\begin{equation}\label{T_disc}
  T^q_\mu(\vec k,t) \equiv -\sum_{\vec x}e^{i\vec k\cdot\vec x}
\Tr\left[\gamma_\mu\gamma_5 D^{-1}_q(x,x)\right],
\end{equation}
where $D_q$ is the lattice Dirac operator with a fixed gauge
background and $x=(\vec x,t)$. We then obtain the disconnected
contribution to $C_\text{3pt}$ from the correlation between this loop
and the nucleon two-point function.

To evaluate the disconnected loop, we generate noise fields
$\eta_{a\alpha}(x)$ that have color, spin, and space-time indices but
with support only on a single timeslice\footnote{In this work we have not 
compared the effectiveness of placing noise on one timeslice against
placing it on all timeslices.}, $t$. We use one $\mathbb{Z}_2+i\mathbb{Z}_2$
noise vector for each chosen timeslice and gauge configuration,
i.e., the components of $\eta$ are randomly chosen from
$\{\frac{1+i}{\sqrt{2}}, \frac{1-i}{\sqrt{2}}, \frac{-1+i}{\sqrt{2}},
\frac{-1-i}{\sqrt{2}}\}$. As a result, the diagonal elements of
$\eta\eta^\dagger$ are equal to 1, and the off-diagonal elements are
random with expectation value zero. To reduce noise by replacing
statistical zeros with exact zeros in targeted off-diagonal components
of $\eta\eta^\dagger$, we use color and spin
dilution~\cite{Wilcox:1999ab, Foley:2005ac}, as well as hierarchical
probing~\cite{Stathopoulos:2013aci}. The former makes use of a
complete set of twelve projectors in color and spin space, $P_d$, such
that $P_d\eta$ has support on only one color and one spin
component. The latter makes use of $N_\text{hvec}$
specially-constructed spatial \emph{Hadamard vectors}, $z_n$, that
provide a scheme for progressively eliminating the spatially
near-diagonal contributions to the noise. Combining these yields
$12N_\text{hvec}$ modified noise fields,
\begin{equation}
  \eta^{[d,n]}_{a\alpha}(\vec x) = \sum_{b,\beta}(P_d)^{b\beta}_{a\alpha}z_n(\vec x)\eta_{b\beta}(\vec x).
\end{equation}
We use these as sources for quark propagators,
$\psi^{[d,n]}_q=D^{-1}_q\eta^{[d,n]}$, and obtain an estimator for
$T_\mu^q(\vec k,t)$:
\begin{equation}
  \frac{-1}{N_\text{hvec}}\sum_{d,n}\sum_{\vec x}e^{i\vec k\cdot\vec x}
  \eta^{[d,n]\dagger}(\vec x,t)\gamma_\mu\gamma_5\psi^{[d,n]}_q(\vec x,t). \label{eq:HP}
\end{equation}

We will separately consider the connected and disconnected contributions
to nucleon matrix elements of the light quark axial current. Although
the individual contributions are unphysical, they can be understood
using partially quenched QCD~\cite{Bernard:1993sv}, by introducing a
third degenerate light quark $r$ and a corresponding ghost quark to
cancel its fermion determinant in the path integral. The disconnected
contribution to a nucleon three-point function with current $A_\mu^u$
or $A_\mu^d$ is equal to a nucleon three-point function with
$A_\mu^r$. Since it was shown in Ref.~\cite{Bernard:2013kwa} that
partially quenched staggered fermions have a bounded transfer matrix,
we expect that for our case as well we can separately isolate the
ground state in the connected and disconnected contributions to
three-point functions, i.e., that Eq.~(\ref{eq:ratio}) applies to
$A_\mu^r$. In Section~\ref{sec:renormalization} we will also discuss
renormalization of $A_\mu^r$.

\subsection{\label{sec:ensemble}Lattice ensemble and calculation setup}

We use a single lattice ensemble with a tree-level Symanzik improved
gauge action ($\beta=6.1$) and 2+1~flavors of clover-improved Wilson
fermions that couple to the gauge links after stout smearing (one step
with $\rho=0.125$). The improvement parameters are set to their
tadpole-improved tree-level values. The lattice size is $32^3\times
96$ and the bare quark masses are $am_s=-0.245$ and $am_{ud}=-0.285$.

Based on the $\Upsilon(2S)-\Upsilon(1S)$ energy splitting computed
using lattice NRQCD, the lattice spacing is $a=0.11403(77)$~fm. The
strange quark mass is close to its physical value: the mass of the
unphysical $\eta_s$ meson is 672(3)(5)~MeV, which is within 5\% of its
value determined for physical quark masses~\cite{Dowdall:2011wh}. The
light quark mass is heavier than physical, producing a pion
mass\footnote{For the pion and $\eta_s$ mass, the second error is from
  uncertainty in the lattice spacing.} of 317(2)(2)~MeV. The volume is
quite large, such that $m_\pi L_s\approx 5.9$, and we thus expect
finite-volume effects to be highly suppressed.

We performed calculations using 1028 gauge configurations, on each of
which we chose six equally-spaced source timeslices. For each source
timeslice $t_0$, we used two positions $(\vec x_1,t_0)$
and $(\vec x_2,t_0)$ as sources for three-point
functions. We placed nucleon sinks in both the forward and backward
directions on timeslices $t_0\pm T$ to double statistics and obtain
a total of 24672 samples,
and used five source-sink separations $T/a\in\{6,8,10,12,14\}$. We
computed disconnected loops on timeslices $t_0+\tau$ displaced only in the
forward direction from each source timeslice, yielding 6168 timeslice
samples; the source-operator separations $\tau$ and number of Hadamard
vectors for each flavor are listed in Tab.~\ref{tab:sep_hvec}. For
each source timeslice, we computed sixteen two-point functions from
source positions $(\vec x_i,t_0)$, $i=1,\dots,16$,
yielding 98688 samples for correlating with the disconnected loops.
We imposed two constraints on our choice of momenta:
$(\vec p\,'-\vec p)^2\leq 10(\frac{2\pi}{L_s})^2$
and $(\vec p)^2,(\vec p\,')^2\leq 6(\frac{2\pi}{L_s})^2$. For the
connected diagrams we used two sink momenta, $\vec p\,'=\vec 0$ and
$\vec p\,'=\frac{2\pi}{L_s}(-1,0,0)$, and all source momenta
compatible with the constraints. For the disconnected diagrams we 
used all combinations of $\vec p$ and $\vec p\,'$ compatible with
the constraints, with the restriction that each $Q^2$ must match
a value available from the connected diagrams.

On each set of four adjacent gauge configurations, we averaged over
all spatially displaced samples of each correlator. This produced 257
blocked samples. Statistical error analysis was done using jackknife
resampling.

\begin{table}
  \centering
  \begin{tabular}{l||r|r|r|r|r}
    \multicolumn{1}{r||}{$\tau/a=$} & 3 & 4 & 5 & 6 & 7 \\\hline
    light & 16 & 128 & 128 & 128 & 16 \\
    strange & & 16 & 128 & 16
  \end{tabular}
  \caption{Number of Hadamard vectors used for disconnected loops of each flavor and source-operator separation $\tau$. Five separations were used for light quarks and three for strange. As shown in Subsec.~\ref{sec:HP}, sixteen Hadamard vectors is generally sufficient for the noise to saturate when using the axial current. Having 128 Hadamard vectors was particularly useful for Ref.~\cite{Green:2015wqa}, which used the vector current.}
  \label{tab:sep_hvec}
\end{table}

The general form for $O(a)$ improvement of quark bilinear operators
with nondegenerate quarks was given in
Ref.~\cite{Bhattacharya:2005rb}. If we simplify the expressions by
keeping only their form at one-loop order in perturbation theory, the
renormalized improved operators take the form
\begin{equation}\label{eq:improvement}
\begin{aligned}
  (A_\mu^q - A_\mu^{q'})^{R,I} &=
  Z_A\left[ A_\mu^q-A_\mu^{q'} + ac_A\partial_\mu (P^q-P^{q'}) + ab_A(m_qA_\mu^q-m_{q'}A_\mu^{q'}) \right], \\
  \left(\sum_q A_\mu^q\right)^{R,I} &=
  \bar Z_A \left[ \sum_qA_\mu^q + a c_A\partial_\mu\sum_q P^q + a b_A \sum_q m_q A_\mu^q \right],
\end{aligned}
\end{equation}
for the flavor nonsinglet and singlet cases, respectively, where $P$
is the pseudoscalar density. Matching with the improvement of the
action, we take the tree-level value $c_A=0$. Note that in nucleon
matrix elements, the term proportional to $c_A$ only contributes to
the $G_P$ form factors and therefore this term is not necessary for
$O(a)$ improvement of $G_A(Q^2)$.\footnote{In practice lattice results
  for $G_A(Q^2)$ could depend on $c_A$ indirectly due to contamination
  from excited states, or from a breakdown of the form factor
  decomposition (\ref{eq:GA_GP}) due to breaking of rotational
  symmetry. The latter can result from either the UV cutoff (an
  $O(a^2)$ effect) or the IR cutoff (suppressed by $e^{-m_\pi L}$).}
The mass-dependent terms can effectively cause a mixing between
singlet and nonsinglet axial currents; rather than determine $b_A$
explicitly, we absorb the mass-dependent terms into the
renormalization factors, which now become a matrix. The
renormalization matrix is determined nonperturbatively using the
Rome-Southampton method, which we discuss in detail in
Section~\ref{sec:renormalization}.

\subsection{\label{sec:HP}Effectiveness of hierarchical probing}

\begin{figure*}[!h]
\begin{center}
 \includegraphics[width=0.4\linewidth]{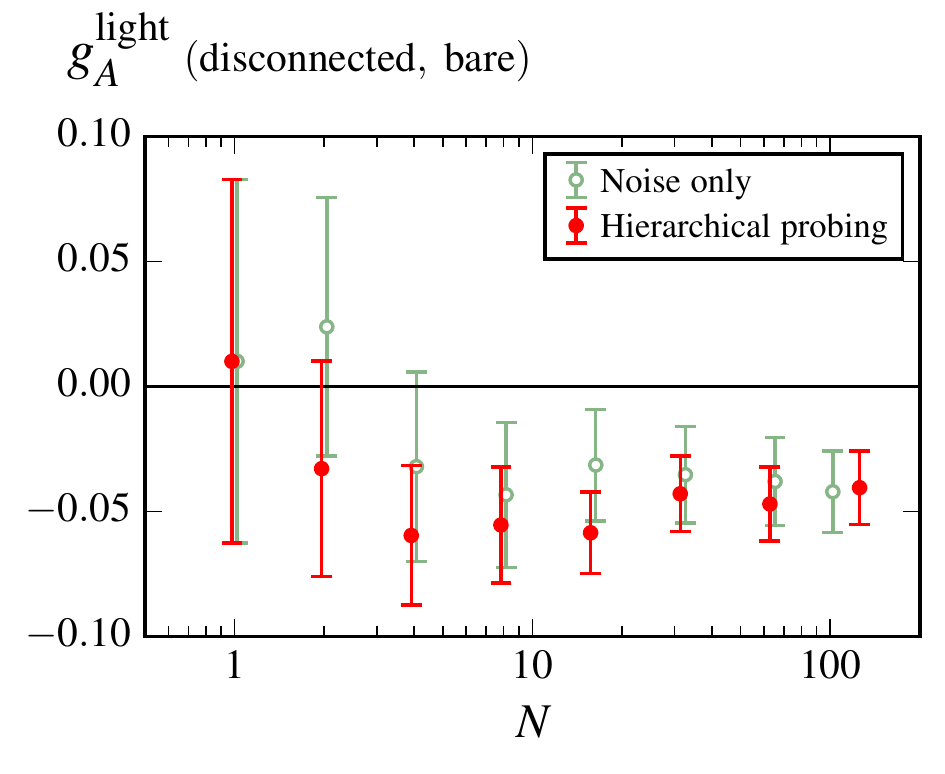} \hspace{0.05\linewidth} \includegraphics[width=0.4\linewidth]{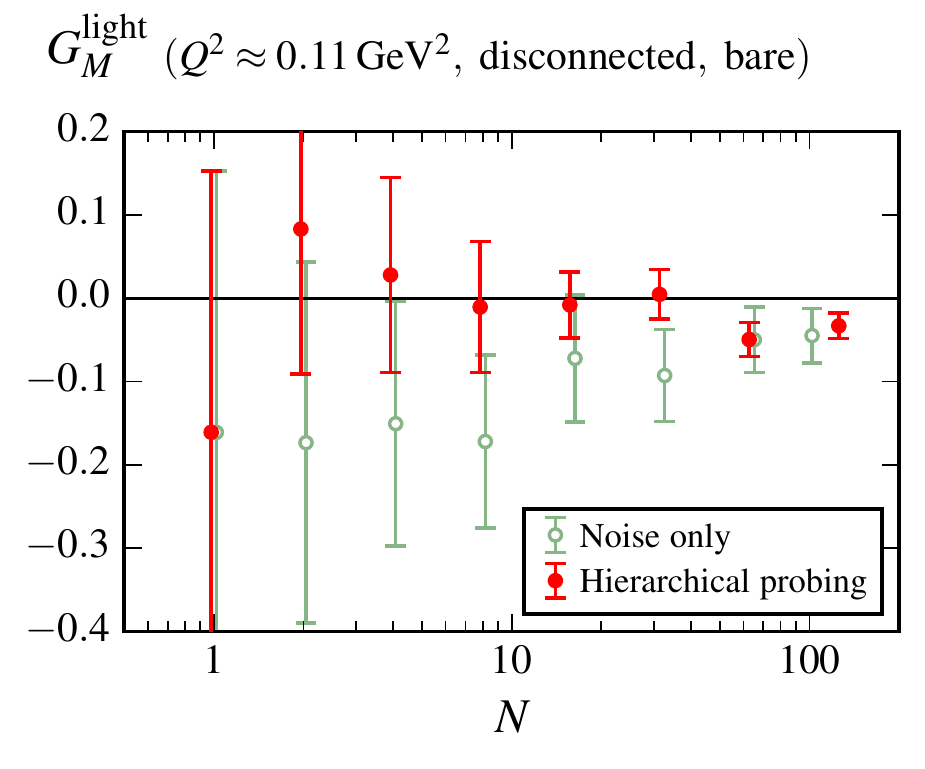}
\end{center}
\caption{\label{fig:HPvsnoise}Comparison of hierarchical probing to the ``Noise only'' method at equal computational cost,
using a reduced set of 366 configurations where we have data from both methods. The plots show results
for the disconnected light-quark $g_A$ (left) and disconnected light-quark magnetic form factor $G_M(Q^2\approx0.11\:\:{\rm GeV}^2)$ (right)
from the ratio method at $T/a=10$, $\tau/a=5$. The results are plotted as a function of $N$, which denotes the number of noise samples or the number of Hadamard vectors
used to estimate each quark loop. Data points (slightly offset horizontally for clarity) are shown for $N=1,2,4,8,16,32,64$ (both methods), $N=100$ (noise only), and $N=128$ (hierarchical probing). }
\end{figure*}

On a reduced set of 366 configurations, we have data for the disconnected light-quark loops from two different methods: hierarchical probing, as used
for the main calculations of this work, and ``Noise only'', where the sum over $n$ in Eq.~(\ref{eq:HP}) is over $N$ random noise samples rather than $N_{\rm hvec}$ Hadamard vectors multiplying
a single noise sample. Note that this means color and spin dilution is used in both cases. Thus, at $N=N_{\rm hvec}$ the computational cost for both methods is the same. Figure \ref{fig:HPvsnoise} shows results from both methods
as a function of $N=N_{\rm hvec}$.  Hierarchical probing is always guaranteed to perform at least as well as the traditional noise method. For our setup we find that the uncertainty in the disconnected light-quark $g_A$ saturates at $N_{\rm hvec}=16$, where it becomes
dominated by gauge noise. For $g_A$ with $N=N_{\rm hvec}=16$,
the reduction in the (combined gauge+stochastic) uncertainty is only by a modest factor of 1.4. The improvement from hierarchical probing is more significant for the disconnected electromagnetic
form factors \cite{Green:2015wqa}, as illustrated in Fig. \ref{fig:HPvsnoise} (right) for the disconnected light-quark contribution to $G_M$ at $Q^2\approx 0.11\:\:{\rm GeV}^2$. In this
case, the stochastic noise dominates over the gauge noise up to a larger value of $N$ (saturation is not yet reached in the range considered), and at large $N$ the improvement from hierarchical probing is more
pronounced, as expected because of the greater ``coloring distance'' \cite{Stathopoulos:2013aci}.

\subsection{Excited-state effects}

It turns out that the different form factors suffer from quite
different amounts of excited-state contamination. In addition, the
available $(T,\tau)$ combinations are quite different between our
connected-diagrams data and our disconnected-diagrams data. In
particular, the former are much better suited for applying the
summation method than the latter. Therefore we choose the best method
for isolating the ground state separately for each form factor. We do
this by examining ``plateau'' plots where, for each $(T,\tau)$ we
determine ``effective'' form factors\footnote{In this subsection we
  show bare form factors, i.e.\ before renormalization.} from the ratios
assuming the absence of excited states. In a region where
excited-state effects are negligible, these effective form factors
will form a stable plateau. In addition to these plateaus from the
ratio method, we also show results from the summation method, taking
the sums with three adjacent points $\{T,T+2a,T+4a\}$ and fitting with
a line to determine the slope.

\begin{figure*}
  \centering
  \includegraphics[width=0.49\textwidth]{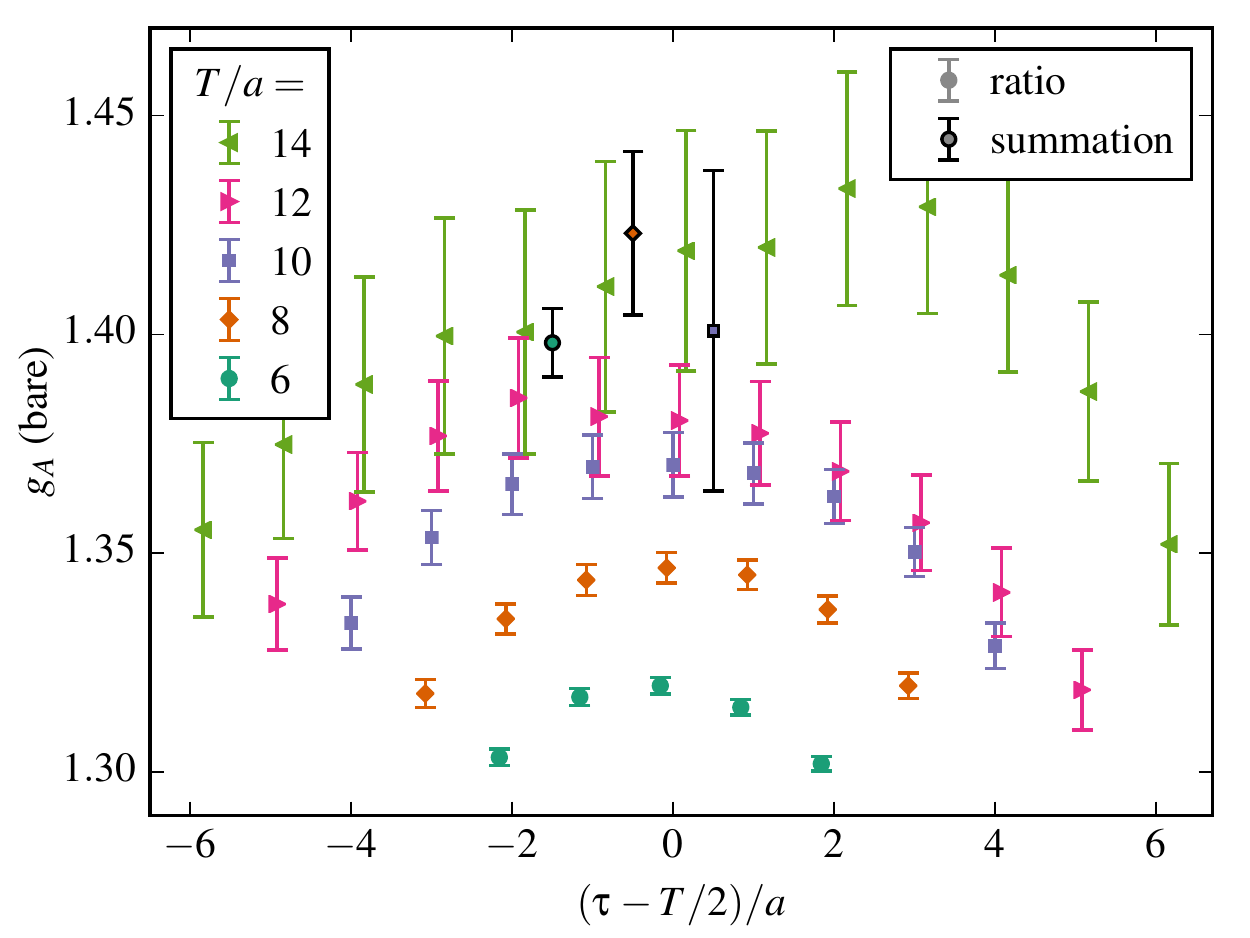}
  \includegraphics[width=0.49\textwidth]{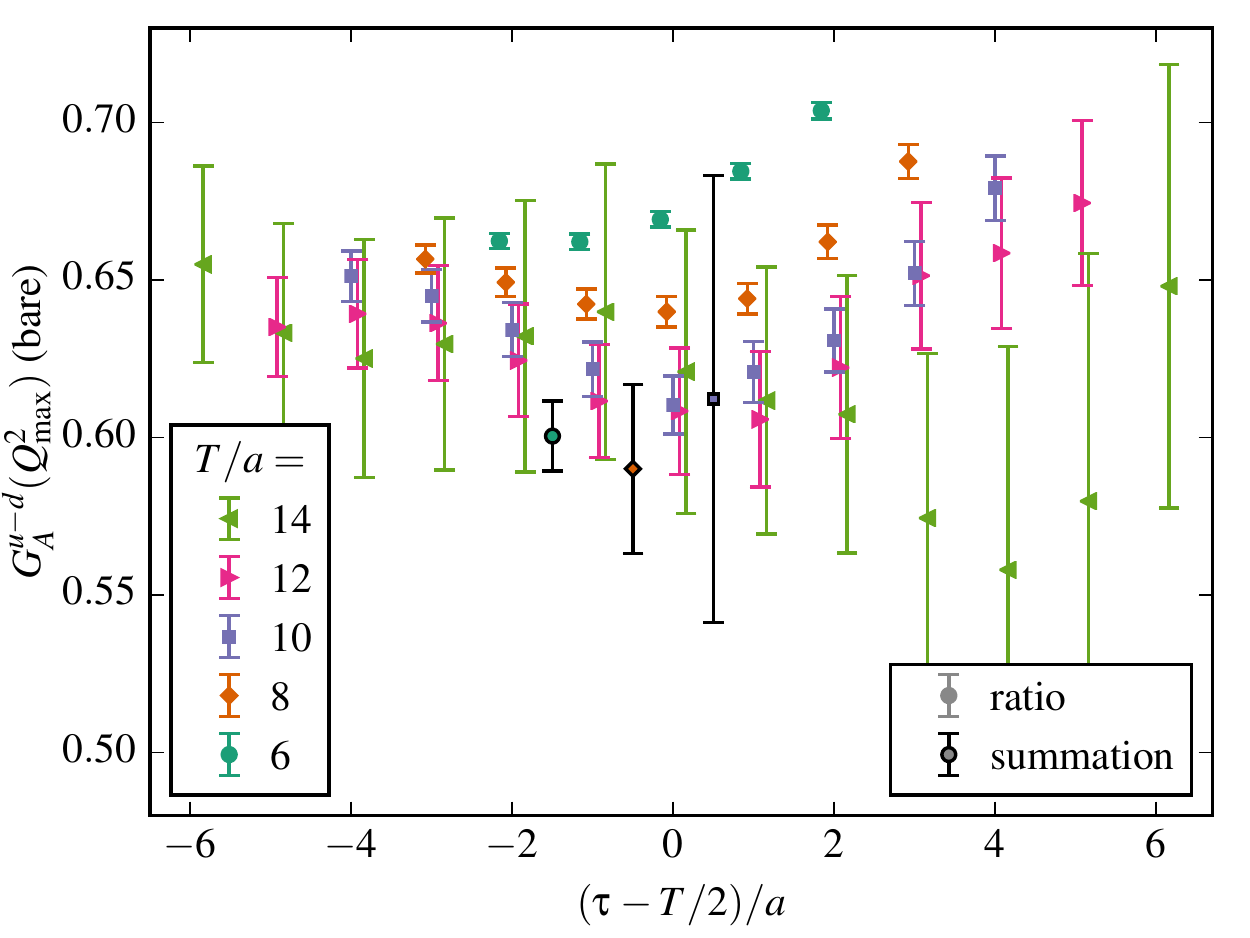}\\
  \includegraphics[width=0.49\textwidth]{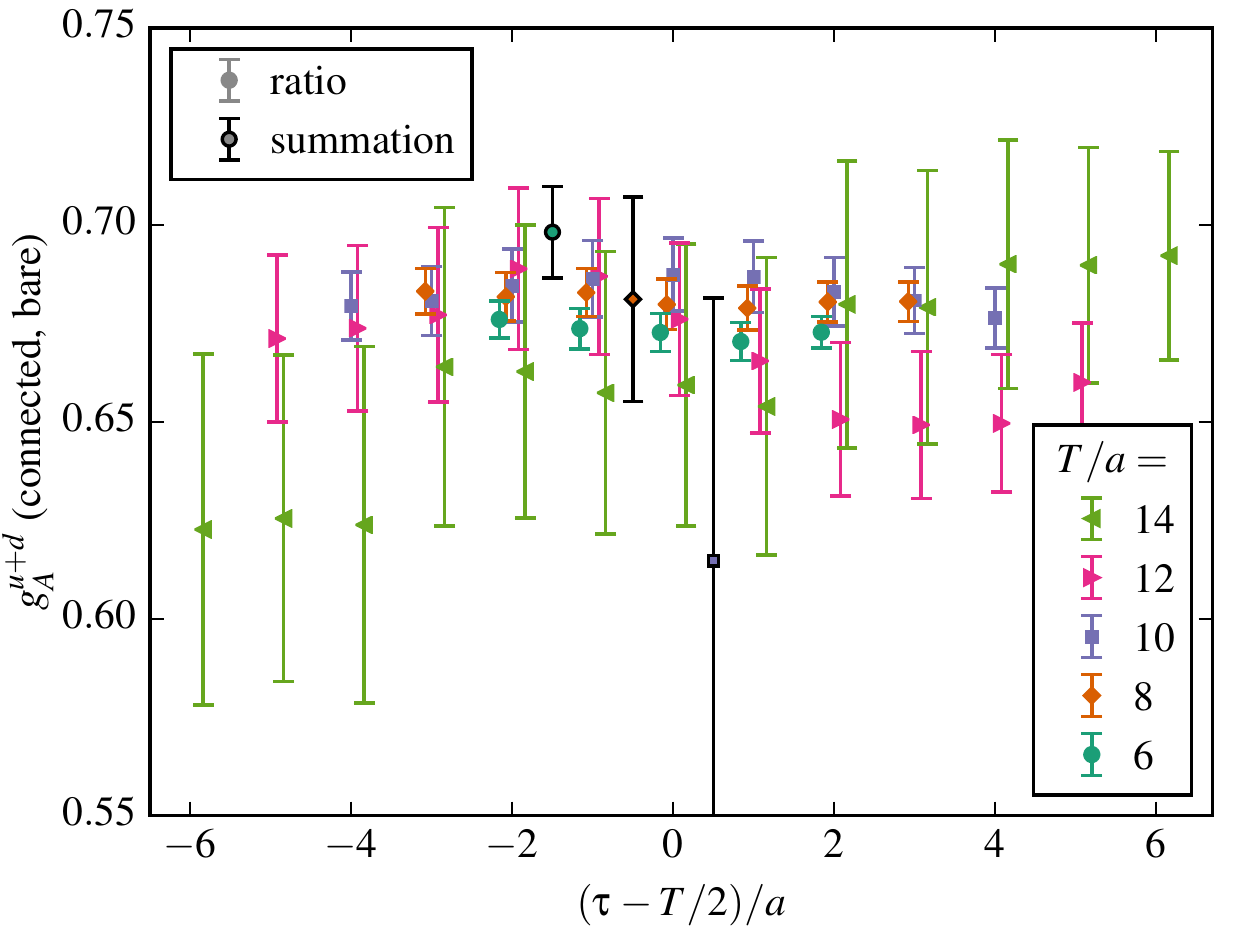}
  \includegraphics[width=0.49\textwidth]{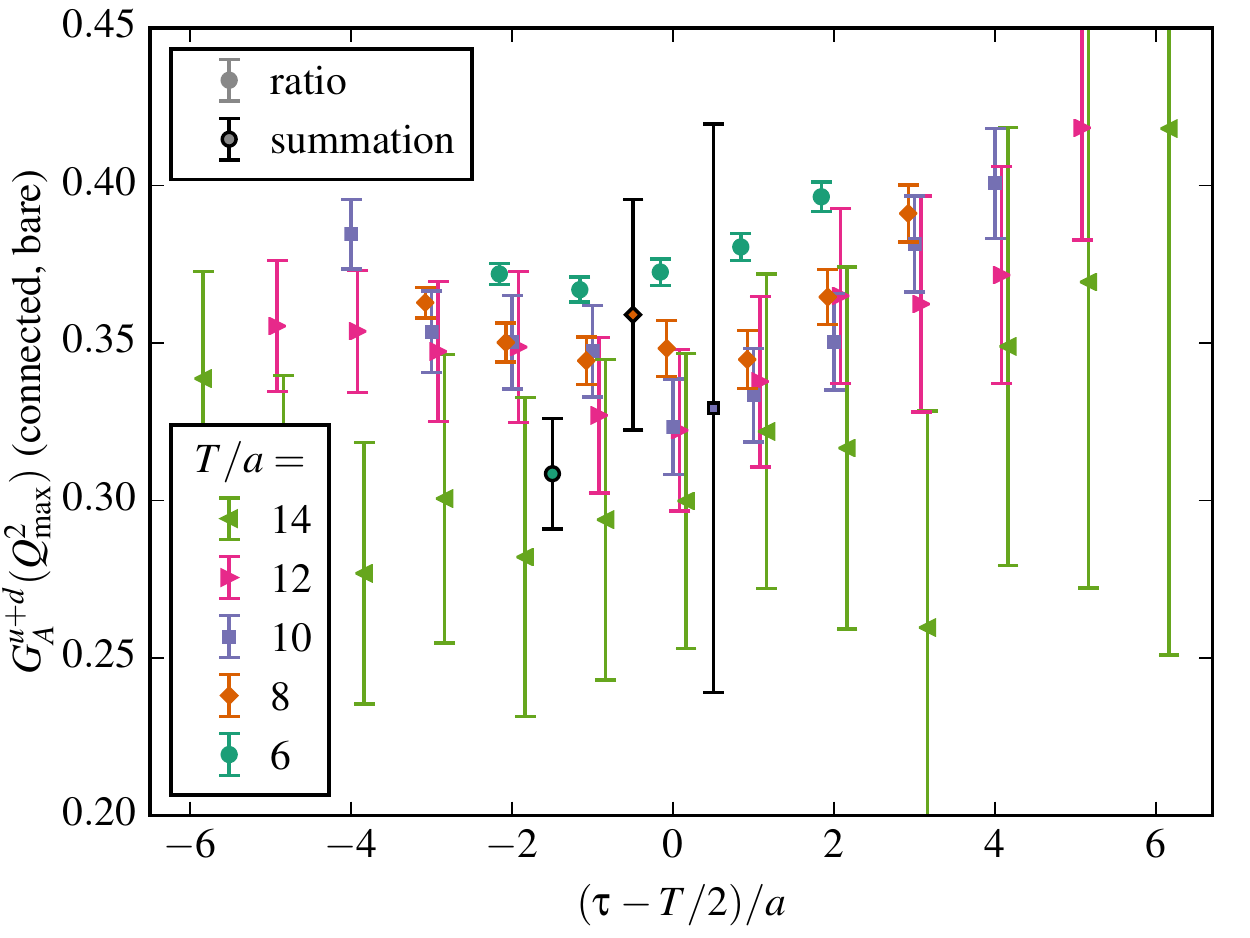}
  \caption{Plateau plots for the bare isovector (top row) and
    connected isoscalar (bottom row) axial form factors at zero (left
    column) and the highest (right column) momentum transfer
    $Q^2$. Solid symbols indicate data computed using the ratio method.
    Symbols with black outlines and black error bars indicate data
    from the summation method and are plotted in open spaces between
    ratio data near the origin for clarity.}
  \label{fig:connected_GA_plateaus}
\end{figure*}

Figure~\ref{fig:connected_GA_plateaus} (top row) shows plateau plots
for the isovector axial form factor $G_A^{u-d}(Q^2)$. For the axial
charge $g_A\equiv G_A^{u-d}(0)$ (top left), the centers of the plateaus
appear stable by $T/a=10$ and 12, which agree within uncertainty. The
center of the plateau for the largest source-sink separation, $T=14a$,
is shifted significantly higher, however its statistical uncertainty
is quite large and the magnitude of the shift goes against
expectations: in the asymptotic regime, as $T$ is increased the shift
between neighboring values of $T$ is expected to decrease. Therefore
we conclude that the shift at $T=14a$ is likely a statistical
fluctuation\footnote{Similar behavior was previously seen in the
  isovector Pauli form factor computed using the same
  dataset~\cite{Green:2013hja}.}
and take the results from $T=12a$ as the best option using
the ratio method. For the summation method, all three points are
consistent within the uncertainty and we conclude that the summation
method has reached a plateau already at the shortest source-sink
separation, $T=6a$ (i.e., from fitting to the sums with
$T/a\in\{6,8,10\}$). We take this as our primary analysis method for
the isovector axial form factor $G_A^{u-d}(Q^2)$. For this form factor
and for any observable derived from it, we estimate systematic
uncertainty due to excited-state effects as the root-mean-square (RMS)
deviation between the primary result (summation with $T=6a$) and two
alternatives: the ratio method with $T=12a$ and the summation method
with $T=8a$. Looking at the corresponding plateau plot (top right) for
the isovector axial form factor at our largest momentum transfer
(about 1.1~GeV$^2$) indicates that this approach is also reasonable at
nonzero $Q^2$. The bottom row of the same figure shows the equivalent
plots for the contribution from quark-connected diagrams to the
isoscalar axial form factor $G_A^{u+d}(Q^2)$. The excited-state
effects appear to be slightly milder than for the isovector case, and
we thus choose to apply the same analysis strategy.

\begin{figure*}
  \centering
  \includegraphics[width=0.49\textwidth]{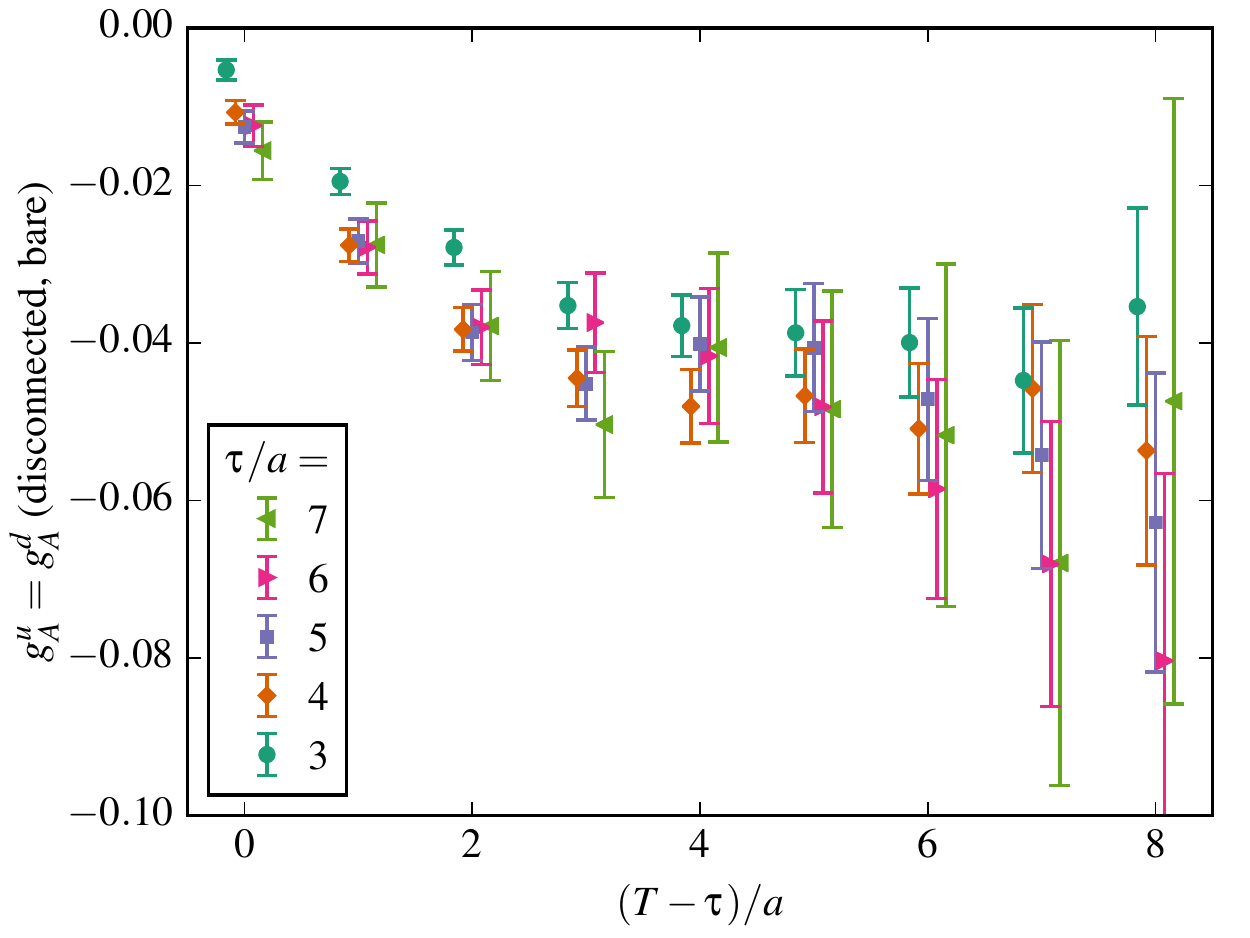}
  \includegraphics[width=0.49\textwidth]{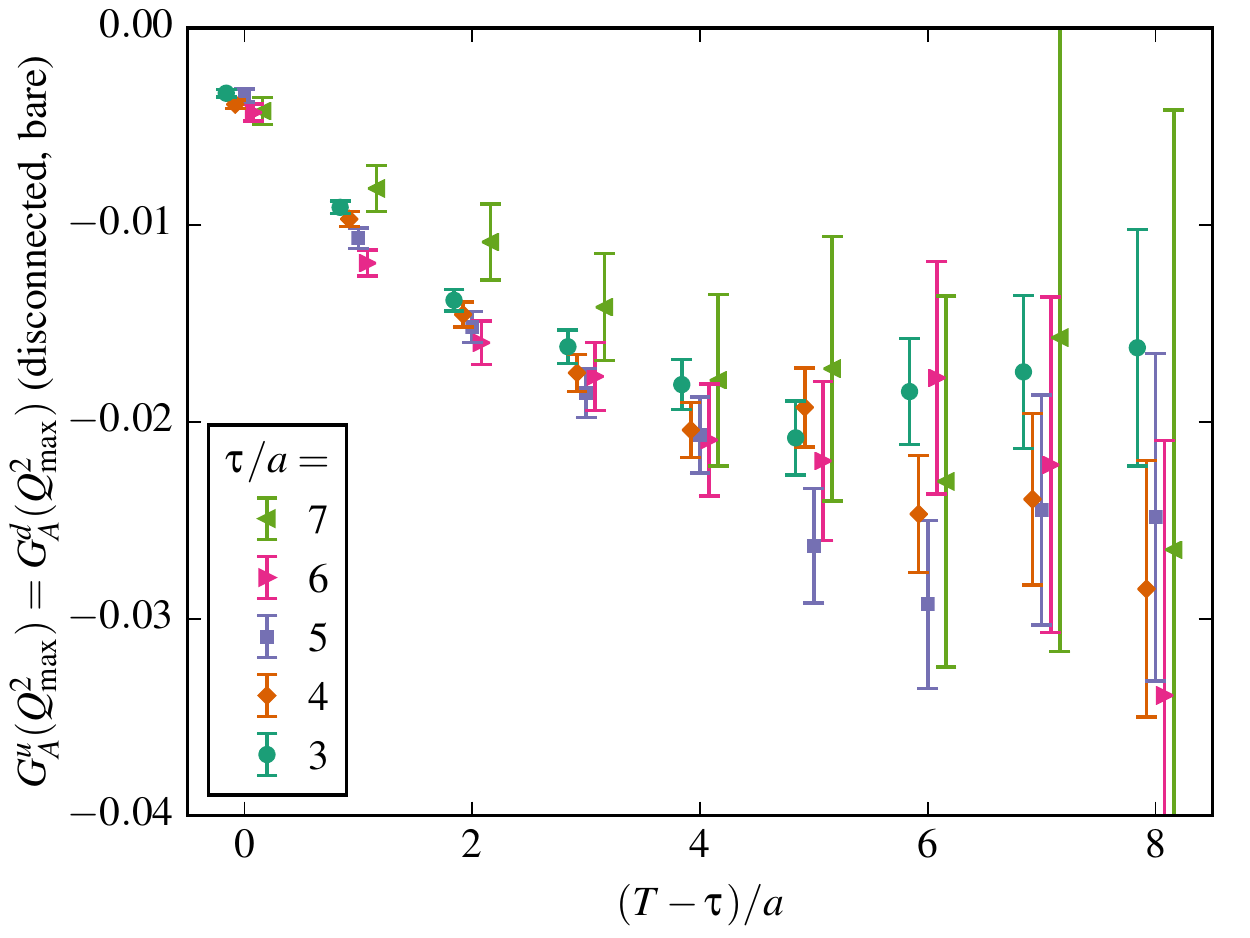}\\
  \includegraphics[width=0.49\textwidth]{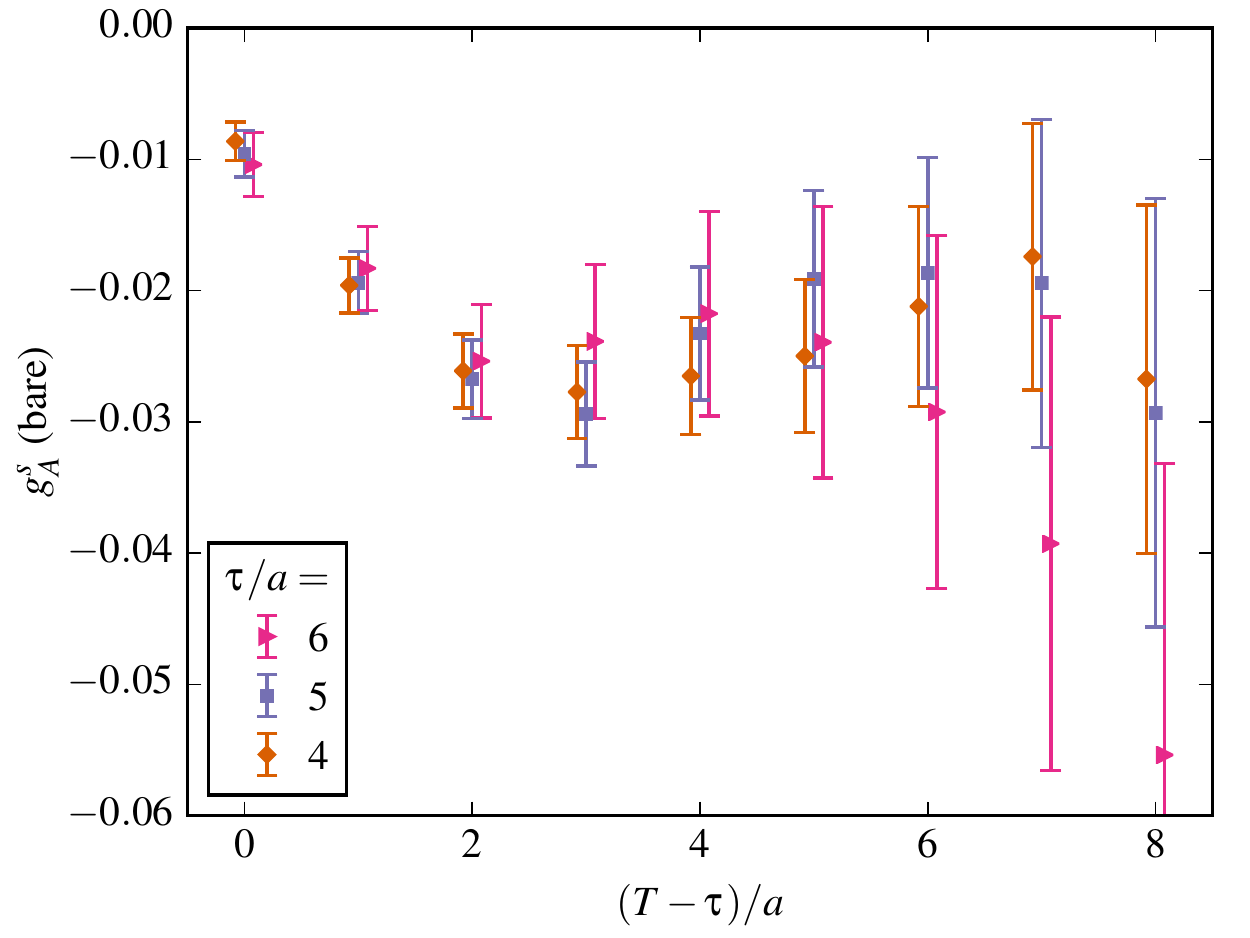}
  \includegraphics[width=0.49\textwidth]{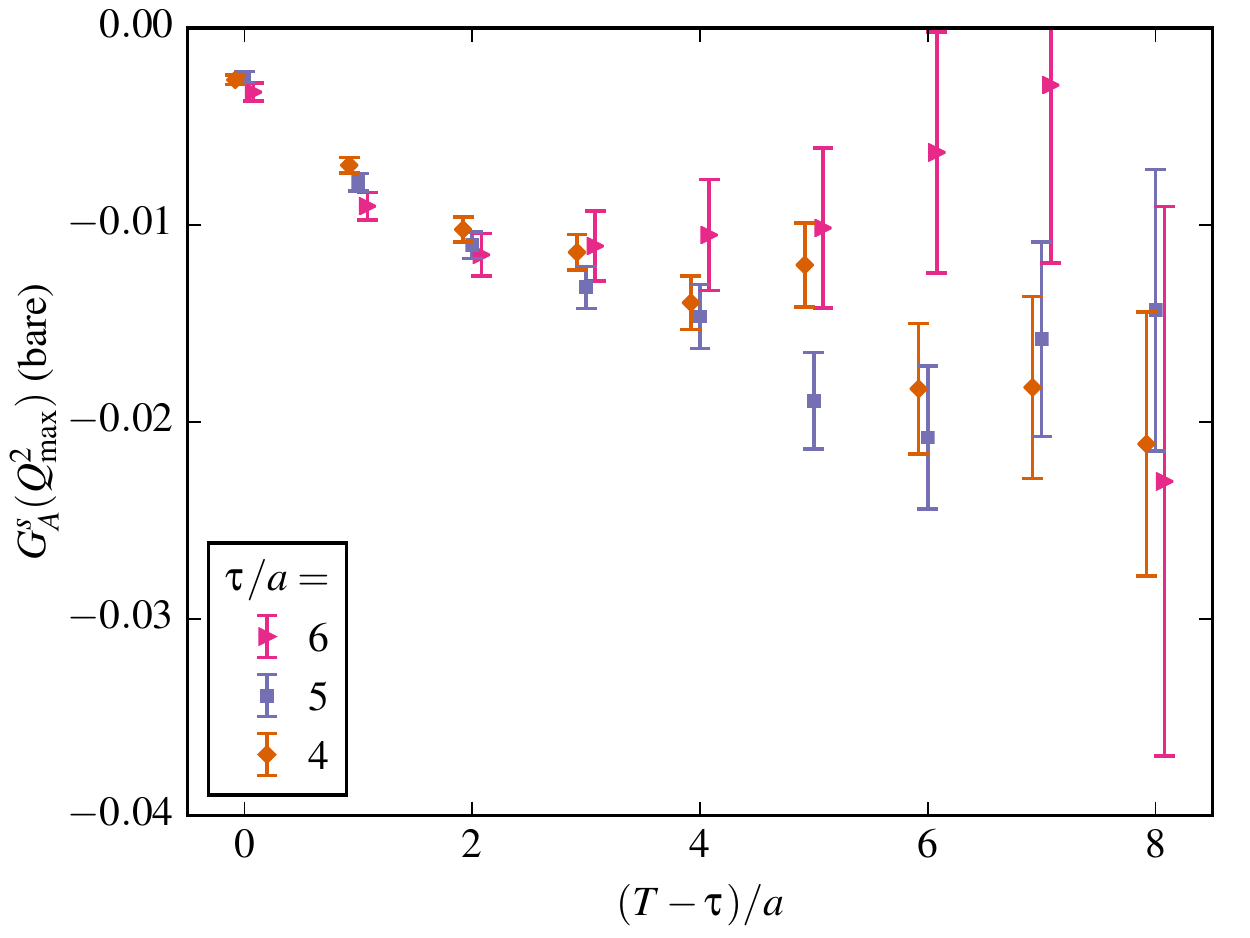}
  \caption{Plateau plots for the bare disconnected light (top row) and
    strange (bottom row) axial form factors at zero (left column) and
    the highest (right column) momentum transfer $Q^2$.}
  \label{fig:disconnected_GA_plateaus}
\end{figure*}

Plateau plots for the contributions from quark-disconnected diagrams
to axial form factors are shown in
Fig.~\ref{fig:disconnected_GA_plateaus}. Note that since these form
factors were computed for several fixed source-operator separations
$\tau$, we choose to use the operator-sink separation $T-\tau$ as the
horizontal axis. The top row shows the light-quark case, where we
computed disconnected loops for five source-operator separations, and
the bottom row shows the strange-quark case where we only computed
three source-operator separations. The left and right columns show
$Q^2=0$ (i.e., the contributions to the nucleon spin) and our largest
momentum transfer, respectively. In general, we do not see any
significant dependence on $T-\tau$ for $T-\tau \gtrsim 5a$. Since the
disconnected data were averaged over the exchange of source and sink
momenta, the effective form factors are expected to be symmetric, and
therefore this corresponds to a source-sink separation of $T=10a$. We
use this for our primary result (averaged over the three points near
$\tau=T/2$, which reduces statistical uncertainty), and use the RMS
deviation with results from $T=8a$ and $T=12a$ as our estimate of
systematic uncertainty due to excited states.

\begin{figure*}
  \centering
  \includegraphics[width=0.49\textwidth]{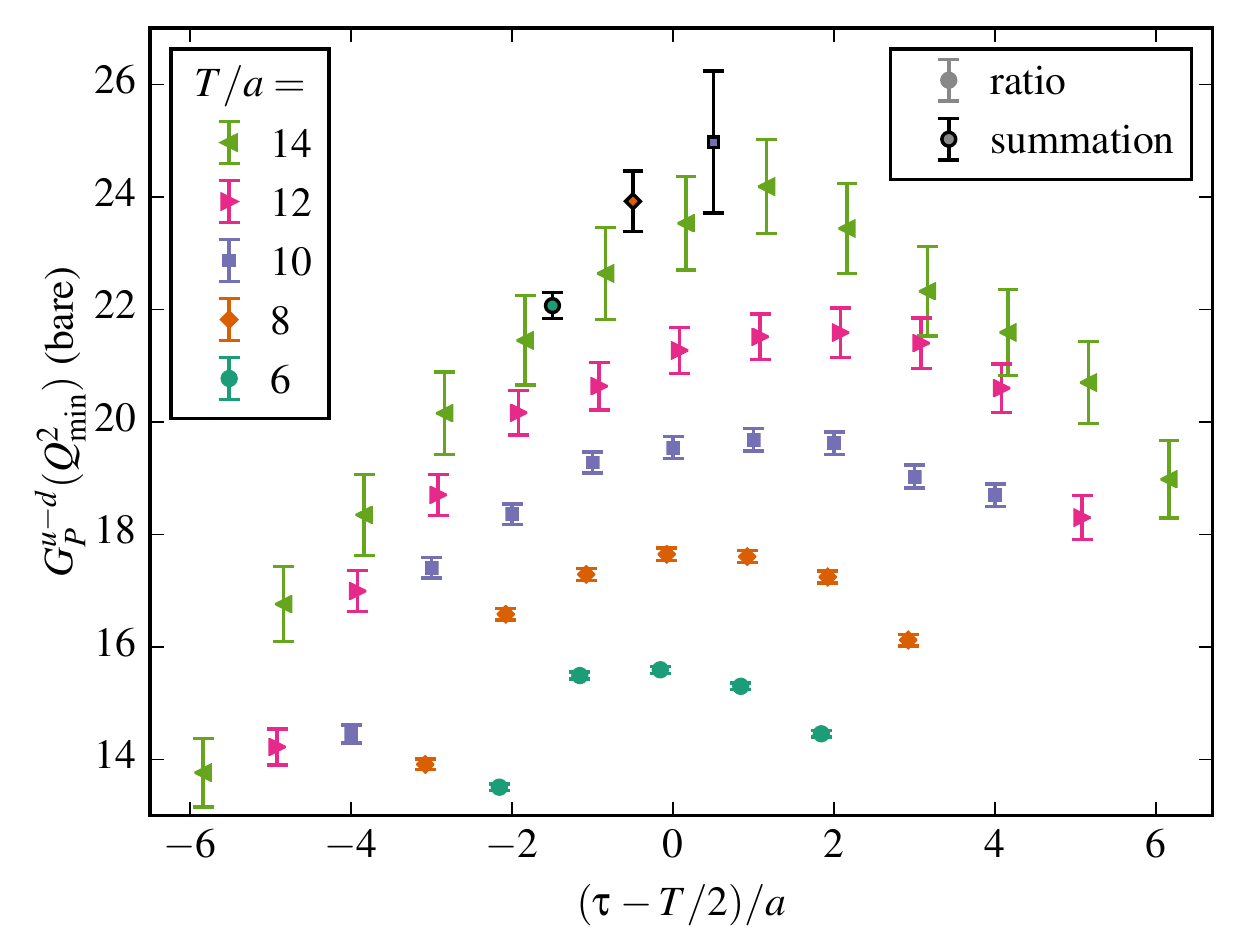}
  \includegraphics[width=0.49\textwidth]{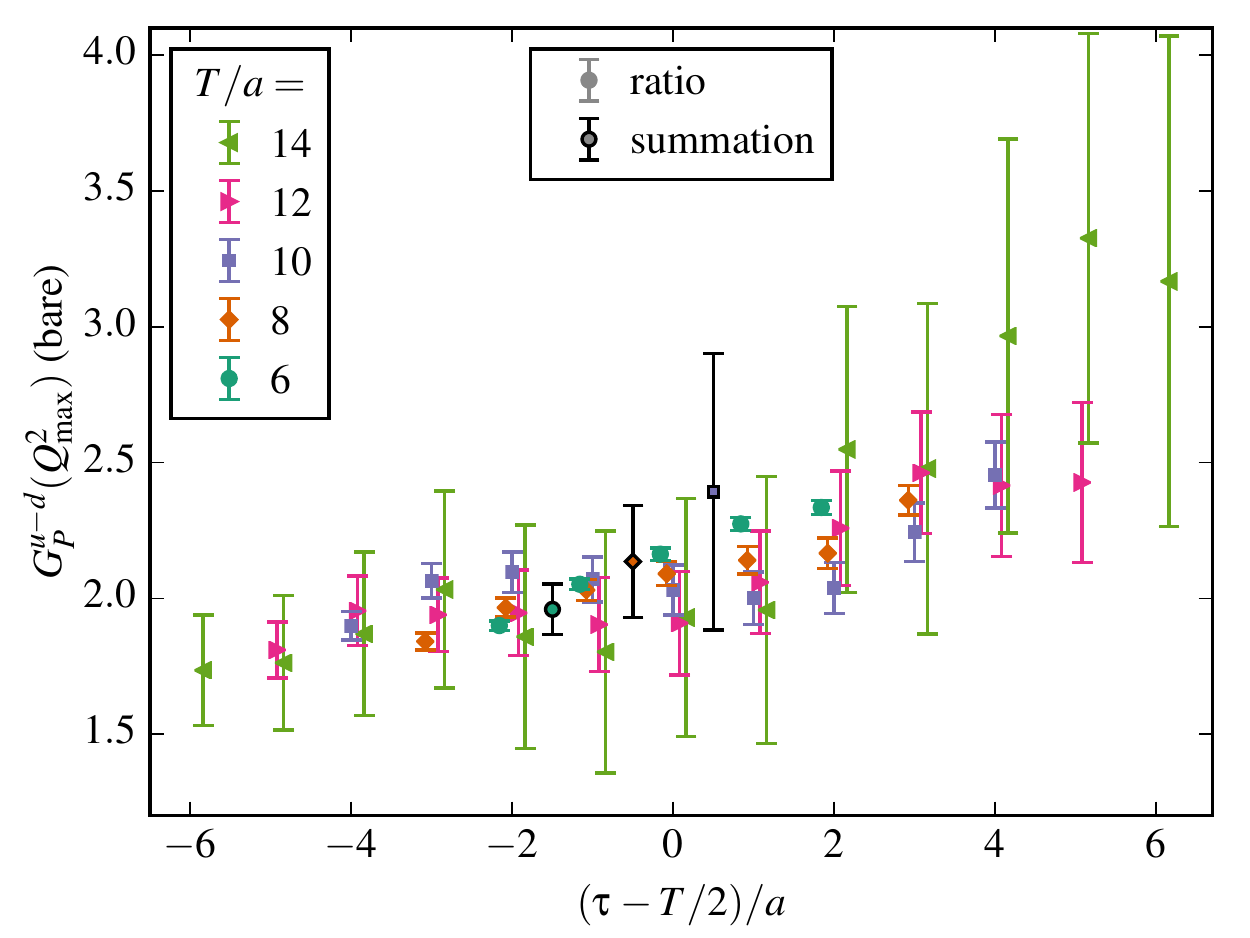}
  \caption{Plateau plots for the bare isovector induced pseudoscalar
    form factor $G_P^{u-d}(Q^2)$ at the lowest (left) and highest
    (right) momentum transfer $Q^2$. Solid symbols indicate data
    computed using the ratio method, and symbols with black outlines
    and black error bars indicate data from the summation method.}
  \label{fig:isovector_GP_plateaus}
\end{figure*}

The isovector induced pseudoscalar form factor $G_P^{u-d}(Q^2)$ at the
lowest available momentum transfer (about 0.1~GeV$^2$) is shown in
Fig.~\ref{fig:isovector_GP_plateaus} (left). This has very large
excited-state effects (there is nearly a factor of two between the
smallest and largest value on the plot), and there is no sign that a
plateau has been reached using the ratio method. For the summation
method, the points with $T/a=8$ and 10 are consistent, suggesting that
a plateau might possibly have been reached. We take the summation
method with $T=8a$ as our primary analysis method for this form factor
and estimate the systematic uncertainty as the RMS deviation between
the primary result and those from the ratio method with $T/a=14$ and
12. Although the latter is clearly not in the plateau regime, we
nevertheless include it in order to reflect the poor control over
excited-state effects that is available in our data. At larger $Q^2$
(right), the excited-state effects are much milder and our
error estimate should be conservative.

\begin{figure*}
  \centering
  \includegraphics[width=0.49\textwidth]{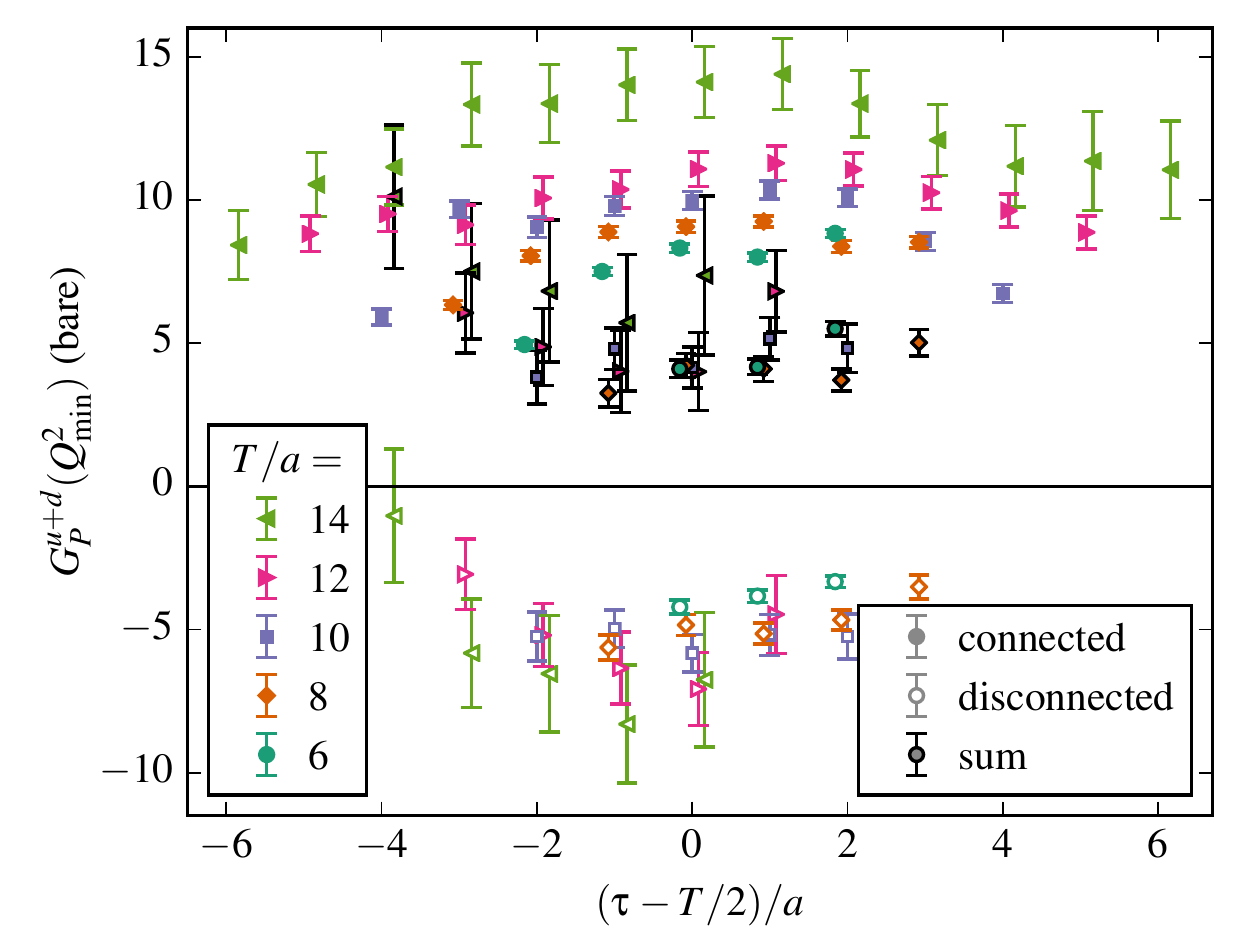}
  \includegraphics[width=0.49\textwidth]{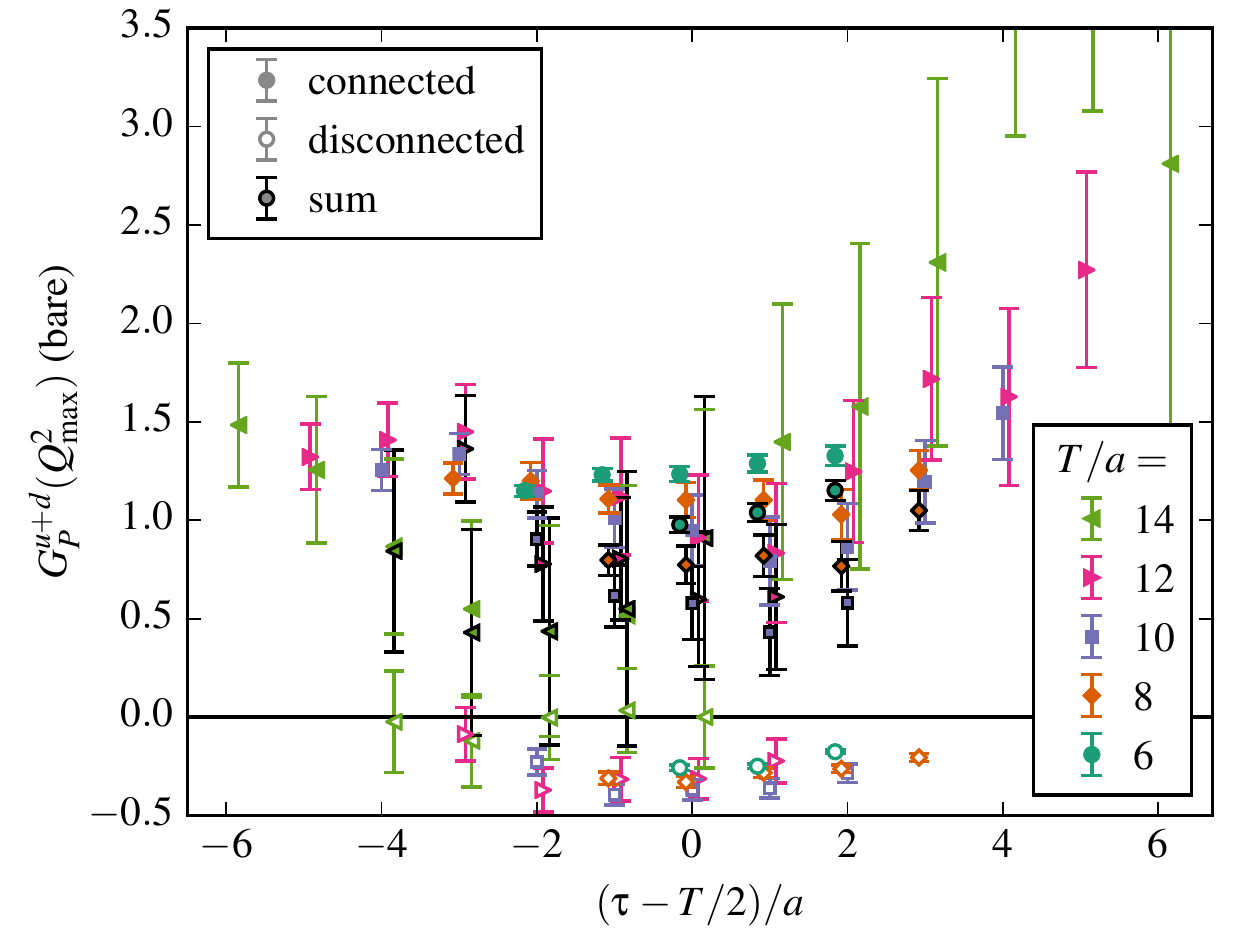}\\
  \includegraphics[width=0.49\textwidth]{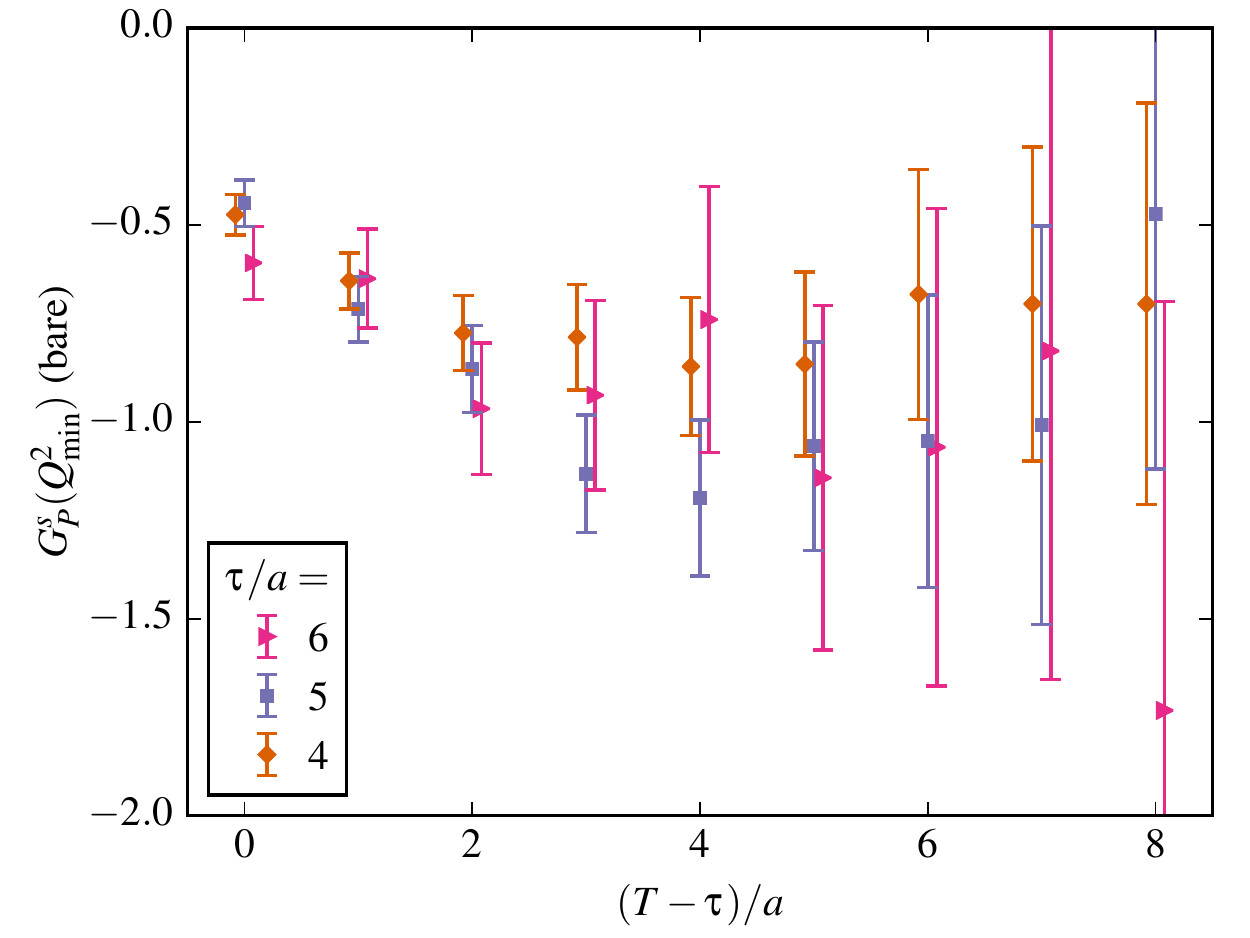}
  \includegraphics[width=0.49\textwidth]{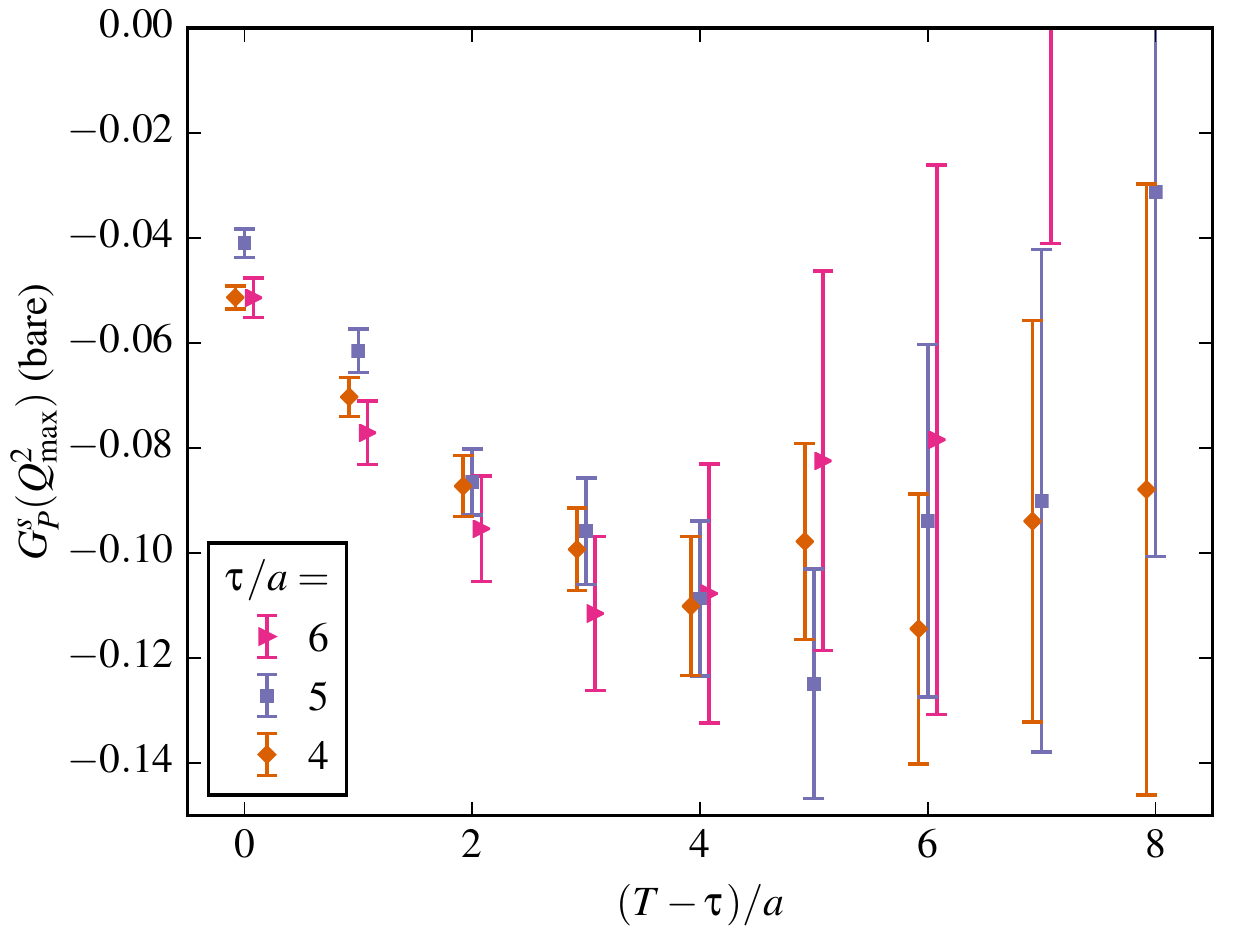}
  \caption{Plateau plots for the bare isoscalar light (top row) and
    strange (bottom row) induced pseudoscalar form factors at the
    lowest (left) and highest (right) momentum transfer $Q^2$. In the
    top row: solid and open symbols indicate the contributions from
    connected and disconnected diagrams, respectively, and symbols
    with black outlines and black error bars indicate their sum.}
  \label{fig:isoscalar_GP_plateaus}
\end{figure*}

Plateau plots for the light and strange isoscalar induced pseudoscalar
form factors are shown in Fig.~\ref{fig:isoscalar_GP_plateaus}. For
$G_P^{u+d}(Q^2)$ at the lowest available momentum transfer (top left),
we again find that the connected contributions have significant
excited-state effects. On the same plot, we show the partial plateaus
(limited to the available values of $\tau$) for the contributions from
disconnected diagrams. Although they are a bit noisier, they also
appear to contain excited-state effects, with the opposite sign. In
fact, the opposite signs cause the sum of connected and disconnected
diagrams to have smaller excited-state contamination. For the sum,
using the ratio method with $T=10a$ appears to be a safe choice, also
at the maximum momentum transfer (top right). When we examine the
individual connected and disconnected contributions, we will make the
same choice, with the understanding that the results include some
contamination from excited states, and can only be studied
qualitatively. This choice also appears safe for $G_P^s(Q^2)$ (bottom
left and right). As for the disconnected $G_A$ form factors, we use
the RMS difference with $T/a=8$ and 12 as our estimate of systematic
uncertainty due to excited states.

\subsection{\label{sec:zexp}Form factor fits using the $z$ expansion}

Having computed nucleon form factors at several discrete values of
$Q^2$, we fit them with curves to characterize their overall shape and
determine observables such as the axial radius from their slope at
$Q^2=0$. It has been common to perform these fits using simple
ansatzes, such as a dipole, which is often used to describe
experimental data for the isovector $G_A(Q^2)$, however these tend to
be highly constrained and introduce a model dependence into the
results.

Instead, we use the model-independent $z$ expansion. This was used in
Refs.~\cite{Bhattacharya:2011ah,Bhattacharya:2015mpa,Meyer:2016oeg}
to study axial
form factors determined from quasielastic (anti)neutrino-nucleon
scattering; it was found that fitting with the $z$ expansion produced
a significantly larger axial radius with a larger uncertainty,
compared with dipole fits. The $z$ expansion makes use of a conformal
mapping from $Q^2$, where the given form factor is analytic on the
complex plane outside a branch cut on the timelike real axis, to the
variable $z$ such that the form factor is analytic for $|z|<1$. We use
\begin{equation}\label{eq:z_Q2}
  z(Q^2) = \frac{\sqrt{t_\text{cut}+Q^2}-\sqrt{t_\text{cut}}}{\sqrt{t_\text{cut}+Q^2}+\sqrt{t_\text{cut}}},
\end{equation}
where we use the particle production threshold for the isovector
form factors, $t_\text{cut}=(3m_\pi)^2$. For the isoscalar form
factors the actual threshold may be higher, but we use the same
$t_\text{cut}$ everywhere for simplicity. We have chosen the mapping
such that $Q^2=0$ maps to $z=0$.

The $G_P$ form factors have an isolated pole below the particle
production threshold at the pseudoscalar meson mass, which we remove
before fitting. We thus perform fits to
\begin{equation}
  G(Q^2) = \begin{cases}
    G_A(Q^2) & \\
    (Q^2+m_\pi^2)G_P(Q^2) & \text{isovector}\\
    (Q^2+m_\eta^2)G_P(Q^2) & \text{isoscalar}
\end{cases}.
\end{equation}
Each form factor can be described by a convergent Taylor series in
$z$. We truncate this series and obtain our fit form,
\begin{equation}
  G(Q^2) = \sum_{k=0}^{k_\text{max}} a_k z(Q^2)^k.
\end{equation}
The first two coefficients, $a_0$ and $a_1$, give the intercept and
slope of the form factor at $Q^2=0$. Specifically, $G(0)=a_0$ and,
for the axial form factors, $r_A^2=-3a_1/(2a_0t_\text{cut})$.
We impose Gaussian priors on the
remaining coefficients, centered at zero with width equal to
$w=5\max\{|a_0|,|a_1|\}$. The series is truncated with $k_\text{max}=5$,
but this is large enough that increasing it further has no effect in
our probed range of $Q^2$; i.e., the priors cause $a_kz^k$ to be
negligible for $k>5$.

We perform correlated fits, minimizing
\begin{equation}
  \chi^2_\text{aug} \equiv \sum_{i,j}\left(G(Q^2_i)-\sum_k a_kz(Q^2_i)^k\right) \Xi_{ij}\left(G(Q^2_j)-\sum_{k'}a_{k'}z(Q^2_j)^{k'}\right)+\sum_{k>1}\frac{a_k^2}{w^2}
\end{equation}
with respect to $\{a_k\}$, where $\Xi$ is an estimator for the
inverse covariance matrix and the last term augments the chi-squared
with the Gaussian priors. With limited statistics it can be difficult
to obtain a reliable estimator, and therefore we choose to reduce
statistical fluctuations by interpolating between the jackknife
estimate of the covariance matrix and a simplified (less noisy but
biased) estimate, and then inverting the resulting matrix. This is in
the spirit of shrinkage estimators~\cite{Ledoit_2004,Schaefer_2005},
however we do not perform an optimization step with respect to the
interpolation parameter.

\begin{figure*}
  \centering
  \includegraphics[width=\textwidth]{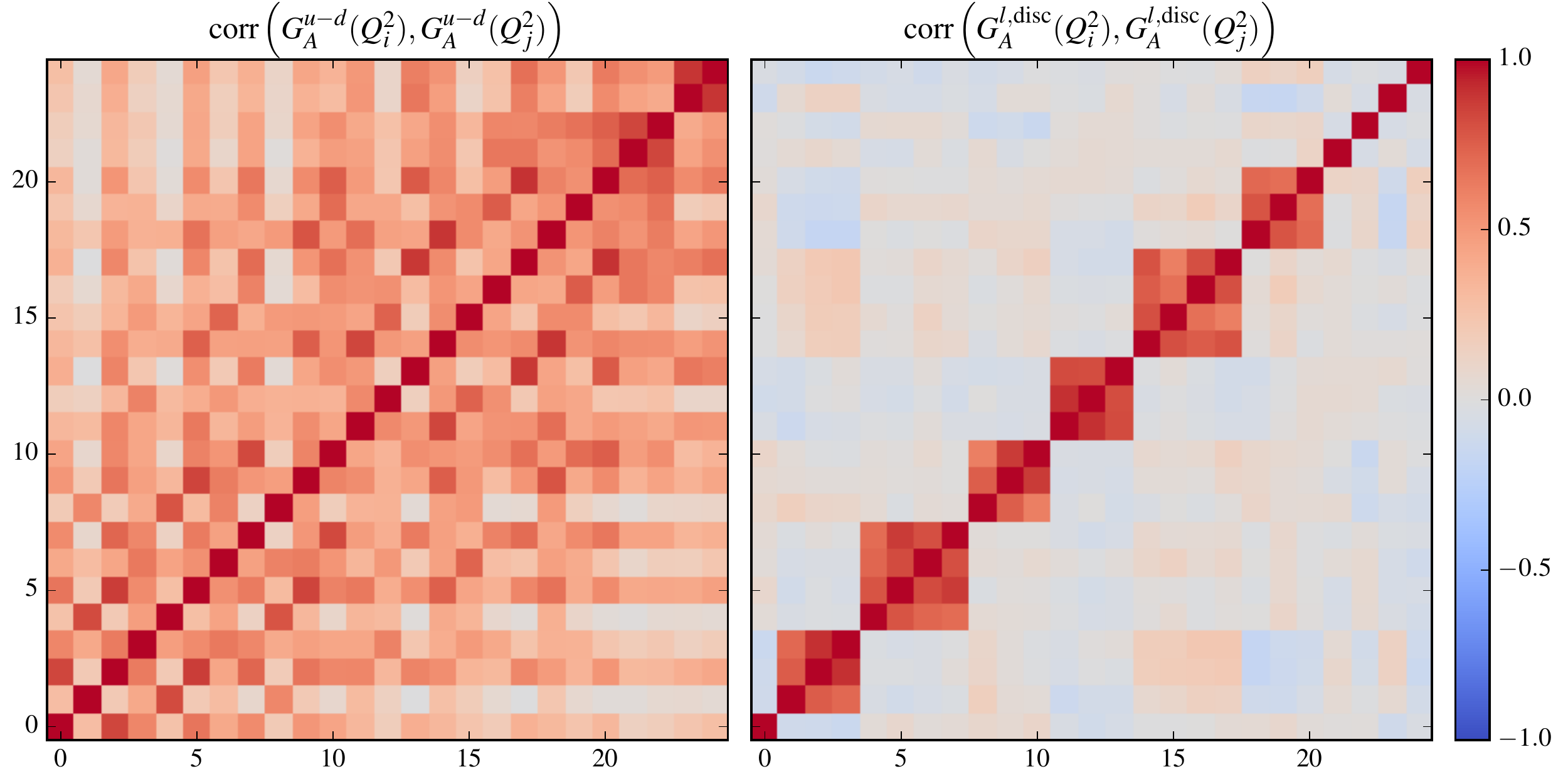}
  \caption{Correlations between data at different $Q^2$. Left: the
    isovector axial form factor $G_A^{u-d}(Q^2)$. Right: the
    quark-disconnected contribution to the light-quark axial form
    factor $G_A^{l,\text{disc}}(Q^2)$. The axes index the different
    momentum transfers, which are sorted in order of increasing
    $Q^2$.}
  \label{fig:correlation_GA}
\end{figure*}

In order to choose the form of the target (simplified) covariance
matrix, we examine the correlation matrix
\begin{equation}
  R_{ij} \equiv \frac{C_{ij}}{\sqrt{C_{ii}C_{jj}}},
\end{equation}
where $C$ is the jackknife estimate of the covariance matrix.  We find
that this has a quite different form between connected diagrams and
disconnected diagrams. Figure~\ref{fig:correlation_GA} shows two
example correlation matrices. For connected diagrams, illustrated with
$G_A^{u-d}(Q^2)$ (left), we find modest correlations between different
values of $Q^2$ but no strong pattern. For disconnected diagrams,
illustrated with the quark-disconnected contribution to the
light-quark $G_A$ form factor (right), the correlation matrix is
nearly block-diagonal. Each block corresponds to values of $Q^2$ that
share the same spatial momentum transfer $(\vec p\,'-\vec p)^2$ and
thus the same Fourier modes of the disconnected loops. There are
strong correlations within each block but weak correlations between
different blocks.

For connected diagrams, we set $\Xi=\left((1-\lambda)C+\lambda
  C_\text{diag}\right)^{-1}$, where $C_\text{diag}$ is the diagonal
part of the covariance matrix. This is equivalent to multiplying the
off-diagonal elements of $C$ by $1-\lambda$. We use the mild value of
$\lambda=0.1$ as our main choice. For disconnected diagrams, we
compute the average $r$ over all elements of $R_{ij}$ where $i$ and
$j$ ($i\neq j$) correspond to the same spatial momentum transfer. We
then use for $\Xi$ the inverse of the matrix
$R^\star_{ij}\sqrt{C_{ii}C_{jj}}$, where
\begin{equation}
  R^\star_{ij} = \begin{cases}
    1 & i=j \\
    (1-\lambda_1)R_{ij} & \text{$i$ and $j$ have different $(\vec p\,'-\vec p)^2$}\\
    (1-\lambda_2)R_{ij}+\lambda_2 r & \text{$i$ and $j$ have the same $(\vec p\,'-\vec p)^2$}
    \end{cases}.
\end{equation}
As our main choice, we use $(\lambda_1,\lambda_2)=(1,\tfrac{1}{2})$.

To estimate systematic uncertainty from fitting, we perform several
alternative fits. We halve the value of $w$. For connected diagrams,
we perform fits with $\lambda=0$ and 1. For disconnected diagrams, we
perform fits with $(\lambda_1,\lambda_2)=(0,0)$, $(1,0)$, and
$(1,1)$. Finally, we take the RMS difference between results from all
of the alternative fits as our estimate.

\section{\label{sec:renormalization}Renormalization}

\newcommand{\MSbar}{\overline{\text{MS}}}
To compare our results with phenomenology, the lattice axial
current needs to be renormalized. We determine the necessary
renormalization factors nonperturbatively using the Rome-Southampton
approach~\cite{Martinelli:1994ty}. Going beyond the usual computation
of the flavor nonsinglet renormalization factor, we also renormalize
the flavor singlet axial current nonperturbatively. This requires
disconnected quark loops but we are able to reuse the same loops that
were computed for nucleon three-point functions. Since we perform
these calculations on just one ensemble without taking the chiral
limit, we effectively absorb the mass-dependent operator improvement
terms into the renormalization (see Subsec.~\ref{sec:ensemble}), which
requires us to determine a matrix of renormalization factors.

The singlet-nonsinglet difference in axial renormalization factors has
been previously studied nonperturbatively by
QCDSF~\cite{Chambers:2014pea} at the $SU(3)$ flavor symmetric point,
using additional lattice ensembles and the Feynman-Hellmann relation
to determine the contributions from disconnected quark loops. For the
case of two degenerate quark flavors, nonperturbative results were
presented by RQCD at the Lattice 2016
conference~\cite{Bali:2017jyw}, using stochastic estimation for
the disconnected loops similarly to this work. The singlet-nonsinglet
difference has also been studied at leading (two-loop) order in
lattice perturbation theory for a variety of improved Wilson-type
actions~\cite{Skouroupathis:2008mf,Constantinou:2016ieh}.

This section is organized as follows: we present the Rome-Southampton
method and the RI$'$-MOM and RI-SMOM schemes for the single-flavor
case in Subsec.~\ref{RSM}, determine the light and strange vector
current renormalization factors in Subsec.~\ref{zv_3pt}, study
discretization effects and breaking of rotational symmetry in
Subsec.~\ref{sec:ptsrc}, and discuss issues of matching to the
$\MSbar$ scheme and running of the flavor singlet axial current in
Subsec.~\ref{sec:running}. Subsections~\ref{RAC} and \ref{VS} explain
our procedure for calculating the $Z_A$ renormalization matrix, and
finally we give the details of the calculation and its results in
Subsec.~\ref{RS}.
\subsection{Rome-Southampton method, RI$'$-MOM, and RI-SMOM}
\label{RSM}
For calculating the axial renormalization constants, we follow the
Rome-Southampton approach in both RI$'$-MOM~\cite{Martinelli:1994ty,
  Gockeler:1998ye} and RI-SMOM schemes~\cite{Sturm:2009kb}. 
In Landau gauge, we compute quark propagators
\begin{equation}\label{prop}
 S(p) = \frac{1}{V} \sum_{x,y} e^{-ip(x-y)} \langle q(x) \bar{q}(y) \rangle,
\end{equation}
Green's functions,
\begin{equation}\label{green_func}
G_{\mathcal{O}}(p,p') = \frac{1}{V} \sum_{x,y,z} e^{-ip'.(x-y)-ip.(y-z)} \langle q(x)\mathcal{O}(y) \bar q(z) \rangle,
\end{equation}
and amputated Green's functions,
\begin{equation}\label{vertex}
\Lambda_{\mathcal{O}} (p,p') = S(p')^{-1} G_{\mathcal{O}}(p,p') S(p)^{-1}.
\end{equation}
The renormalized quantities are defined as $S_R(p) = Z_q S(p)$ and
$\Lambda_{\mathcal{O}}^R(p,p') = Z_q^{-1} Z_{\mathcal O}
\Lambda_{\mathcal O}(p,p')$.  In RI$'$-MOM, renormalization conditions
are imposed for $p'=p$, at scale $p^2=p'^2=\mu^2$. For the quark field
and vector and axial currents\footnote{This combination of conditions
  for $Z_q$, $Z_V$, and $Z_A$ has also been called the MOM scheme or
  the RI$'$ scheme. Note that the name RI$'$-MOM has also been used to
  refer to the combination of this condition for $Z_q$ and the
  original RI-MOM conditions~\cite{Martinelli:1994ty} for $Z_V$ and
  $Z_A$, even though this is not compatible with the vector and axial
  Ward identities.}:
\begin{equation}
\begin{gathered}
\lim_{m\to 0} \frac{-i}{12p^2} \Tr\left[S_R^{-1}(p) \slashed p\right] = 1,\\
\lim_{m\to 0} \frac{1}{36} \Tr\left[\Lambda_{V_\mu}^R (p,p) \left(\gamma_\mu - \frac{p_\mu \slashed p}{p^2}\right)\right] = 1, \\
\lim_{m\to 0} \frac{1}{36} \Tr\left[\Lambda_{A_\mu}^R (p,p) \gamma_5\left(\gamma_\mu - \frac{p_\mu \slashed p}{p^2}\right)\right] = 1 .
\end{gathered}
\end{equation}
RI-SMOM conditions are imposed at the symmetric point
$p^2=p'^2=q^2=\mu^2$, where $q=p'-p$. The quark-field renormalization
is the same as RI$'$-MOM, whereas for the vector and axial currents:
\begin{equation}
\label{ri-smom}
\begin{gathered}
\lim_{m \to 0} \frac{1}{12q^2} \Tr\left[q_\mu \Lambda_{V_\mu}^R (p,p') \slashed q\right] = 1,\\
\lim_{m \to 0} \frac{1}{12q^2} \Tr\left[q_\mu \Lambda_{A_\mu}^R (p,p') \gamma_5 \slashed q\right] = 1.
\end{gathered}
\end{equation}

As stated previously, in our calculations we do not take the chiral limit.
We also avoid directly determining the quark-field
renormalization. Instead, we impose the above renormalization
conditions on the vector current, which gives $Z_q/Z_V$, and
independently obtain $Z_V$ from three-point functions of pseudoscalar
mesons. Our estimate for $Z_q$ in RI$'$-MOM is then obtained using
\begin{equation}
(Z_q)_\text{RI$'$-MOM} = \frac{Z_V}{36} \Tr\left[\Lambda_{V_\mu} (p,p) \left(\gamma_\mu - \frac{p_\mu \slashed p}{p^2}\right)\right].
\end{equation}
In RI-SMOM, we estimate $Z_q$ in the same way using Eq.~\eqref{ri-smom}.

The renormalization scale $\mu$ should be chosen such that it is much larger
than $\Lambda_{\text{QCD}}$, in order to be able to connect the nonperturbative
renormalization schemes to $\MSbar$ using perturbation theory
(in our case, this is needed for the flavor-singlet axial current), and much
smaller than the inverse lattice spacing to avoid large discretization errors:
\begin{equation}
\Lambda_{\text{QCD}} \ll \mu \ll \pi/a.
\end{equation}
As our lattice spacing is fairly coarse, we do not expect to find a
stable plateau region in this window. Instead, we will perform fits to
remove the leading $O(a^2\mu^2)$ artifacts, and make use of the two
different schemes to estimate unaccounted-for systematic
uncertainties. 

\subsection{Vector current renormalization}
\label{zv_3pt}
\begin{figure*}
  \centering
  \includegraphics[width=0.495\textwidth]{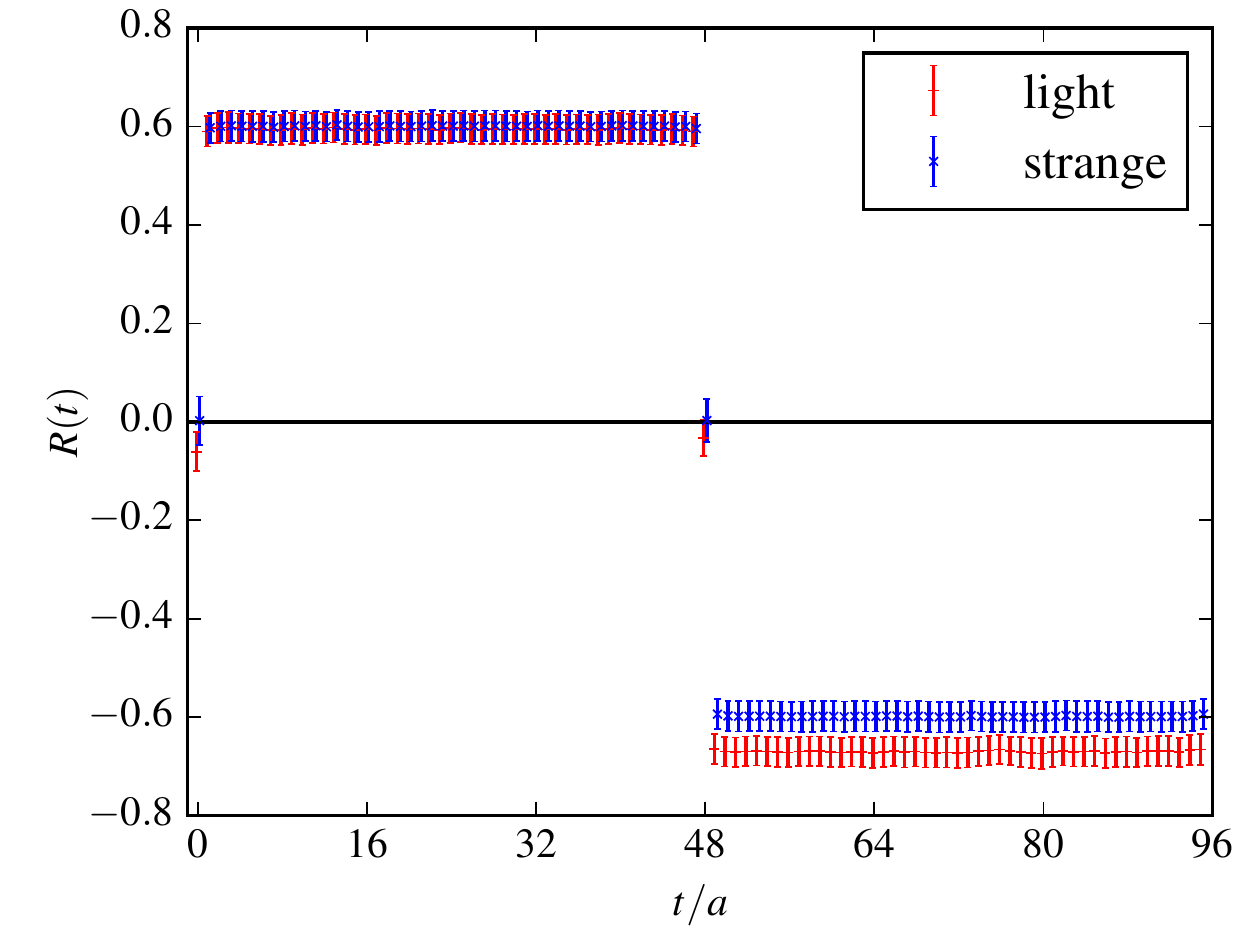}
  \includegraphics[width=0.495\textwidth]{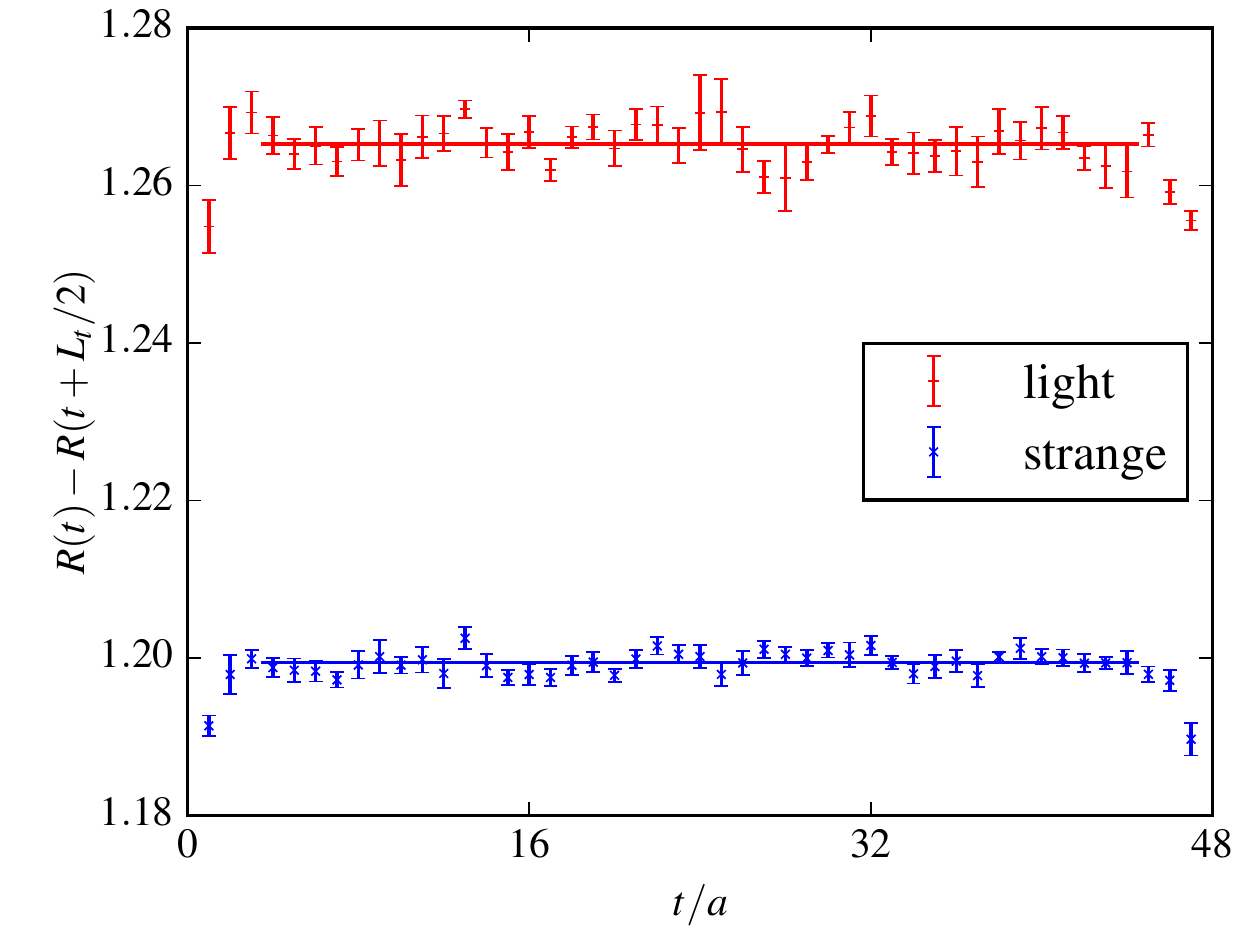}
  \caption{Determination of the vector current renormalization
    factors. Left: ratio of pseudoscalar three-point to two-point
    functions. Right: difference of the ratio on opposite sides of the
    interpolating operator. The horizontal lines indicate the plateau
    averages.}
  \label{fig:ZV}
\end{figure*}

We obtain the mass-dependent light and strange vector current
renormalization factors from matrix elements of pseudoscalar mesons
following, e.g., Ref.~\cite{Durr:2010aw}. For $\pi$ and $\eta_s$
states, we compute zero-momentum two-point functions $C_2(t)$ as well
as three-point functions $C_3(t)$ with source-sink separation
$T=L_t/2$ and an operator insertion of the time component of the local
(light or strange) vector current at source-operator separation
$t$. We form the ratio $R(t)=C_3(t)/C_2(T)$, so that the charge of the
interpolating operator gives the renormalization condition
\begin{equation}
  Z_V \left( R(t_1) - R(t_2) \right) = 1,
\end{equation}
for $0 < t_1 < T < t_2 < L_t$. Taking the difference $R(t)-R(t+T)$
results in a large cancellation of correlated statistical
uncertainties. Results are shown in Fig.~\ref{fig:ZV}. We average over
the long plateau, excluding three points at each end, and obtain
$Z_V^l=0.7903(2)$ and $Z_V^s=0.8337(2)$.

\subsection{\label{sec:ptsrc}Study of discretization effects}

We perform a dedicated study of discretization effects and breaking of
rotational symmetry, for the isovector case in the RI$'$-MOM
scheme. Using translation invariance to remove the sum over $y$ in
Eqs.~\eqref{prop} and \eqref{green_func}, we compute point-source
quark propagators from a fixed point $y$, which allows us to
efficiently obtain the gauge-averaged quark propagator and Green's
functions for a large set of momenta. Specifically, we save data for
all momenta in the inner 1/16 of the lattice Brillouin zone, i.e.,
with $|p_\mu|\leq\frac{\pi}{2a}$. After checking that the breaking of
hypercubic symmetry due to the different lattice temporal and spatial
extents is negligible, we averaged the estimates for the isovector
$Z_A$ over all hypercubic equivalent momenta.

\begin{figure}
  \centering
  \includegraphics[width=0.5\textwidth]{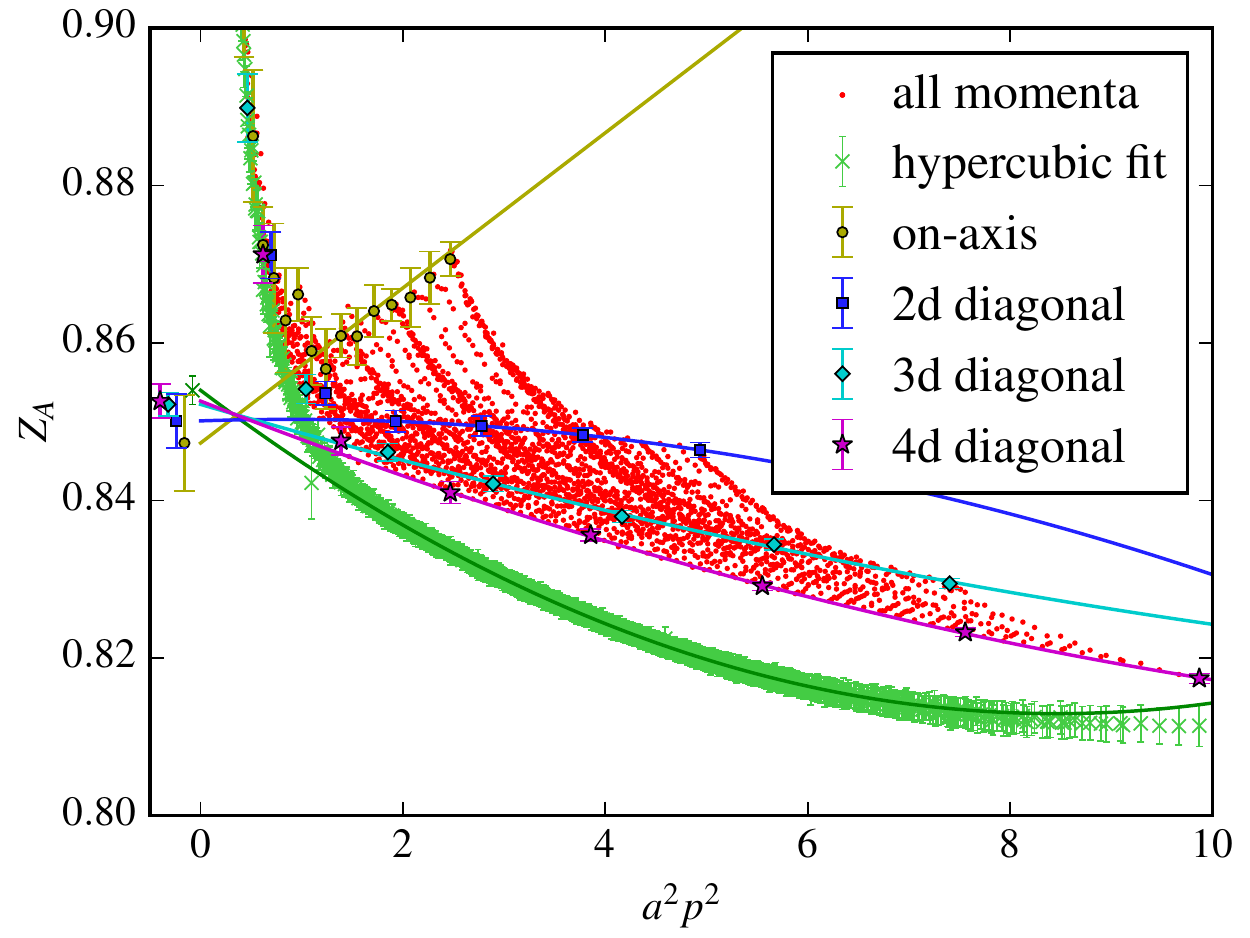}
  \caption{Isovector axial renormalization factor in the RI$'$-MOM
    scheme, computed for all lattice momenta with
    $|p_\mu|\leq\frac{\pi}{2a}$. The raw data for all momenta are
    shown without error bars to reduce clutter. The points that
    correspond to momenta that are on-axis or along one of the
    diagonals are highlighted and shown with error bars, as are the
    points that result from the hypercubic fit. The curves show the
    fits that extrapolate $a^2p^2$ to zero to remove rotationally
    invariant lattice artifacts, and the points at $a^2p^2\lesssim 0$
    show the results of the extrapolations.}
  \label{fig:ZA_ptsrc}
\end{figure}

Since the lattice breaks rotational symmetry, estimates of $Z_A$ will
depend not only on $p^2$, but also the hypercubic invariants
$p^{[2n]}\equiv\sum_\mu (p_\mu)^{2n}$. We make use of the hypercubic
fit form from Refs.~\cite{Boucaud:2003dx,Boucaud:2005rm} to remove the
leading terms that break rotational symmetry and collapse the data to
a single function of $p^2$:
\begin{equation}
  Z_A(p^2,p^{[4]},p^{[6]},\dots) = Z_A^0(p^2)
  + c_1 \frac{a^2p^{[4]}}{p^2} + c_2 \left(\frac{a^2p^{[4]}}{p^2}\right)^2
  + c_3 \frac{a^4p^{[6]}}{p^2} + c_4 a^4p^{[4]}.
\end{equation}
The fit parameters are the four $c_i$ that control breaking of
hypercubic symmetry and a separate $Z_A^0(p^2)$ for each $p^2$.  The
data $Z_A(p^2,p^{[4]},\dots)$ and the fit result $Z_A^0(p^2)$ are
shown in Fig.~\ref{fig:ZA_ptsrc}. This is effective at producing a
smooth curve that depends only on $p^2$ and not the other hypercubic
invariants. The resulting curve still contains $O(a^2p^2)$
rotationally invariant lattice artifacts, so we perform a second fit
in the range $a^2p^2\in[2,6]$ assuming a quadratic dependence on
$a^2p^2$, and extrapolate to $a^2p^2=0$; this is also shown in
Fig.~\ref{fig:ZA_ptsrc}.

An alternative approach is to pick an initial direction $p_*$ and
restrict our analysis to points $p=\lambda p_*$. Then the hypercubic
invariants have the form $p^{[2n]}=c_{2n}p^{2n}$ for some fixed $c_{2n}$
that depend on $p_*$. Thus, for this set of points along a fixed
direction, the dependence on hypercubic invariants reduces to
dependence only on $p^2$. We choose four sets of points: on-axis
momenta, and momenta along 2, 3, or 4-dimensional diagonals, i.e.,
$p_*=(0,0,0,1)$, $(0,0,1,1)$, $(0,1,1,1)$, and $(1,1,1,1)$. For each
set of points, we again do a fit to extrapolate $a^2p^2$ to
zero. Because in this case there are fewer points available, we expand
the fit range to be $a^2p^2\in[1.5,10]$. For on-axis points we use a linear
fit because $a^2p^2$ does not reach very high, and for the
$n$-dimensional diagonals we use a quadratic fit. The points from each
set and the fit curves are shown in Fig.~\ref{fig:ZA_ptsrc}.

We find that the $Z_A$ determined from the hypercubic fit and from the
fits along different diagonals are all consistent with one
another. This indicates that we can reliably control these lattice
artifacts by choosing only points along a fixed direction, which is
the approach that we will use for our main results for the axial
renormalization matrix.

 \subsection{\label{sec:running}Matching to $\MSbar$ and running of the singlet axial current}
 We consider the singlet and nonsinglet axial currents,
\begin{equation}
 \label{Eq:amu}
 A_\mu^0 = \frac{1}{\sqrt{N_f}}\bar \psi \gamma_\mu \gamma_5 \psi,  \qquad   A_\mu^a = \bar \psi  \gamma_\mu \gamma_5 \lambda^a \psi,
\end{equation}
where $\psi$ is the fermionic field and $\lambda^a$ is an $SU(N_f)$
generator acting in flavor space. The nonsinglet current should be
renormalized such that it satisfies the axial Ward identity associated
with chiral symmetry, and the renormalized singlet current should
satisfy the one-loop form of the axial anomaly.
The nonsinglet axial current has no anomalous dimension and is
appropriately renormalized to all orders in perturbation theory in 
$\MSbar$ (using dimensional regularization with a
naive anticommuting version of $\gamma_5$), RI$'$-MOM and
RI-SMOM schemes. Thus the matching factor between these schemes
is 1, and $Z_A=1$ when using a chiral regulator.

For the singlet current, dimensional regularization with a naive
$\gamma_5$ is inappropriate since the anomaly is not reproduced, and
thus the 't~Hooft-Veltman prescription for $\gamma_5$ is
necessary. Using it in $\MSbar$, an additional finite matching factor
$Z_5^s$ is needed for the renormalized current to satisfy the one-loop
form of the axial anomaly~\cite{Larin:1993tq}.  Thus renormalized, the
singlet current has an anomalous dimension, $\gamma =
(\frac{\alpha}{4\pi})^2(-6C_FN_f) +
O(\alpha^3)$~\cite{Kodaira:1979pa}, where the $O(\alpha^3)$ term is
given in Ref.~\cite{Larin:1993tq}. Using the same dimensional
regularization, it was shown in Ref.~\cite{Bhattacharya:2015rsa} that
the conversion factor between $\MSbar$ (including the finite factor
$Z_5^s$) and RI-SMOM is $1+O(\alpha^2)$.

For computing the matching between RI$'$-MOM and RI-SMOM, at one-loop
order there should be no distinction between singlet and nonsinglet
currents. Since the matching factor is $1$ for nonsinglet currents, we
conclude that the conversion factor for the singlet axial current in
RI$'$-MOM is $1+O(\alpha^2)$.

We remove the running of the singlet $Z_A$ 
by evolving to a fixed scale. The evolution is given by
\begin{align}
\mu^2 \frac{d}{d \mu^2}\ \log\left(Z_5^s Z_A^{\MSbar,\text{HV}}\right) &= \gamma(\alpha) = -\sum_i \gamma_i \alpha^{i+1}, \\
\mu^2 \frac{d}{d \mu^2} \alpha &= \beta(\alpha) = - \sum_i \beta_i \alpha^{i+2},
\end{align}
where the relevant coefficients are
\begin{equation}
\begin{gathered}
\begin{aligned}
\beta_0 &= \frac{1}{4\pi}\;\left(\frac{11}{3} C_A - \frac{4}{3} T_F N_f \right) = \frac{1}{4\pi} \left(11-\frac{2}{3} N_f\right),\\
\beta_1 &= \frac{1}{(4\pi)^2}\;\left(\frac{34}{3} C_A^2 - \frac{20}{3} C_A T_F N_f - 4 C_F T_F N_f\right) = \frac{1}{(4\pi)^2} \left(102 - \frac{38}{3} N_f\right),\\
\gamma_0 &= 0,\\
\gamma_1 &=  \frac{1}{(4\pi)^2} (6C_F N_f) =  \frac{1}{(4\pi)^2} 8N_f, 
\end{aligned}
\end{gathered}
\end{equation}
using $C_A = 3$, $C_F = 4/3$, and $T_F = 1/2$. 
At two-loop order, the evolution of $\alpha$ is given by~\cite{Prosperi:2006hx}:
\begin{equation}
\alpha(\mu) = -\frac{\beta_0}{\beta_1} \frac{1}{1+W_{-1}(\zeta)}, \qquad \zeta = -\frac{\beta_0^2}{e\beta_1} \left( \frac{\Lambda^2}{\mu^2}\right)^{\beta_0^2/\beta_1},
\end{equation}
where $W_k$ is the many-valued Lambert function 
defined by $W_k(\zeta)e^{W_k(\zeta)} = \zeta$.
We use the PDG value, $\Lambda_3^{\MSbar} = 332(19)$~MeV~\cite{Olive:2016xmw}.
Using $\gamma_0=0$, the evolution of the renormalization factor at 
two-loop order is given by
\begin{equation}\label{pert}
\frac{Z(\mu)}{Z(\mu_0)} = \left( \frac{\beta_0 + \beta_1 \alpha(\mu)}{\beta_0+\beta_1\alpha(\mu_0)}\right)^{\gamma_1/\beta_1}.
\end{equation}
\subsection{Renormalization of the axial current: $N_f = 2+1$}
\label{RAC}
Consider the flavor-diagonal axial currents, Eq.~\eqref{Eq:amu}, with $\psi=(u\;d\; s)^T$.
We take $a=3,8,0,$ with $\Tr(\lambda^a\lambda^b) = \delta^{ab}$,
\begin{equation}
\lambda^3 = \frac{1}{\sqrt 2}\begin{pmatrix} 1&0&0 \\0&-1&0\\ 0&0&0 \end{pmatrix},\; \lambda^8 = \frac{1}{\sqrt 6}\begin{pmatrix} 1&0&0 \\0&1&0\\ 0&0&-2 \end{pmatrix},\; \lambda^0 = \frac{1}{\sqrt 3}\begin{pmatrix} 1&0&0 \\0&1&0\\ 0&0&1 \end{pmatrix}.
\end{equation}
Using $i,j$ to label quark flavors, we compute the quark propagator $S_i(p)$ [Eq.~\eqref{prop}] for quark flavor-$i$, nonamputated and amputated Green's functions [Eq.~\eqref{green_func}, Eq.~\eqref{vertex}] for mixed quark flavors-$i$ and -$j$, $G_{i,j}^\mathcal {O}(p',p)$, and $\Lambda_{ij}^\mathcal O(p',p)$, respectively. These renormalize as 
\begin{equation}
\Lambda_{R,ij}^{A_\mu^a}(p',p) = \frac{Z_A^{ab}}{\sqrt{Z_q^iZ_q^j}} \Lambda_{ij}^{A_\mu^b}(p',p).
\end{equation}
For $N_f = 2+1$, the renormalization pattern is
\begin{equation}
Z_A = \begin{pmatrix} Z_A^{33} & 0&0 \\ 0 & Z_A^{88} & Z_A^{80}  \\ 0 & Z_A^{08} & Z_A^{00}\end{pmatrix},
\end{equation}
and for $N_f = 3$, this reduces to two independent factors since $Z_A^{88}
= Z_A^{33}$ and $Z_A^{80} = Z_A^{08} = 0$.

In a RI$'$-MOM or RI-SMOM scheme, the renormalization condition for
$Z_A$ involves tracing $\Lambda^{A_\mu}$ with some projector $P_\mu$
at kinematics corresponding to the scale $\mu^2$ (see
Subsec.~\ref{RSM}). In the case of multiple flavors, this becomes
\begin{equation}
\sum_{ij} \lambda_{ji}^a \Tr \left [ \Lambda_{R,ij}^{A_\mu^b} P_\mu\right]_{\mu^2} = \delta^{ab},
\end{equation}
so that we get
\begin{equation}
(Z_A^{-1}(\mu))^{ba} = \sum_{ij} \lambda_{ji}^a \Tr \left[ \frac{1}{\sqrt {Z_q^i Z_q^j}}\Lambda_{ij}^{A_\mu^b} P_\mu\right]_{\mu^2}.
\end{equation}
Specifically, this yields for $N_f = 2+1$
\begin{align}\label{eq:za_33}
(Z_A^{-1})^{33} &= \frac{1}{2Z_q^l} \Tr \left[\left( \Lambda_{u,u}^{A_\mu^{u-d}} - \Lambda_{d,d}^{A_\mu^{u-d}}\right) P_\mu \right] = \frac{1}{Z_q^l} \Sigma_{l,\text{conn}},
\end{align}
where $\Sigma_{l,\text{conn}}$ is the connected contribution 
to the ($u$ or $d$)-quark amputated axial vertex function, 
traced with $P_\mu$. This corresponds to the usual isovector 
result. Writing $\Sigma_{j,\text{disc}}^i$ for the disconnected contribution 
to the amputated vertex function with the flavor-$i$ axial current 
and flavor-$j$ external quark states, traced with $P_\mu$, we get
\begin{align}
(Z_A^{-1})^{88} &= \frac{1}{6} \Tr \left[\left( \frac{1}{Z_q^l} \Lambda _{u,u}^{A_\mu^{u+d-2s}} +  \frac{1}{Z_q^l} \Lambda _{d,d}	^{A_\mu^{u+d-2s}} -  \frac{2}{Z_q^s} \Lambda _{s,s}^{A_\mu^{u+d-2s}}   \right) P_\mu \right] \nonumber \\
&=  \frac{1}{3}\left(  \frac{1}{Z_q^l} \Sigma_{l,\text{conn}} +  \frac{2}{Z_q^s} \Sigma_{s,\text{conn}} \right) +
\frac{2}{3} \left(  \frac{1}{Z_q^l} \Sigma_{l,\text{disc}}^{l-s} -  \frac{1}{Z_q^s} \Sigma_{s,\text{disc}}^{l-s} \right), \label{eq:za_88}\\
(Z_A^{-1})^{80} &= \frac{1}{3\sqrt 2} \Tr \left[\left( \frac{1}{Z_q^l} \Lambda _{u,u}^{A_\mu^{u+d-2s}} +  \frac{1}{Z_q^l} \Lambda _{d,d}^{A_\mu^{u+d-2s}} +  \frac{1}{Z_q^s} \Lambda _{s,s}^{A_\mu^{u+d-2s}} \right) P_\mu \right] \nonumber \\
&=\frac{\sqrt 2}{3}\left(  \frac{1}{Z_q^l} \Sigma_{l,\text{conn}} -  \frac{1}{Z_q^s} \Sigma_{s,\text{conn}} \right) +
\frac{\sqrt 2}{3} \left(  \frac{2}{Z_q^l} \Sigma_{l,\text{disc}}^{l-s} +\frac{1}{Z_q^s} \Sigma_{s,\text{disc}}^{l-s} \right),\\
(Z_A^{-1})^{08} &= \frac{1}{3\sqrt 2} \Tr \left[\left( \frac{1}{Z_q^l} \Lambda _{u,u}^{A_\mu^{u+d+s}} +  \frac{1}{Z_q^l} \Lambda _{d,d}^{A_\mu^{u+d+s}} -  \frac{2}{Z_q^s} \Lambda _{s,s}^{A_\mu^{u+d+s}} \right) P_\mu \right] \nonumber \\
&=\frac{\sqrt 2}{3}\left(  \frac{1}{Z_q^l} \Sigma_{l,\text{conn}} -  \frac{1}{Z_q^s} \Sigma_{s,\text{conn}} \right) +\frac{\sqrt 2}{3} \left(  \frac{1}{Z_q^l} \Sigma_{l,\text{disc}}^{2l+s} -  \frac{1}{Z_q^s} \Sigma_{s,\text{disc}}^{2l+s} \right), \\
(Z_A^{-1})^{00} &= \frac{1}{3} \Tr \left[\left( \frac{1}{Z_q^l} \Lambda _{u,u}^{A_\mu^{u+d+s}} +  \frac{1}{Z_q^l} \Lambda _{d,d}^{A_\mu^{u+d+s}} +  \frac{1}{Z_q^s} \Lambda _{s,s}^{A_\mu^{u+d+s}} \right) P_\mu \right] \nonumber \\
&= \frac{1}{3}\left(  \frac{2}{Z_q^l} \Sigma_{l,\text{conn}} +  \frac{1}{Z_q^s} \Sigma_{s,\text{conn}} \right) +
\frac{1}{3} \left(  \frac{2}{Z_q^l} \Sigma_{l,\text{disc}}^{2l+s} +  \frac{1}{Z_q^s} \Sigma_{s,\text{disc}}^{2l+s} \right).\label{eq:za_00}
\end{align}
It is clear that $(Z_A^{-1})^{80}$ and $(Z_A^{-1})^{08}$ 
vanish when $N_f=3$, and the disconnected contribution 
to $(Z_A^{-1})^{88}$ is doubly suppressed by approximate 
$SU(3)_f$ symmetry.

Having evaluated an effective $Z_A^{-1}$ in some scheme 
at a scale $\mu$, we can invert the matrix and evolve to the
target scale of 2~GeV:
\begin{equation}\label{2gev}
Z_A^{8i}(2 \text{ GeV}) = Z_A^{8i}(\mu), \qquad Z_A^{0i}(2 \text{ GeV}) = \left( \frac{Z_A^0(2 \text{ GeV})}{Z_A^0(\mu)}\right)_{\text{pert}} Z_A^{0i}(\mu),
\end{equation}
where the perturbative flavor-singlet evolution is given 
by Eq.~\eqref{pert}. Finally, we fit with a polynomial in 
$a^2\mu^2$ to remove lattice artifacts.
If we want to obtain a single-flavor axial current, 
such as the strange, we can write, e.g.,
\begin{align}
A_\mu^{R,s} &= \frac{1}{\sqrt 3} A_\mu^{R,0} - \sqrt{\frac{2}{3}} A_\mu^{R,8}\nonumber\\
&= \frac{1}{3} \left( Z_A^{00} + 2Z_A^{88} - \sqrt 2 Z_A^{80} - \sqrt 2 Z_A^{08}\right) A_\mu^s
+ \frac{1}{3} \left( Z_A^{00} - Z_A^{88} + \frac{1}{\sqrt 2} Z_A^{08} - \sqrt 2 Z_A^{80}\right) A_\mu^{u+d} \label{eq:za_s} \\
&\equiv Z_A^{s,s} A_\mu^s + Z_A^{s,u+d} A_\mu^{u+d}. \nonumber
\end{align}
Similarly, we can evaluate the renormalized $u+d$ current,
\begin{align}
A_\mu^{R,u+d} &= \frac{2}{\sqrt 3} A_\mu^{R,0} + \sqrt{\frac{2}{3}} A_\mu^{R,8}\nonumber\\
&= \frac{1}{3} \left( 2Z_A^{00} + Z_A^{88} + \sqrt 2 Z_A^{80} + \sqrt 2 Z_A^{08}\right) A_\mu^{u+d}
+ \frac{2}{3} \left( Z_A^{00} - Z_A^{88} + \frac{1}{\sqrt 2} Z_A^{08} - \sqrt 2 Z_A^{80}\right) A_\mu^{s}  \label{eq:za_ud} \\
&\equiv Z_A^{u+d,u+d} A_\mu^{u+d} + Z_A^{u+d,s} A_\mu^{s}. \nonumber
\end{align}

In order to study the disconnected light-quark current by itself, 
as described in Subsection~\ref{disc}, we introduce a quenched third light quark $r$,
degenerate with $u$ and $d$. Then the \emph{connected} contribution to the matrix 
elements of the $u+d$ current is the same as matrix elements of the 
$u+d-2r$ current. Since this is a nonsinglet flavor combination formed 
from degenerate light quarks, it has the same renormalization factor as 
the isovector current. To find the \emph{disconnected} light-quark contribution, 
we take the difference,
\begin{equation}\label{diff}
 \begin{aligned}
 A_\mu^{R,l,\text{disc}}  = A_\mu^{R,r} & = \frac{1}{2} (A_\mu^{R,u+d} - A_\mu^{R,u+d-2r}) \\
 &= \frac{1}{2} (Z_A^{u+d,u+d}A_\mu^{u+d} + Z_A^{u+d,s}A_\mu^s - Z_A^{33} A_\mu^{u+d,\text{conn}}) \\
 &=Z_A^{u+d,u+d} A_\mu^{l,\text{disc}} + \frac{1}{2} \left((Z_A^{u+d,u+d} - Z_A^{33}) A_\mu^{u+d,\text{conn}} + Z_A^{u+d,s} A_\mu^s\right).
 \end{aligned}
\end{equation}
\subsection{Volume-source approach and reuse of disconnected diagrams}
\label {VS}
We evaluate our observables using quark propagators with four-dimensional volume
plane-wave sources $D_q^{-1}(x|p)\equiv \sum_y D_q^{-1}(x,y)e^{ipy}$.
For a quark-bilinear operator $\mathcal{O} = \bar q\Gamma q$ 
($\Gamma = \gamma_\mu\gamma_5$ for the axial current), 
the connected contribution to the Green's function is obtained using
\begin{equation}
G_{\mathcal{O},\text{conn}}(p',p) = \frac{1}{V} \left\langle \sum_y e^{i(p'-p)y}  \gamma_5 D_q^{-1}(y|p')^\dagger \gamma_5 \Gamma D_q^{-1}(y|p) \right\rangle_U,
\end{equation}
where $\langle\dots\rangle_U$ denotes the average over gauge configurations.
We obtain the disconnected contribution by correlating the plane-wave-source
propagators with the previously-computed disconnected loops\footnote{Recall
that the loops are gauge invariant and thus do not need to be transformed
to Landau gauge.} $T_\mu^q(\vec k,t)$ [Eq.~\eqref{T_disc}]:
\begin{equation}
G_{\mathcal{O},\text{disc}}(p',p)
= \frac{L_t}{V} \left \langle \sum_x  e^{-ip'x} D_{q'}^{-1}(x|p) e^{ik_4t} T_\mu^q(\vec 0, t)\right \rangle_U,
\end{equation}
where $q$ and $q'$ are the quark flavors of the operator and the
external quark states, and we choose $p'-p=(\vec 0,k_4)$. Translation
invariance implies that this expression is independent of $t$, and we
average over all timeslices on which the disconnected loops were
computed.

\subsection{Results}
\label{RS}

\begin{figure*}
  \centering
  \includegraphics[width=0.49\textwidth]{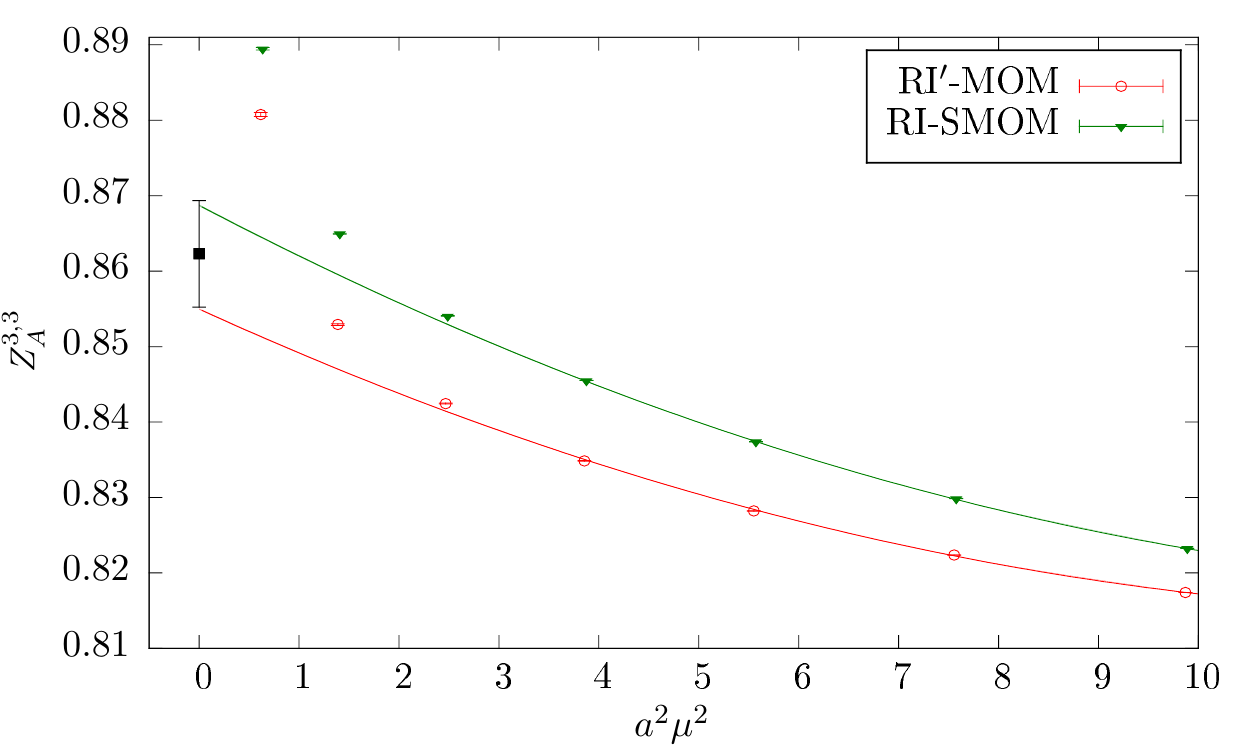}\\
  \includegraphics[width=0.49\textwidth]{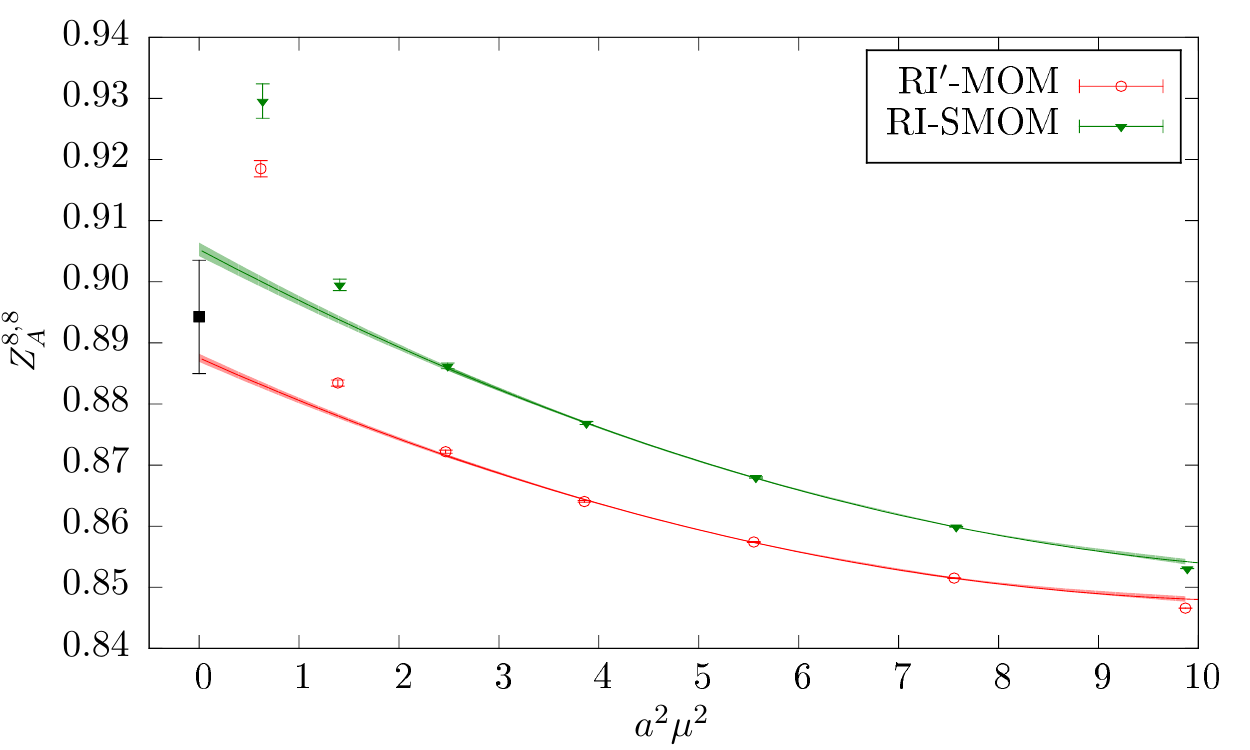}
  \includegraphics[width=0.49\textwidth]{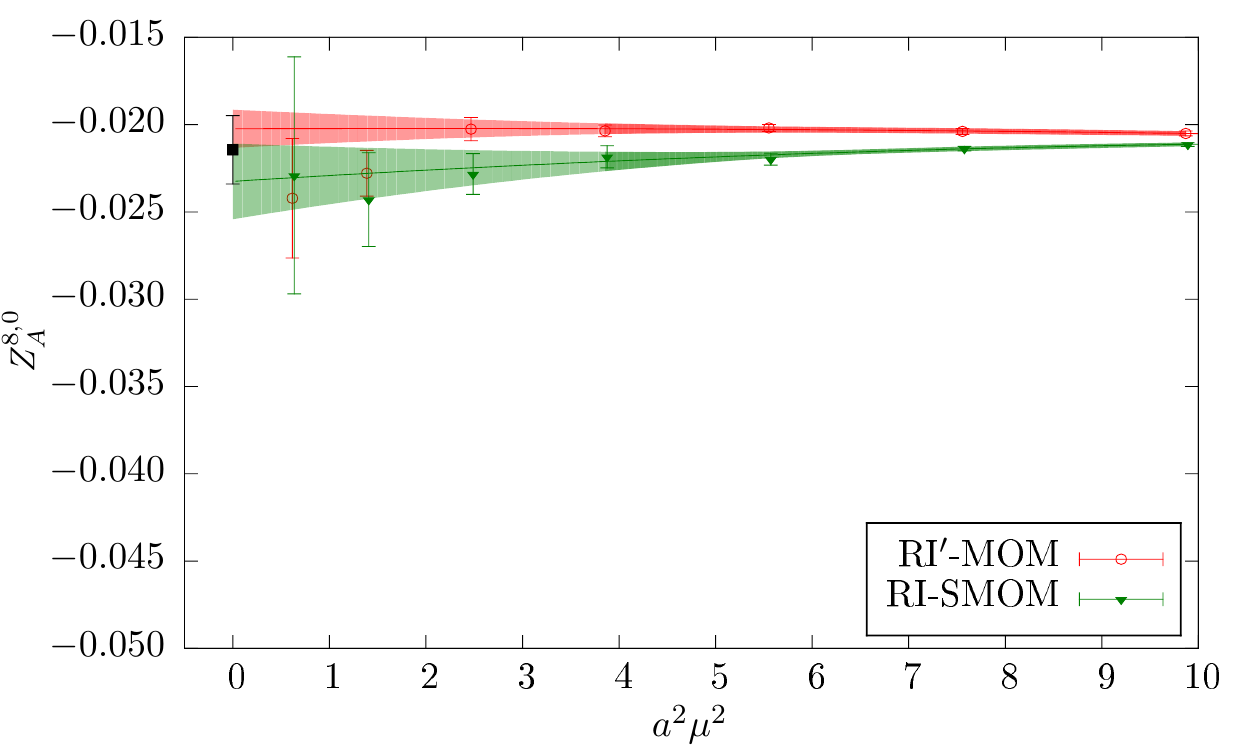}\\
  \includegraphics[width=0.49\textwidth]{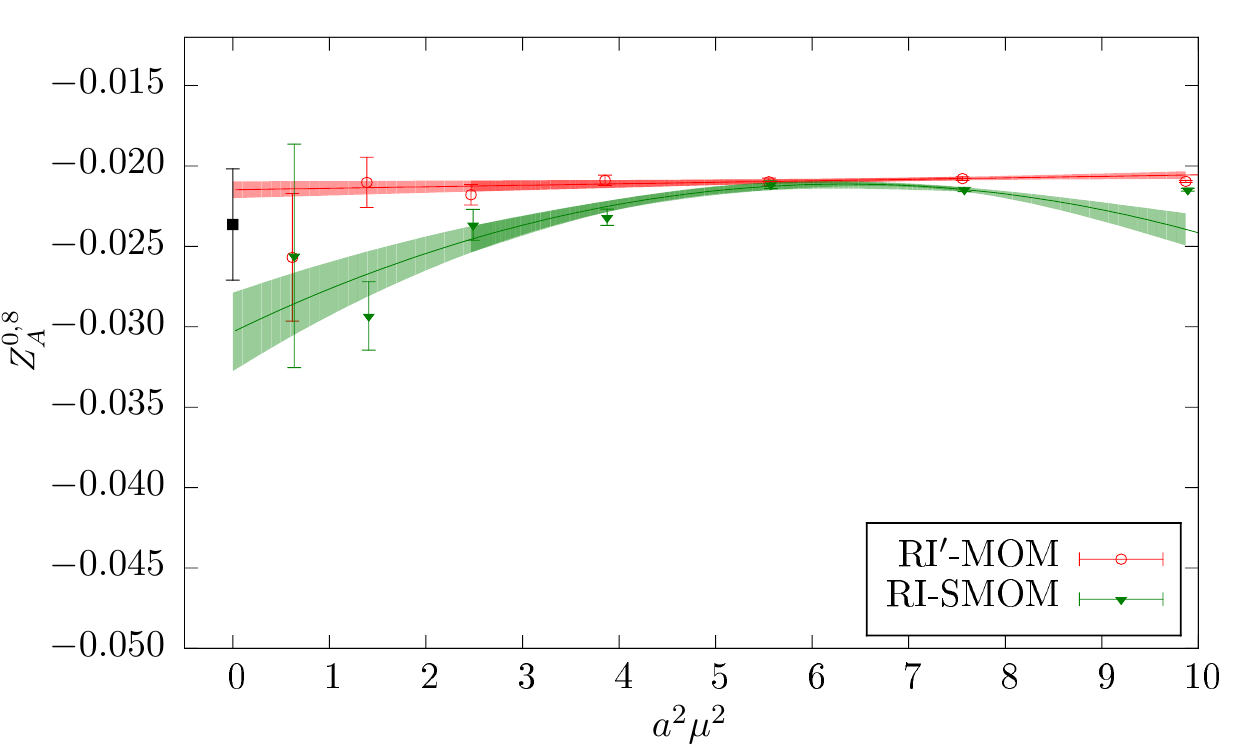}
  \includegraphics[width=0.49\textwidth]{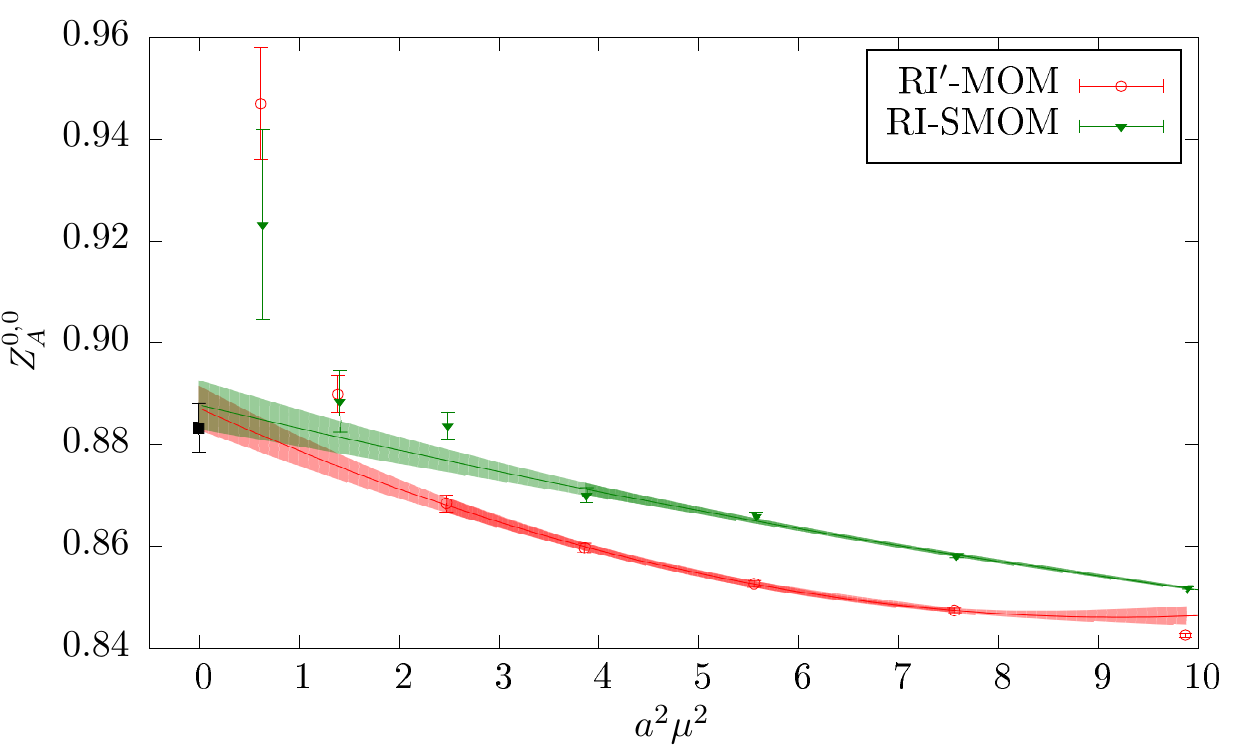}
  \caption{$Z_A$ matrix elements for the $\{A_\mu^3,A_\mu^8,A_\mu^0\}$
    basis, in the $\MSbar$ scheme at scale 2~GeV. Each plot shows the
    data versus the matching point $a^2\mu^2$ for the two intermediate
    schemes, as well as an illustrative fit curve for each scheme used
    to extrapolate to $a^2\mu^2=0$. The black point at $a^2\mu^2=0$
    shows the value and the combined statistical and systematic
    uncertainty, based on these and other fits.}
  \label{me08}
\end{figure*}

\begin{figure*}
  \centering
  \includegraphics[width=0.49\textwidth]{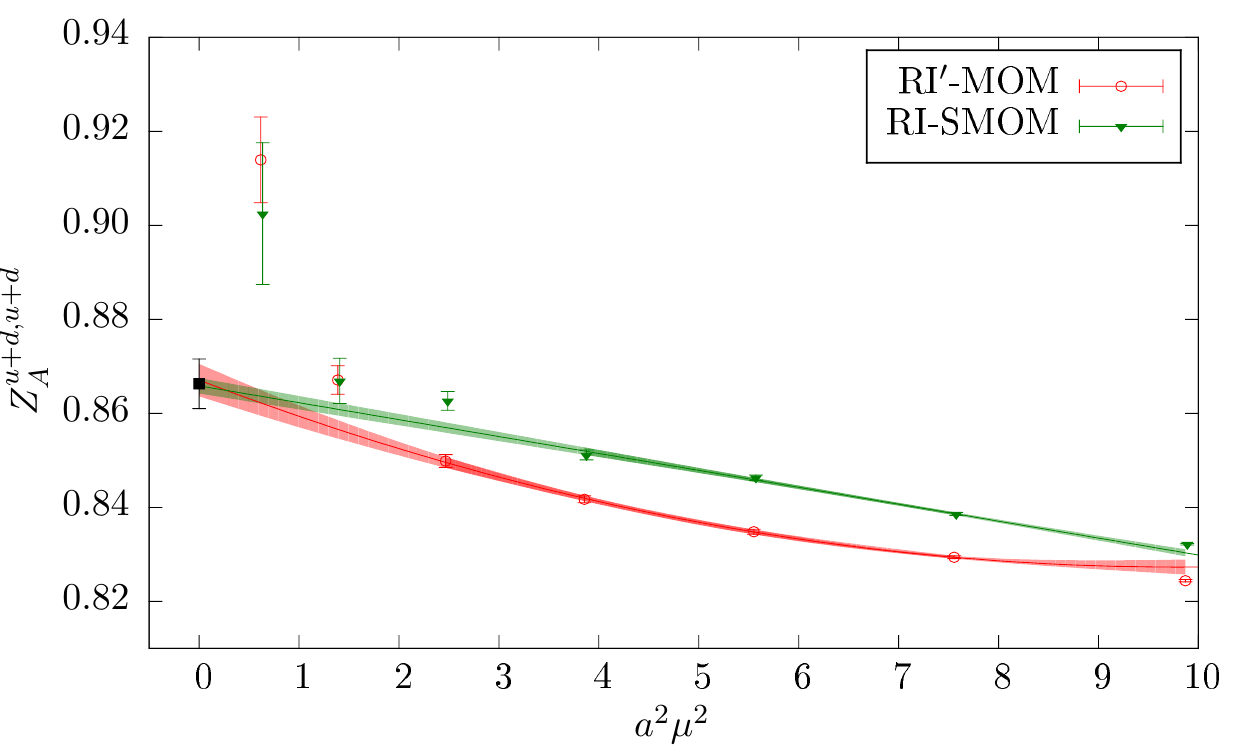}
  \includegraphics[width=0.49\textwidth]{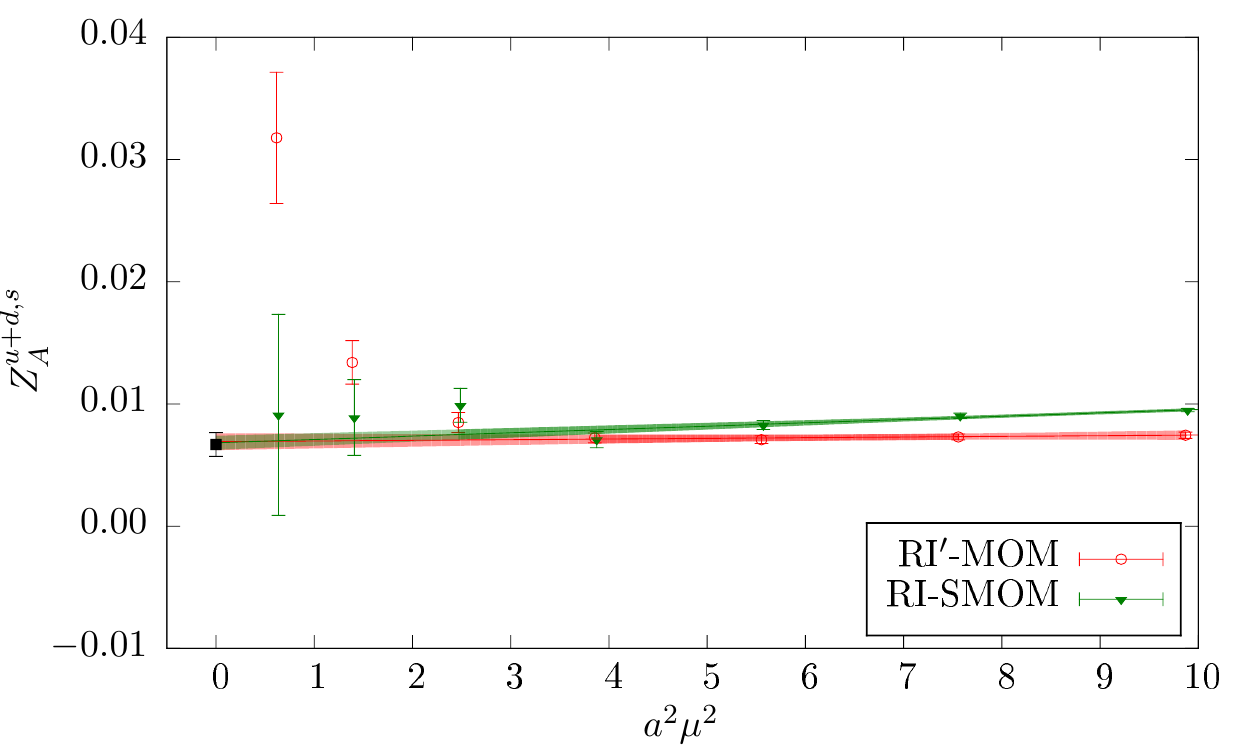}\\
  \includegraphics[width=0.49\textwidth]{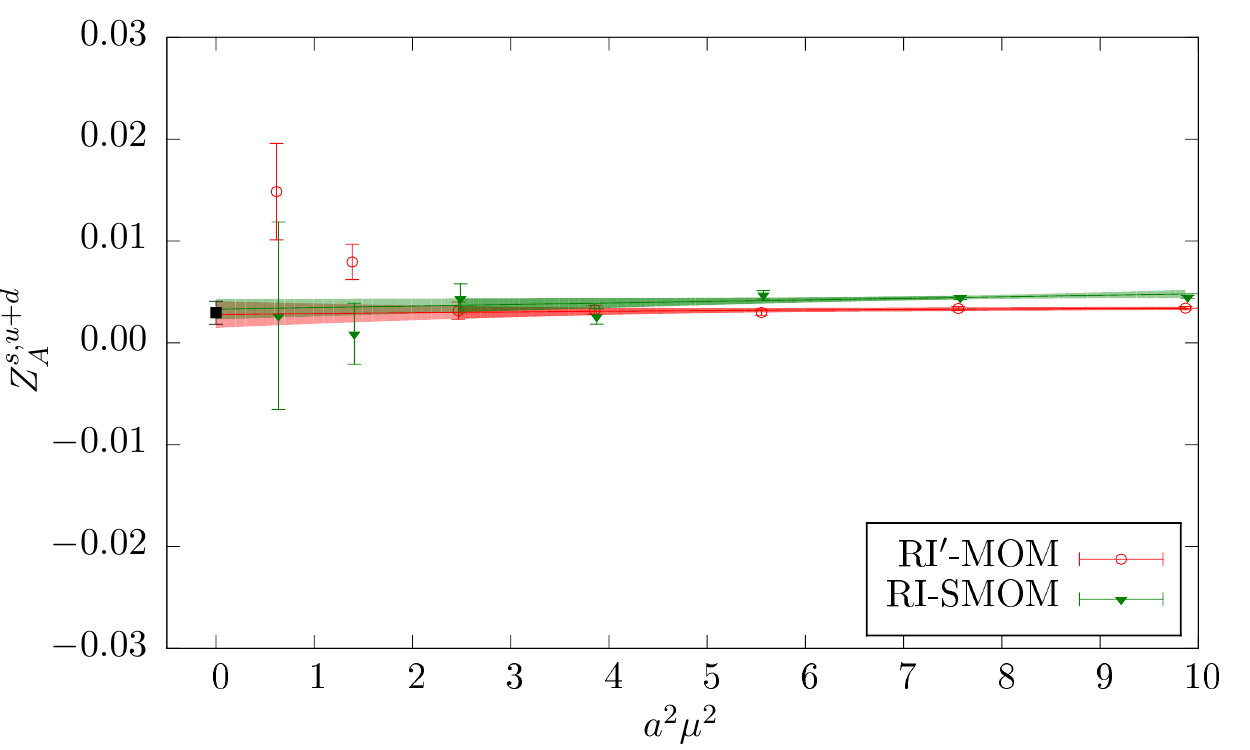}
  \includegraphics[width=0.49\textwidth]{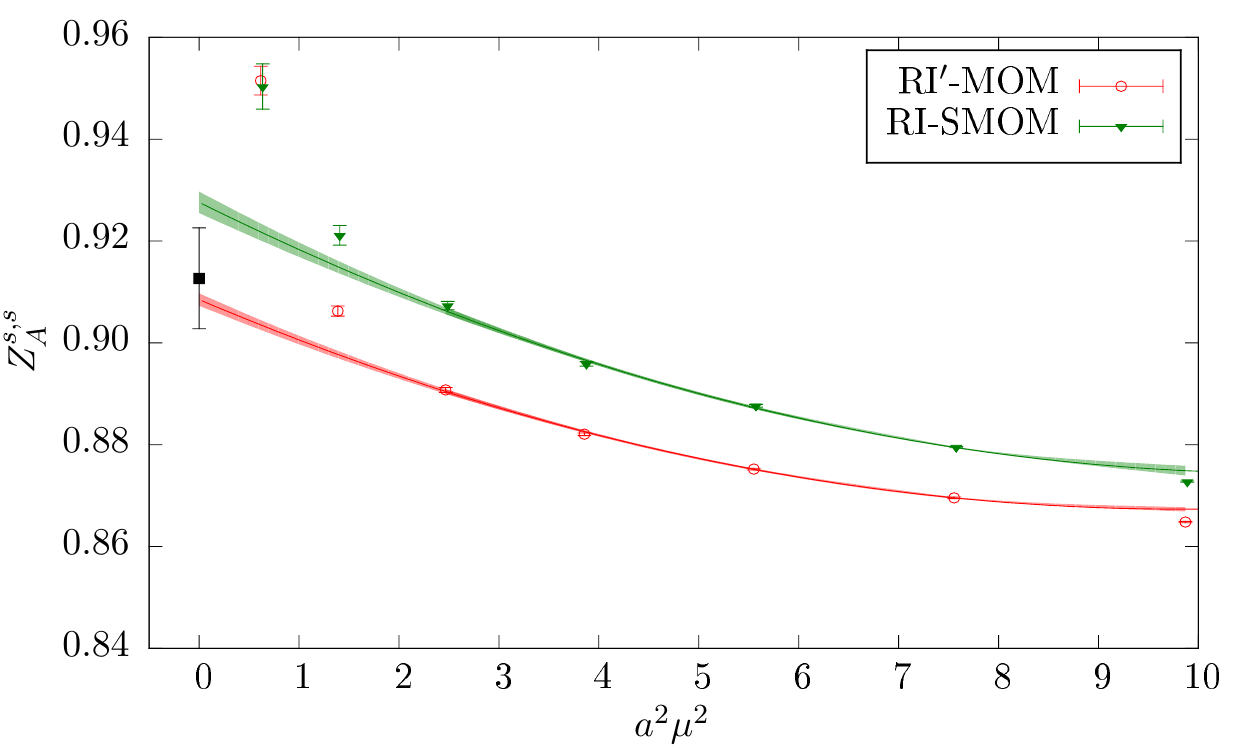}
  \caption{$Z_A$ matrix elements for $\{A_\mu^{u+d},A_\mu^s\}$. See
    the caption of Fig.~\ref{me08}.}
  \label{meuds}
\end{figure*}

In order to minimize cut-off effects we choose momenta on the diagonal
of the Brillouin zone $p,p' \in \frac{2\pi k}{L_s} (1,1,1,\pm 1)$ for
$k \in \{2,3,...,8\}$.  Therefore, our momenta span the range $0.6<
a^2 \mu^2<10$.  We used for this calculation about 200 gauge
configurations. This procedure involves the following steps:
\begin{enumerate}
\item Compute Landau gauge-fixed quark propagators and Green's
  functions for both light and strange quarks as outlined in the
  previous section. Form the amputated vertex functions.
\item On the connected diagrams, impose the RI$'$-MOM or RI-SMOM
  vector current renormalization conditions, together with the
  renormalization factors from Subsec.~\ref{zv_3pt}, to find estimates
  for $Z_q^l$ and $Z_q^s$ at each scale $\mu$.
\item Trace the axial amputated vertex functions with $P_\mu$ to
  obtain $\Sigma^l_\text{conn}$, $\Sigma^s_\text{conn}$,
  $\Sigma^l_{l,\text{disc}}$, $\Sigma^s_{l,\text{disc}}$,
  $\Sigma^l_{s,\text{disc}}$, and $\Sigma^s_{s,\text{disc}}$ at each
  scale $|p|$. By combining the different $\Sigma$ following
  Eqs.~(\ref{eq:za_33}--\ref{eq:za_00}), form the matrix $Z_A^{-1}$.
\item Invert the matrix and evolve from scale $\mu$ to 2~GeV.
\item Optionally, convert the $Z_A$ matrix from the basis
  $\{A_\mu^3,A_\mu^8,A_\mu^0\}$ to
  $\{A_\mu^{u-d},A_\mu^{u+d},A_\mu^s\}$, using Eqs.~\eqref{eq:za_s} and
  \eqref{eq:za_ud}.
\item Extrapolate $\mu$ to zero to remove $O(a^2\mu^2)$ lattice artifacts.
\end{enumerate}
For estimating the statistical and systematic errors in removing the
$O(a^2\mu^2)$ artifacts, we apply linear and quadratic fits 
for each matrix element, $Z_A^{ij} = c_0^{ij}+c_1^{ij}(a\mu)^2$ and
$Z_A^{ij} = c_0^{ij}+c_1^{ij}(a\mu)^2 + c_2^{ij}(a\mu)^4$.
We apply these fits in different ranges of $a^2\mu^2$, all of which
lie within the range $[2.5,10]$, i.e., always excluding the first two
points. This fit procedure is applied to results from both RI$'$-MOM
and RI-SMOM schemes. We take then three best fits in each scheme
(yielding six values), average all of them to get the central value
and statistical uncertainty, and use the root-mean-square difference between
the six values and the average to get the systematic uncertainty.
Figures~\ref{me08} and \ref{meuds} show illustrative fits for
obtaining the matrix elements in the different bases from both
RI$'$-MOM and RI-SMOM schemes. We obtain the following $Z_A$ matrices:
\begin{gather}
\begin{pmatrix}
A_\mu^{R,3} \\[0.5em]
A_\mu^{R,8}\\[0.5em]
A_\mu^{R,0}
\end{pmatrix}
= 
\begin{pmatrix}
 0.8623(1)(71) & 0 & 0 \\[0.5em]
0 & 0.8942(6)(93) &-0.0214(13)(14)\\[0.5em]
0 &-0.0236(1)(33)& 0.8832(30)(36)
\end{pmatrix}
\begin{pmatrix}
A_\mu^3\\[0.5em]
A_\mu^8 \\[0.5em]
A_\mu^0
\end{pmatrix},\\
\begin{pmatrix}
A_\mu^{R,u-d} \\[0.5em]
A_\mu^{R,u+d}\\[0.5em]
A_\mu^{R,s}
\end{pmatrix}
= 
\begin{pmatrix}
 0.8623(1)(71) & 0 & 0 \\[0.5em]
0 &0.8662(26)(45) & 0.0067(8)(5)\\[0.5em]
0 &0.0029(10)(5)& 0.9126(11)(98)
\end{pmatrix}
\begin{pmatrix}
A_\mu^{u-d}\\[0.5em]
A_\mu^{u+d} \\[0.5em]
A_\mu^s
\end{pmatrix}.
\end{gather}
Note that these two different matrices were obtained from independent
fits to remove $O(a^2\mu^2)$ artifacts, and thus they are not related
exactly by Eqs.~\eqref{eq:za_s} and \eqref{eq:za_ud}. For
renormalizing our nucleon form factor data, we use the latter
matrix. Finally, the contribution from the bare connected light axial
current to the renormalized disconnected light axial current depends
on the difference $Z_A^{u+d,u+d} - Z_A^{33}$, as shown in
Eq.~\eqref{diff}. In order to reduce uncertainties, we computed this
difference by itself using the above procedures, and found
$Z_A^{u+d,u+d} -Z_A^{33} = 0.0061(18)(10)$.

From Eq.~\eqref{eq:improvement} and the full mass-dependent $O(a)$
improvement in Ref.~\cite{Bhattacharya:2005rb}, $Z_A^{u+d,s}$ and
$Z_A^{s,u+d}$ first appear at two-loop order in lattice perturbation
theory; since the mass-dependent part is further suppressed by $am_s$,
it follows that these are largely sensitive to the singlet-nonsinglet
difference\footnote{This is in contrast with, e.g.,
  $Z_A^{0,0}-Z_A^{8,8}$, which has a contribution at tree level
  proportional to $ab_A(m_s-m_{ud})$.}. These elements are less than
one percent of the diagonal ones, indicating a small difference, which
is consistent with previous studies. For example,
Ref.~\cite{Chambers:2014pea} found a singlet-nonsinglet difference
$\bar Z_A-Z_A=0.020(3)$, using a similar lattice action. In the
$SU(3)$ flavor limit, this corresponds to $\bar
Z_A-Z_A=3Z_A^{s,u+d}=\frac{3}{2}Z_A^{u+d,s}$, so that those mixing
factors are about twice as large as ours.

\section{\label{sec:formfacs}Axial form factors}

\subsection{$G_A$ form factors}

\begin{figure*}
  \centering
  \includegraphics[width=0.495\textwidth]{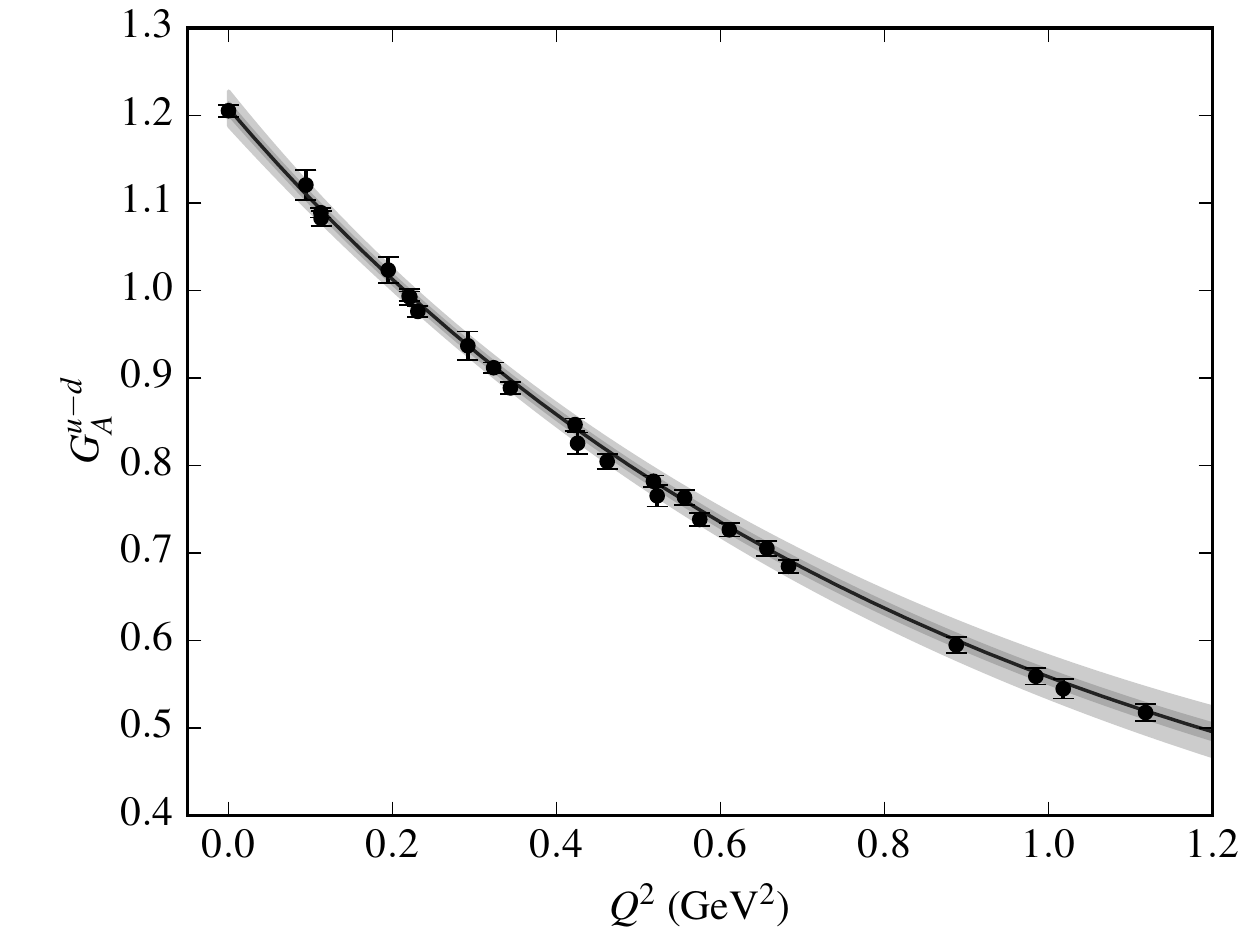}
  \includegraphics[width=0.495\textwidth]{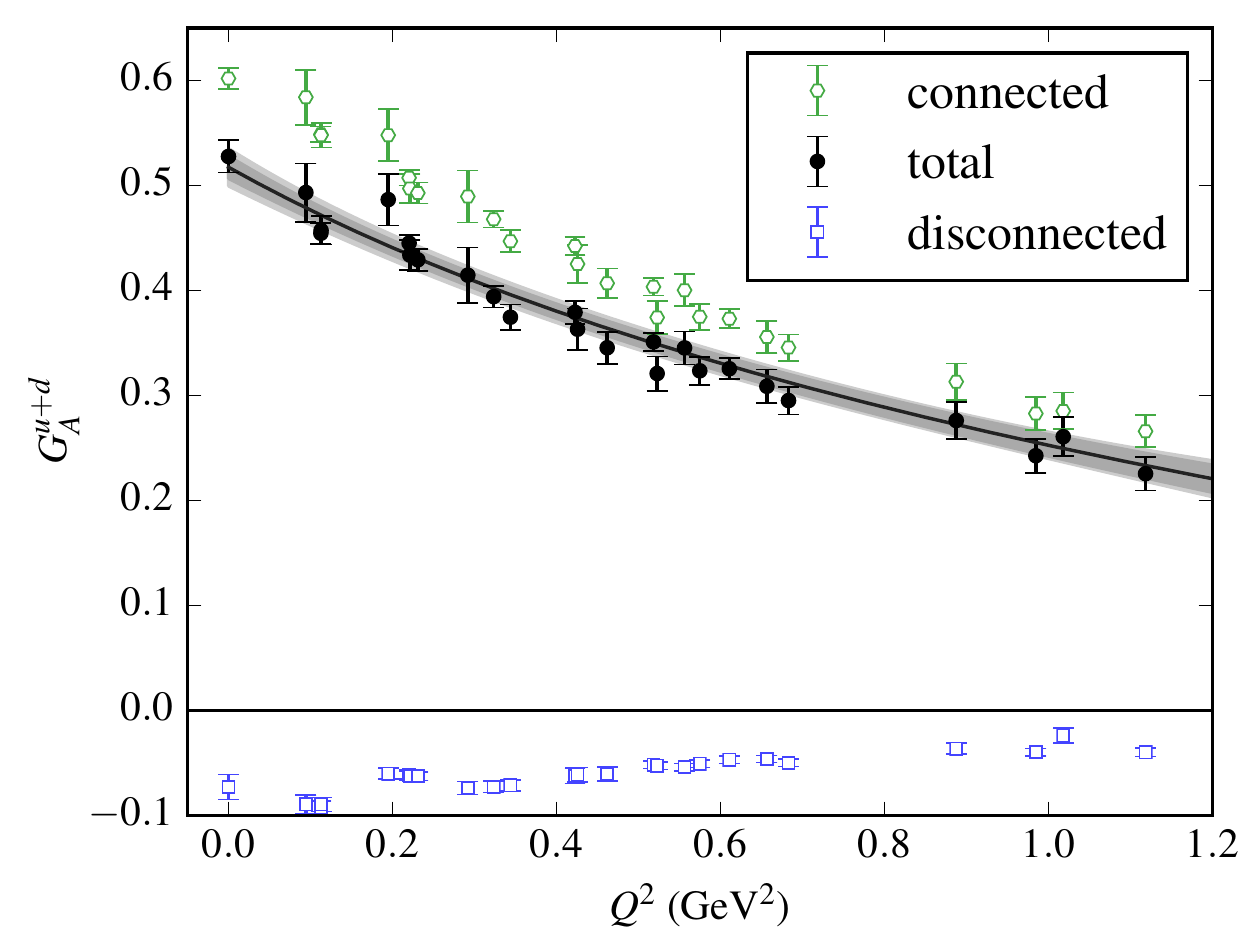}
  \caption{Isovector and light isoscalar axial form factors
    $G_A^{u-d}(Q^2)$ (left) and $G_A^{u+d}(Q^2)$ (right), and
    $z$-expansion fits to them. The lattice data and the inner error
    band for the fit show statistical uncertainties, whereas the outer
    error band for the fit shows the quadrature sum of statistical and
    systematic uncertainties. In addition, for the light isoscalar
    axial form factor, the corresponding form factors from the
    renormalized connected and disconnected diagrams are also shown.}
  \label{fig:fit_GA_UmD_UpD}
\end{figure*}

The isovector axial form factor is shown in
Fig.~\ref{fig:fit_GA_UmD_UpD} (left). From the fit, we find
$g_A=1.208(6)(16)(1)(10)$ and $r_A^2=0.213(6)(13)(3)(0)\text{ fm}^2$,
where the uncertainties are due to statistics, excited states,
fitting, and renormalization, respectively. The dominant uncertainty
is excited-state effects. The fitted value of $g_A$ is quite
compatible with the value taken from the form factor at $Q^2=0$,
$1.206(7)(19)(0)(10)$, with slightly smaller uncertainties. The axial
charge was recently determined in a mostly independent calculation
using the same ensemble~\cite{Yoon:2016jzj}, with somewhat higher
statistics and different methodology. If we examine the bare quantity
to avoid differences in renormalization factors, we get
$g_A^\text{bare}=1.401(7)(18)(2)$, which differs from the result in
Ref.~\cite{Yoon:2016jzj}, $g_A^\text{bare}=1.431(15)$, by slightly
more than one standard deviation. We can compare the axial radius
with the recent reanalysis of neutrino-deuteron scattering
data~\cite{Meyer:2016oeg} that found $r_A^2=0.46(22)\text{ fm}^2$. Our
result is slightly more than one standard deviation smaller.

Figure~\ref{fig:fit_GA_UmD_UpD} (right) shows the light-quark
isoscalar form factor $G_A^{u+d}(Q^2)$. The fit yields
$g_A^{u+d}=0.517(11)(14)(1)(3)$ and
$(r_A^2)^{u+d}=0.197(21)(21)(4)(0)\text{ fm}^2$. The statistical
errors are relatively much larger than for the isovector case, and the
dominant source of these errors is the connected diagrams. The
uncertainty due to renormalization in $g_A^{u+d}$ is mostly due to the
diagonal element of the renormalization matrix; the effect of mixing
with strange quarks is very small.

\begin{figure*}
  \centering
  \includegraphics[width=0.495\textwidth]{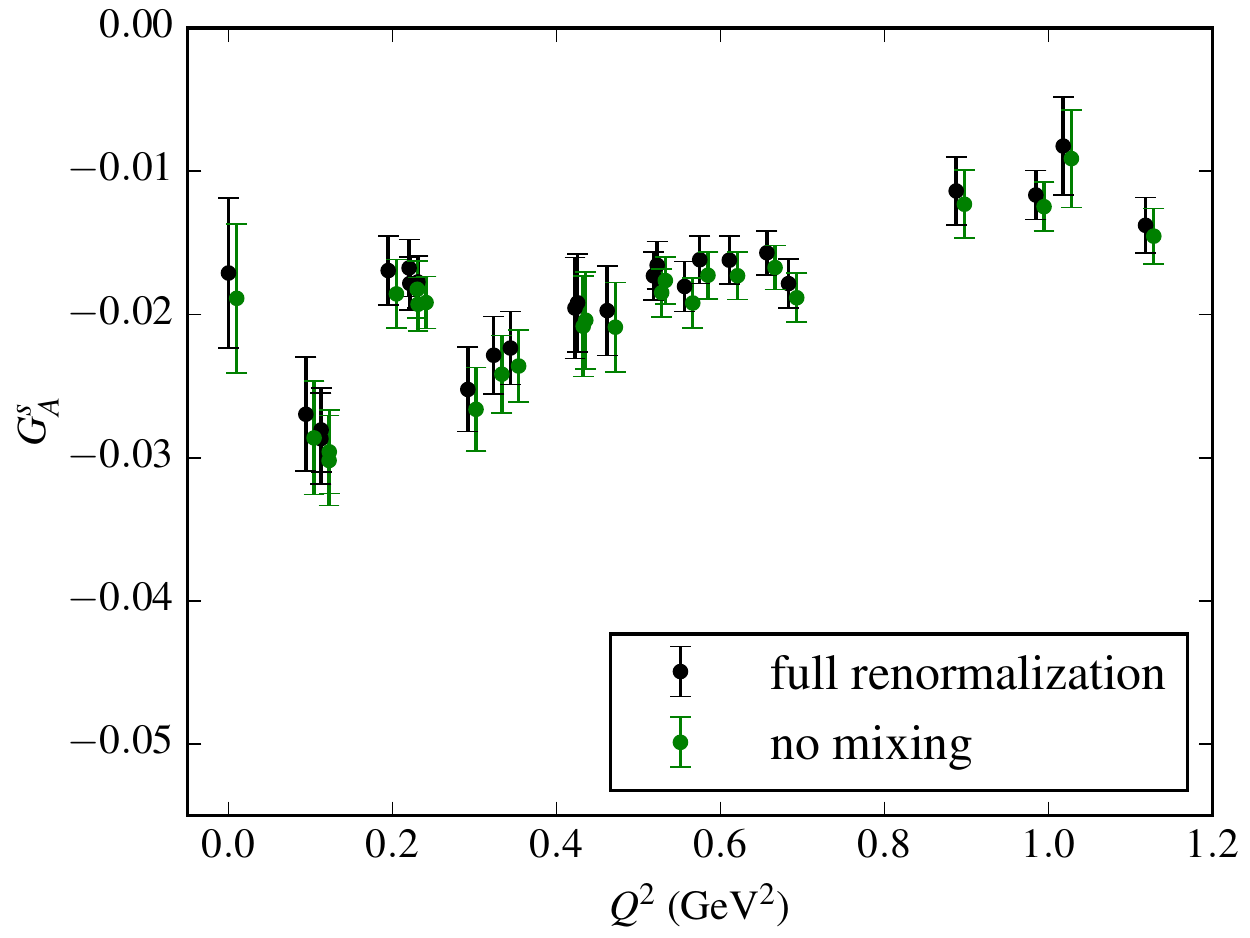}
  \includegraphics[width=0.495\textwidth]{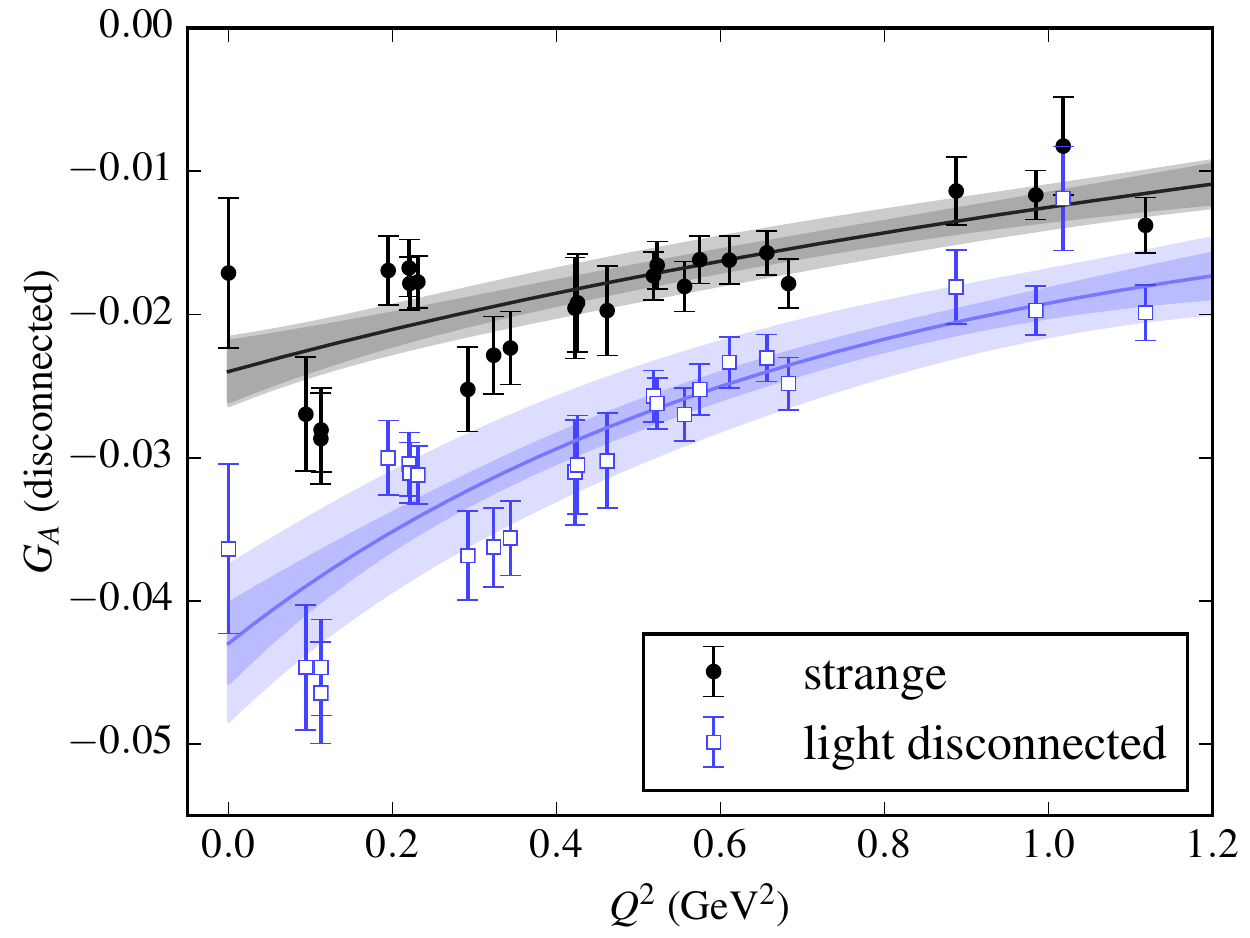}
  \caption{Disconnected axial form factors. Left: strange form factor,
    both with the full renormalization matrix and after setting the
    mixing with light quarks to zero. Right: strange and disconnected
    light-quark axial form factors, including $z$-expansion fits to
    them. See the caption of Fig.~\ref{fig:fit_GA_UmD_UpD}.}
  \label{fig:fit_GA_disc}
\end{figure*}

In Fig.~\ref{fig:fit_GA_disc} we show the strange and light
disconnected axial form factors. The strange axial form factor
$G_A^s(Q^2)$ is the most important case for mixing between light and
strange axial currents, since it is small and it mixes under
renormalization with $G_A^{u+d}(Q^2)$, which has a contribution from
connected diagrams and is much larger. The effect of this mixing is
shown in the left plot: it reduces the magnitude of the form factor by
up to 10\%, although this effect is smaller than the total statistical
uncertainty. In these plots the block-correlated nature of the
statistical uncertainties is clearly visible, particularly at low
$Q^2$: the data that are strongly correlated form clusters of nearby
points, but there are large fluctuations between different
clusters. This effect was previously seen in the disconnected
electromagnetic form factors computed using the same
dataset~\cite{Green:2015wqa}. Fits using the $z$ expansion to the
strange and light disconnected form factors are shown in the right
plot. From these fits we obtain $g_A^s=-0.0240(21)(8)(2)(7)$ and
$g_A^{l,\text{disc}}=-0.0430(28)(46)(6)(8)$. The fit has the effect of
averaging over several uncorrelated clusters of data, and produces a
considerably smaller uncertainty than the value taken directly from
the form factor at $Q^2=0$. The leading uncertainties are statistical
and (for the light-quark case) excited-state effects. The uncertainty
due to renormalization is dominated by uncertainty in the off-diagonal
part of the renormalization matrix. We also obtain the radii
$(r_A^2)^s=0.155(73)(57)(7)(2)\text{ fm}^2$ and
$(r_A^2)^{l,\text{disc}}=0.248(57)(28)(18)(0)\text{ fm}^2$. Within
their uncertainties, all of the squared axial radii are compatible
with 0.2~fm$^2$.

\subsection{Quark spin contributions}

The axial form factors at zero momentum transfer,
$g_A^q\equiv G_A^q(0)$, determined in the previous subsection, give
the contribution from the spin of quarks $q$ to the proton spin. We
can compare against standard experimental inputs used for
phenomenological determinations of these quark spin
contributions. Using isospin symmetry, the $u-d$ combination is
determined from the axial charge in neutron beta decay,
$g_A^{u-d}=1.2723(23)$~\cite{Olive:2016xmw}. Our result is about 5\%
lower, which could be attributed to our heavier-than-physical pion
mass.

The flavor nonsinglet combination $u+d-2s$ is typically obtained from
semileptonic decays of octet baryons, assuming SU(3) flavor
symmetry. Although there have been efforts to improve this
determination using chiral perturbation theory (dating back to the
original paper on the heavy baryon approach~\cite{Jenkins:1990jv}), it
was shown in Ref.~\cite{Ledwig:2014rfa} that at full next-to-leading
order, there is a new low-energy constant that contributes to
$g_A^{u+d-2s}$ but not to the octet baryon decays. Thus, in the
absence of additional input, this combination cannot be predicted at
NLO. The leading-order fit to octet baryon decay
data~\cite{Ledwig:2014rfa} yields $g_A^{u+d-2s}=3F-D=0.608(30)$. It is
therefore useful to have a lattice QCD calculation of this quantity,
even for a heavy pion mass, since it will enable full NLO chiral
perturbation theory analyses to be done. Our result is
$g_A^{u+d-2s}=0.565(11)(13)$.

\begin{table}
  \centering
  \begin{tabular}{c|D{.}{.}{3.11}}
    $q$ & \multicolumn{1}{c}{$g_A^q$} \\\hline
    $u$ &  0.863(7)(14)\\
    $d$ & -0.345(6)(9)\\
    $s$ & -0.0240(21)(11)
  \end{tabular}
  \caption{Quark spin contributions to the nucleon spin.}
  \label{tab:quark_spin}
\end{table}

We find the total contribution from quark spin to the nucleon spin at
$\mu=2$~GeV is $g_A^{u+d+s}=0.494(11)(15)$, about half. The other half
must come from gluons and from quark orbital angular momentum. This is
somewhat larger than results from phenomenological determinations of
polarized parton distribution functions: recent
analyses~\cite{deFlorian:2009vb,Nocera:2014gqa,Sato:2016tuz} give
values from 0.18 to 0.28, with an uncertainty ranging from 0.04 to
0.21. There are a few possible sources for this discrepancy. First,
that this is caused by our heavier-than-physical pion mass. This would
require that the flavor singlet axial case be more sensitive than the
isovector one to the pion mass. Second, that the unaccounted-for
systematic uncertainties at this pion mass are large. These include
effects due to finite lattice spacing and $O(\alpha^2)$ corrections to
the matching of the flavor singlet axial current to $\MSbar$. In
particular, the latter does not affect the flavor nonsinglet
combinations, which are in better agreement with phenomenology. A
third possibility is that the phenomenological values are
incorrect. The behavior at small momentum fraction $x$ is poorly
constrained, and a recent estimate~\cite{Kovchegov:2016weo} in the
large-$N_c$ limit of the small-$x$ asymptotics suggests that improved
results at small $x$ would lead to higher values of $g_A^{u+d+s}$.

\begin{figure}
  \centering
  \includegraphics[width=0.5\textwidth]{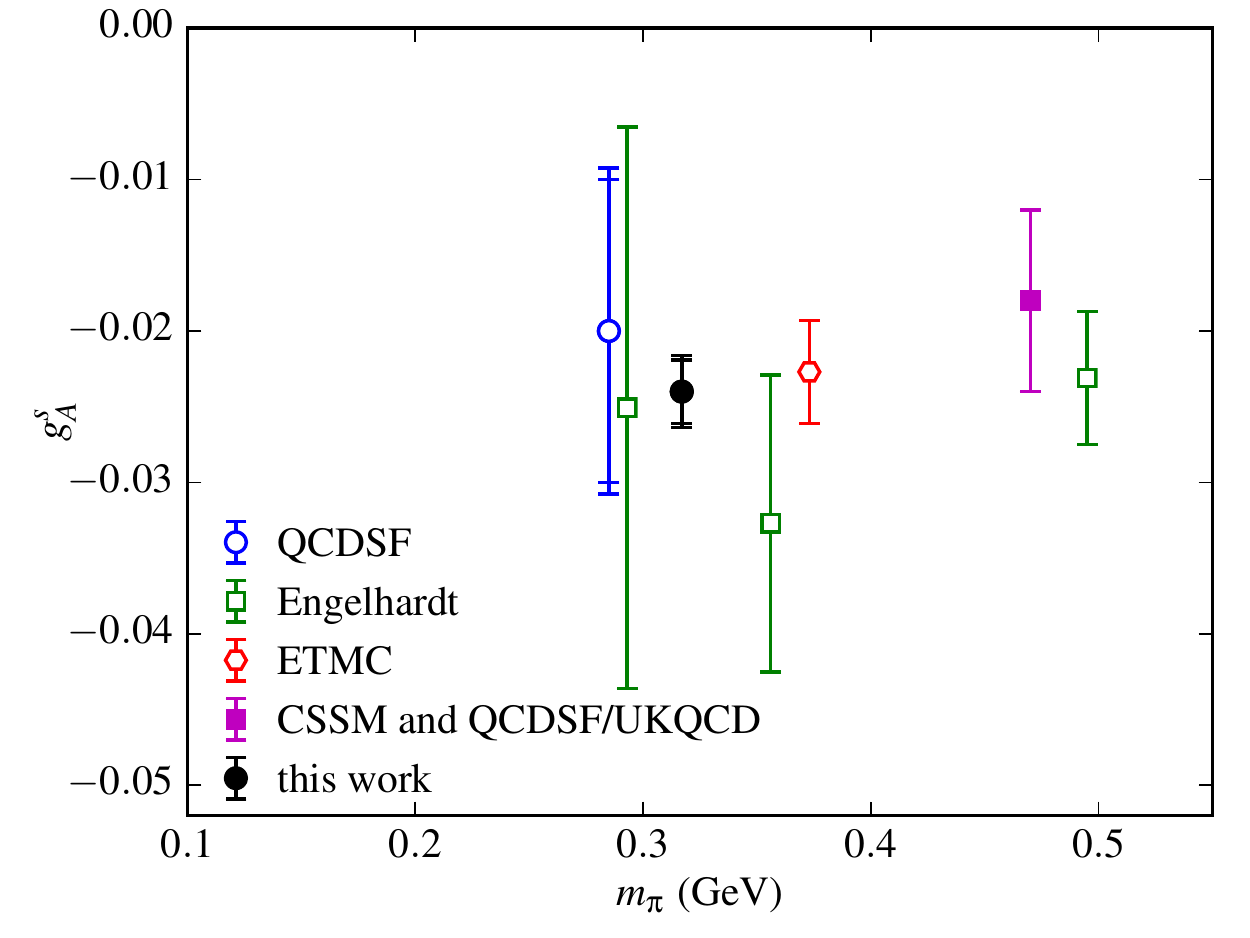}
  \caption{Lattice QCD values for $g_A^s$~\cite{QCDSF:2011aa,
      Engelhardt:2012gd, Abdel-Rehim:2013wlz, Chambers:2015bka},
    keeping only peer-reviewed results that use dynamical fermions and
    nonperturbative renormalization for at least the nonsinglet
    $Z_A$.}
  \label{fig:global_delta_s}
\end{figure}

The individual quark contributions are summarized in
Tab.~\ref{tab:quark_spin}. Our result for $g_A^s$ is compared with
other lattice QCD results in Fig.~\ref{fig:global_delta_s}. The results
are all mutually consistent, and ours is the most precise. Our improved
precision is due to much higher statistics than most previous
calculations, as well as the use of a large volume and the additional
constraints from data at nonzero $Q^2$ in the $z$-expansion fits.
We also note the calculation at the
physical pion mass by ETMC that was presented at Lattice
2016~\cite{Alexandrou:2016tuo}, which found $g_A^s=-0.042(10)$. This
differs from our result by almost two standard deviations, suggesting
that the strange spin contribution to the nucleon spin becomes larger
(more negative) as the light quark mass is decreased.

\subsection{$G_P$ form factors}

\begin{figure*}
  \centering
  \includegraphics[width=0.495\textwidth]{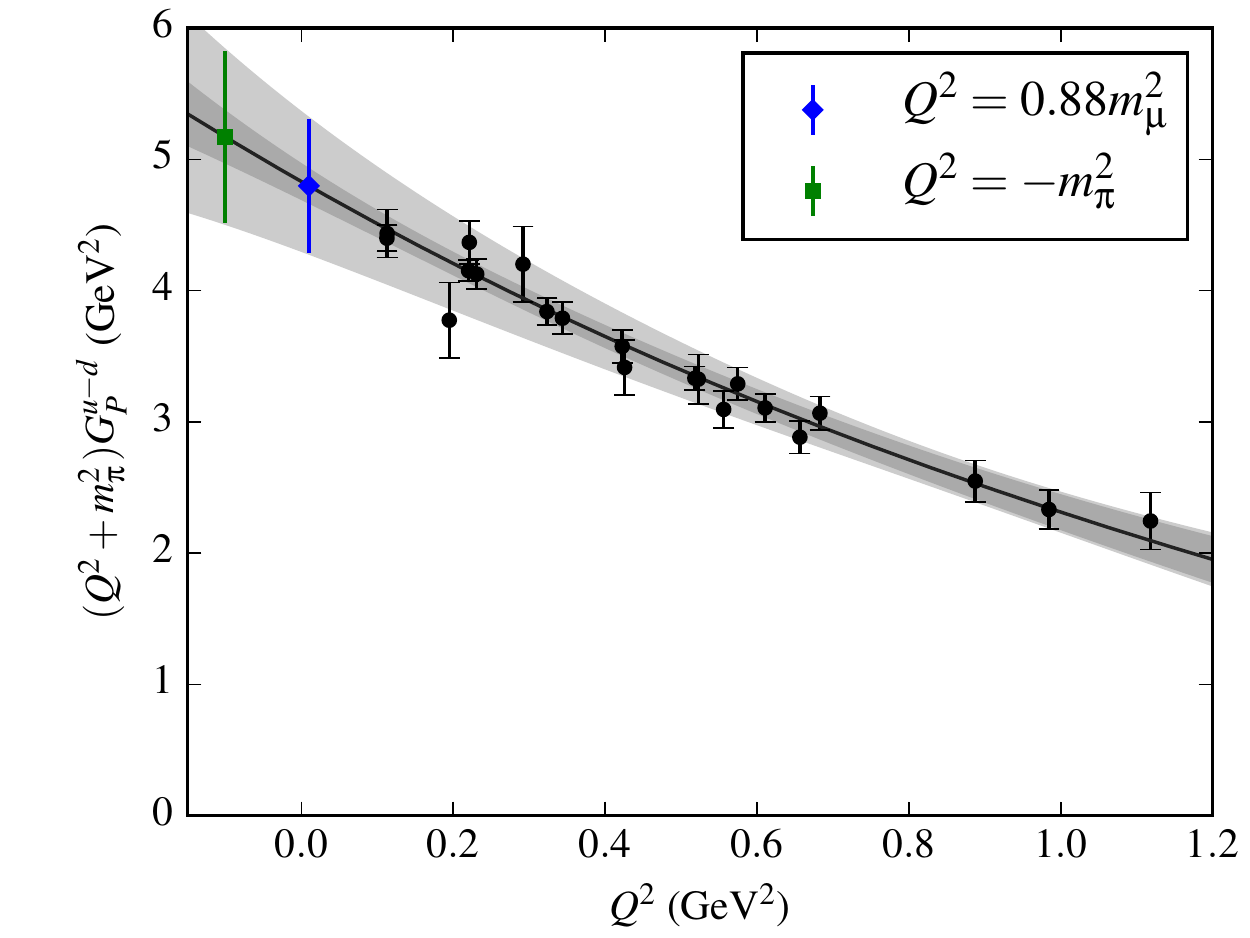}
  \includegraphics[width=0.495\textwidth]{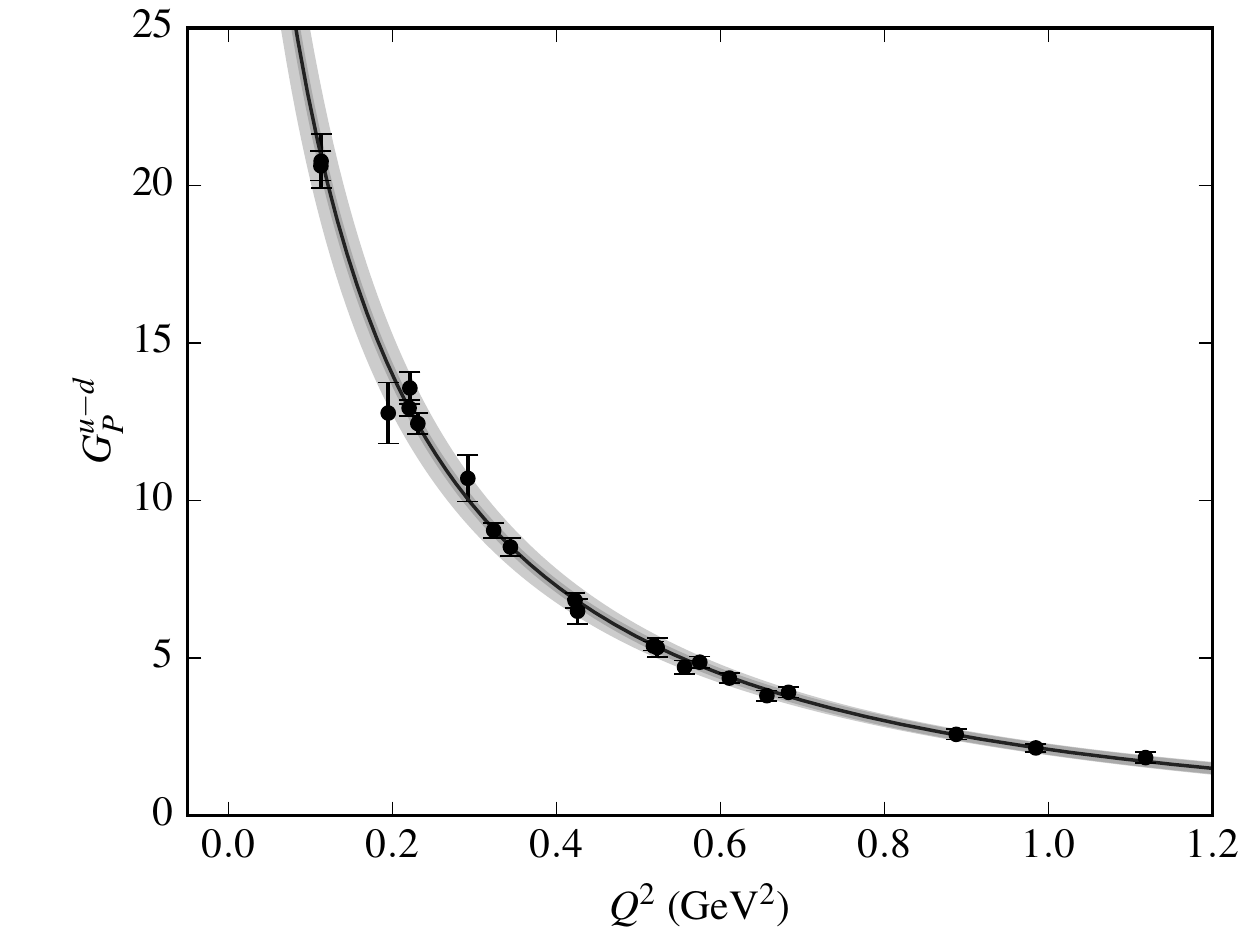}
  \caption{Isovector induced pseudoscalar form factor $G_P^{u-d}(Q^2)$
    and the $z$-expansion fit to it. The left plot shows the form
    factor with the pion pole removed (which is directly fitted using
    the $z$ expansion), and the right plot has the pole restored in
    the fit curve.  The left plot also shows the extrapolations needed
    to obtain $g_P^\text{norm}$ and $g_{\pi NN}$.  See the caption of
    Fig.~\ref{fig:fit_GA_UmD_UpD}.}
  \label{fig:fit_GP_UmD}
\end{figure*}

Figure~\ref{fig:fit_GP_UmD} shows the isovector induced pseudoscalar
form factor $G_P^{u-d}(Q^2)$. As discussed in Subsection~\ref{sec:zexp},
we remove the pion pole that is present in this form factor before
fitting using the $z$ expansion. With the pion pole removed, the
dependence on $Q^2$ is much weaker. At low $Q^2$, there is a large
systematic uncertainty from excited-state contributions. For
comparison with experiment, we consider ordinary muon capture of
muonic hydrogen, which (assuming isospin symmetry) is sensitive to
$g_P^*\equiv \frac{m_\mu}{2m_N}G_P^{u-d}(Q_*^2)$, where
$Q_*^2=0.88m_\mu^2$. To remove the strong dependence on the pion
mass arising from the pion pole, we consider~\cite{Aoki:2010xg}
\begin{equation}
  g_P^\text{norm} \equiv \frac{m_\mu}{2m_N} \frac{Q_*^2+m_\pi^2}{Q_*^2+m_{\pi,\text{phys}}^2} G_P^{u-d}(Q_*^2)\xrightarrow{m_\pi\to m_{\pi,\text{phys}}}g_P^*.
\end{equation}
Using a modest extrapolation of our fit, we find
$g_P^\text{norm}=8.47(21)(87)(2)(7)$, which is consistent with the
measurement by the MuCap experiment~\cite{Andreev:2012fj},
$g_P^*=8.06(55)$. We can also determine the residue of the pion pole:
this is related to the pion decay constant $F_\pi$ and the
pion-nucleon coupling constant $g_{\pi NN}$~\cite{Goldberger:1958vp},
\begin{equation}\label{eq:pipole}
\lim_{Q^2\to -m_\pi^2} (Q^2+m_\pi^2)G_P^{u-d}(Q^2) = 4m_N F_\pi g_{\pi NN}.
\end{equation}
The required extrapolation in $Q^2$ is about twice as far as was required
for $g_P^*$, but is still small compared with our probed range of
$Q^2$. Using $F_\pi=106$~MeV computed on this ensemble, we obtain
$g_{\pi NN}=11.5(4)(1.4)(1)(0)$. This is slightly more than one
standard deviation below the recent result~\cite{Baru:2010xn}
determined using pion-nucleon scattering lengths from measurements of
pionic atoms: $g_{\pi NN}^2/(4\pi)=13.69(20)$, or
$g_{\pi NN}=13.12(10)$. In the chiral limit, the pion-nucleon coupling
constant is related to the axial charge via the Goldberger-Treiman
relation, $g_{\pi NN}=g_A^{u-d}m_N/F_\pi$; on our ensemble the right
hand side equals 12.1, and thus our precision is insufficient to
resolve a nonzero Goldberger-Treiman discrepancy.

\begin{figure*}
  \centering
  \includegraphics[width=0.495\textwidth]{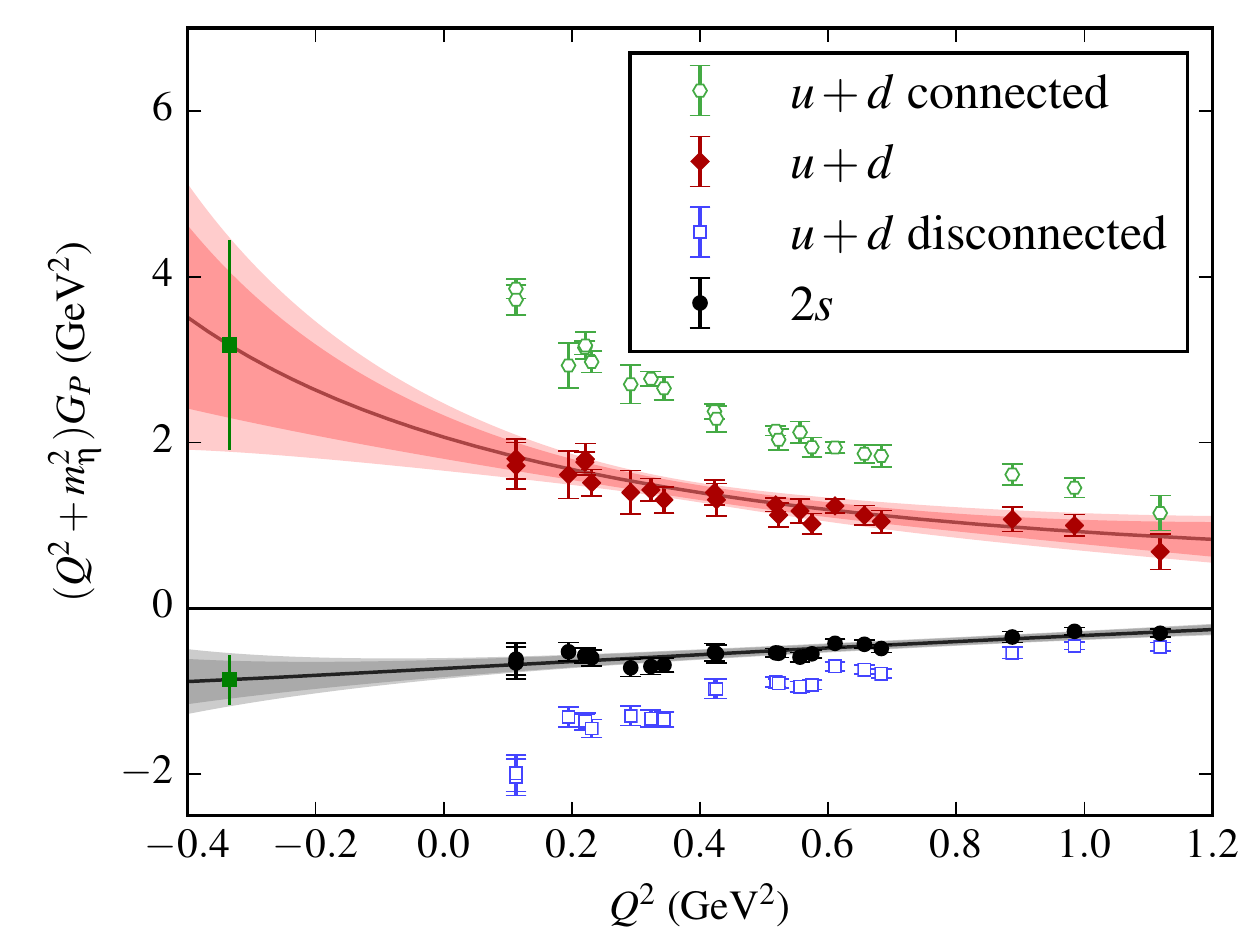}
  \includegraphics[width=0.495\textwidth]{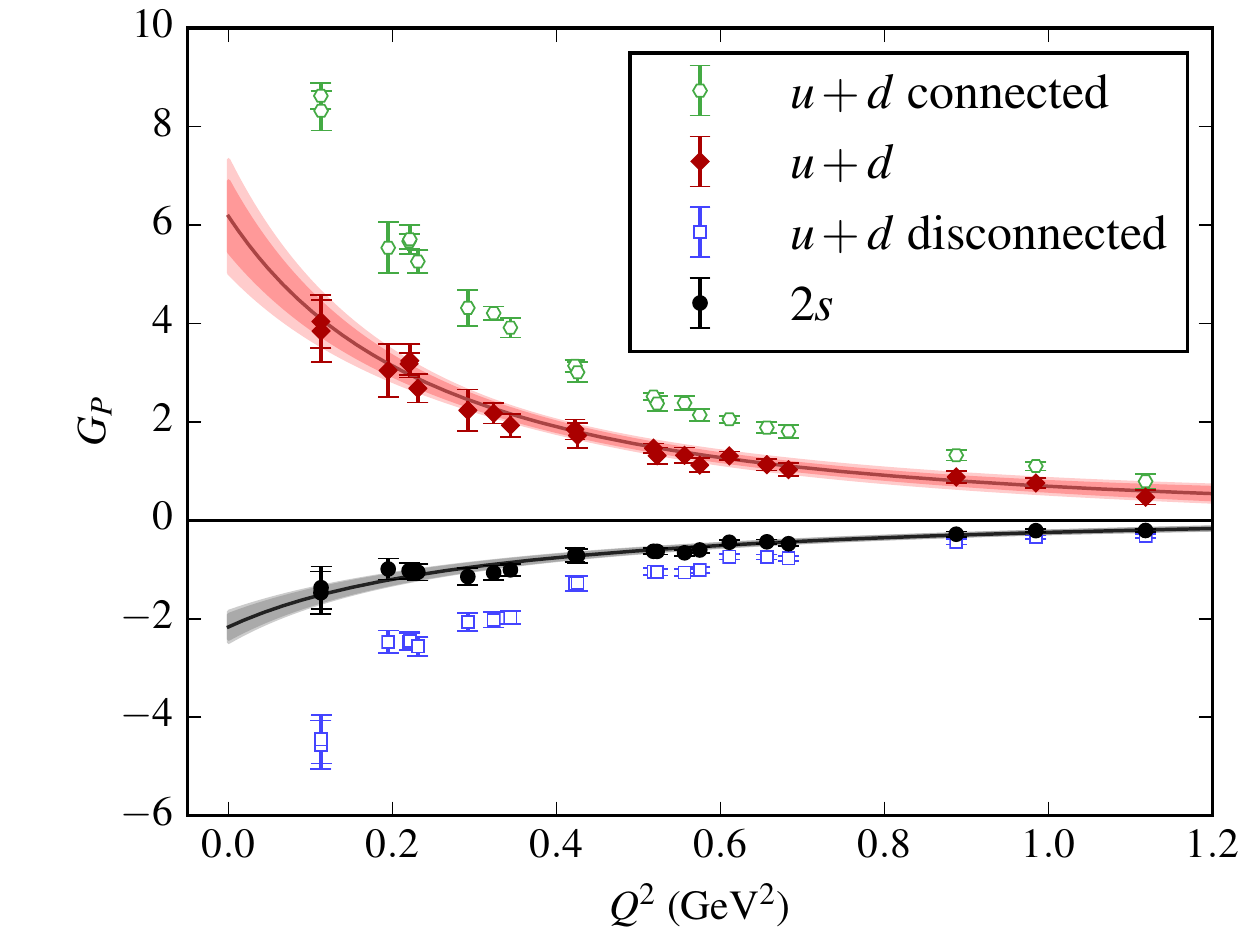}
  \caption{Light and strange isoscalar induced pseudoscalar form
    factors $G_P^{u+d}(Q^2)$ and $G_P^s(Q^2)$ and the $z$-expansion
    fits to them. In addition, for the light isoscalar form factor,
    the corresponding form factors for the renormalized connected and
    disconnected diagrams are also shown. The left plot shows the
    form factors with the eta pole removed (which is directly fitted
    using the $z$ expansion), and the right plot has the pole restored
    in the fit curves. The left plot also shows the extrapolations
    to the eta pole.  See the caption of
    Fig.~\ref{fig:fit_GA_UmD_UpD}.}
  \label{fig:fit_GP_isoscalar}
\end{figure*}

The isoscalar induced pseudoscalar form factors are shown in
Fig.~\ref{fig:fit_GP_isoscalar}. As these contain an eta pole, we
again remove the pole before fitting with the $z$ expansion. The eta
mass is estimated using the leading-order relation from partially
quenched chiral perturbation theory,
$m_\eta^2=(m_\pi^2+2m_{\eta_s}^2)/3$, yielding
$m_\eta\approx 578$~MeV. Relative to the connected diagrams, the
contributions from disconnected diagrams are not small, which is in
contrast with what we saw for the $G_A$ form factors. This can be
understood by considering the partially quenched theory, under which
the connected contributions to $G_P^{u+d}(Q^2)$ are equal to
$G_P^{u+d-2r}(Q^2)$, where $r$ is a third valence light quark,
degenerate with $u$ and $d$. We would expect that this form factor has
a pseudoscalar pole from the $\pi_8$ meson\footnote{The presence of
this pole was already argued in Ref.~\cite{Liu:1991ni}.}
(which is part of the octet
of pseudo-Goldstone bosons under the exact $SU(3)$ symmetry of the
valence $u$, $d$, and $r$ quarks) at $Q^2=-m_\pi^2$. The sharp rise of
this form factor at low $Q^2$ is consistent with this
expectation. Since the physical isoscalar form factor does not contain
a pole at $Q^2=-m_\pi^2$, the pole must be canceled by the
disconnected diagrams, which explains why the disconnected
contribution to $G_P^{u+d}$ must also rise sharply (with opposite
sign) at low $Q^2$. Similarly, the expectation that the octet axial
current $A_\mu^8$ couples more strongly than the singlet current
$A_\mu^0$ to the eta meson suggests that $G_P^s$ and $G_P^{u+d}$
should have opposite sign, as seen in the data.

We can attempt to quantify the couplings to the eta meson by studying
the generalization of Eq.~\eqref{eq:pipole}:
\begin{equation}
\lim_{Q^2\to -m_\eta^2} (Q^2+m_\eta^2)G_P^a(Q^2) = 2 m_N f_\eta^a g_{\eta NN},
\end{equation}
where the eta decay constants are defined\footnote{Note that using
  this definition for the pion decay constant would yield
  $f_\pi^3=\sqrt{2}F_\pi$, where the physical value is
  $f_\pi^3\approx 130$~MeV.} by
$\langle 0|A_\mu^a|\eta(p)\rangle = f_\eta^a
p_\mu$~\cite{Feldmann:1998vh}. As Fig.~\ref{fig:fit_GP_isoscalar}
shows, the extrapolation to the eta pole is rather difficult and the
results have a large uncertainty. Since we have not separately
computed the eta decay constants on this ensemble, we cannot determine
the eta-nucleon coupling constant in this way. However, we can take
the singlet-octet ratio $f_\eta^0/f_\eta^8$, which we find to be
$0.96(16)(21)(4)(1)$. This is larger than expected, and three standard
deviations above the value obtained from the phenomenological
parameters in Ref.~\cite{Feldmann:1998vh},
$f_\eta^0/f_\eta^8=0.16(3)$. In particular, since our pion mass is
heavier than physical, we would expect the reduced breaking of flavor
$SU(3)$ symmetry to yield a value closer to zero. This unexpected
behavior is likely caused by the difficulty in such a large
extrapolation in $Q^2$; direct calculations of these decay constants
such as in Ref.~\cite{Michael:2013vba} are much more reliable since
they do not require a kinematical extrapolation.  If we ignore this
issue, and assume the $SU(3)$ relation $f_\eta^8=f_\pi^3$, then from
$G_P^8\equiv (G_P^{u+d}-2G_P^s)/\sqrt{6}$ we obtain an estimate for
the eta-nucleon coupling constant,
$g_{\eta NN}=5.2(1.0)(1.0)(0.2)(0)$.

\begin{figure*}
  \centering
  \includegraphics[width=0.495\textwidth]{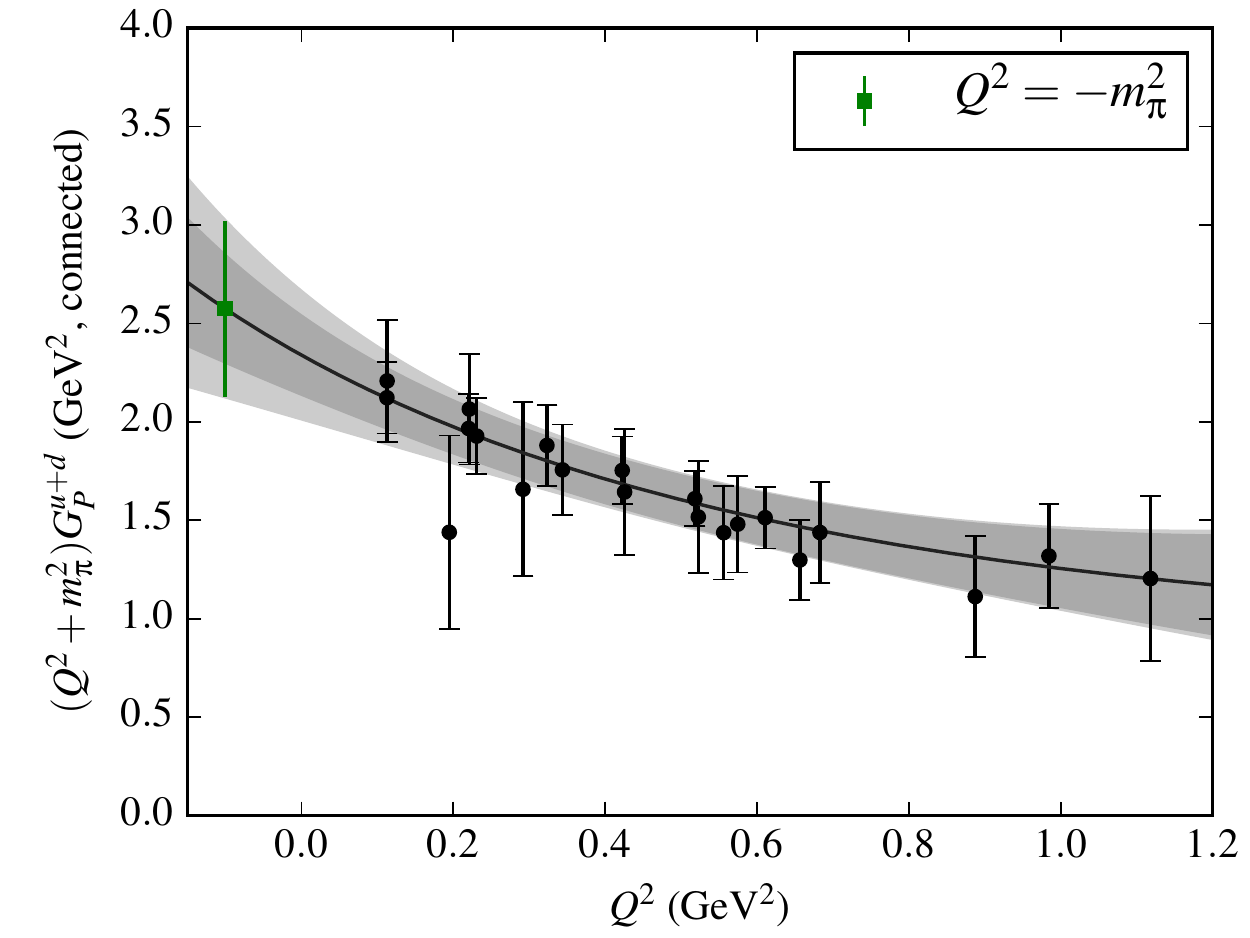}
  \includegraphics[width=0.495\textwidth]{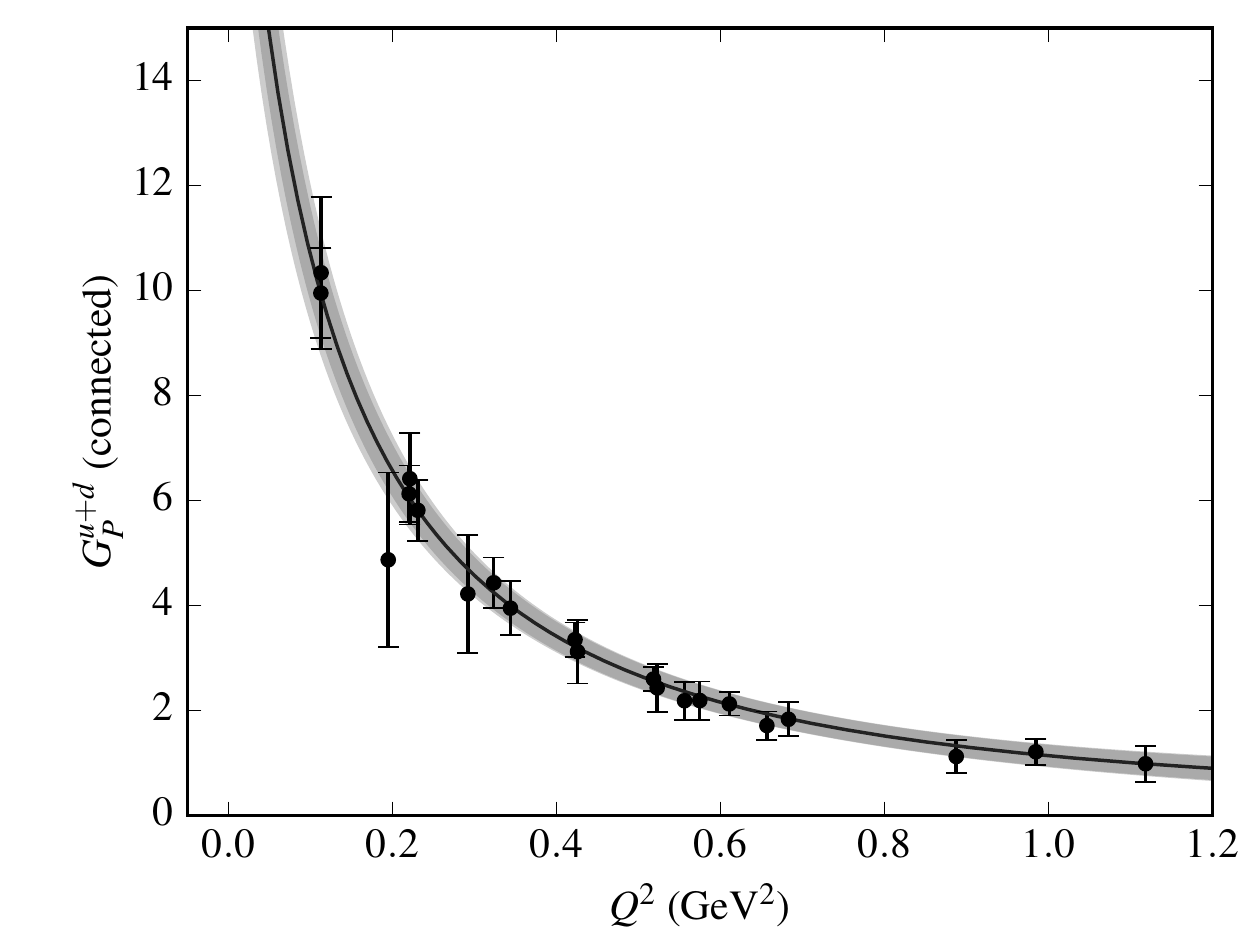}
  \caption{Connected light isoscalar induced pseudoscalar form factor
    $G_P^{u+d,\text{conn}}(Q^2)$ and the $z$-expansion fit to it. See
    the caption of Fig.~\ref{fig:fit_GP_UmD}.}
  \label{fig:fit_GP_UpDconn}
\end{figure*}

Assuming flavor $SU(3)$ symmetry, the eta-nucleon coupling constant
can also be obtained from the connected contribution to $G_P^{u+d}$.
Provided that the considerations from the partially quenched theory
are valid, the residue of the pion pole is proportional to
$F_\pi g_{\pi_8 NN}$, where the $\pi_8$-nucleon coupling constant is
equal (up to $SU(3)$ breaking corrections) to $g_{\eta NN}$. Alone,
the connected contribution does not benefit from the cancellation of
excited-state effects with the disconnected contribution that we have
seen. Therefore, to better control these effects, we determine this
form factor using the summation method in the same way as $G_P^{u-d}$;
this is shown in Fig.~\ref{fig:fit_GP_UpDconn}. We obtain
$g_{\pi_8 NN}=3.29(35)(45)(3)(0)$. The eta-nucleon coupling constant
is not so well known phenomenologically, but both of these estimates
are compatible with the value obtained using a generalized
Goldberger-Treiman relation,
$g_{\eta NN}=3.4(5)$~\cite{Feldmann:1999uf}.

\section{\label{sec:conclusions}Discussion and conclusions}

As with our previous study of electromagnetic form
factors~\cite{Green:2015wqa}, our approach of using hierarchical
probing for disconnected loops and high statistics for nucleon
two-point functions is effective at producing a good signal for
disconnected nucleon axial form factors. In contrast with the previous
study, however, we find that the gauge noise is dominant over the
noise from stochastic estimation of the loops, so that further
improvements in the latter would be of limited value.

A useful feature of disconnected loops is that they can be reused for
calculating many different observables. We did this for computing the
axial renormalization factors nonperturbatively, and we were again
able to obtain a reasonable signal. At the scale $\mu=2$~GeV, the
effect of mixing between light and strange axial currents is small:
$G_A^s(Q^2)$, which is most affected, is reduced in magnitude by up to
10\%. The accuracy of our renormalization is limited by the unknown
$O(\alpha^2)$ term in the matching of the flavor singlet axial current
to the $\MSbar$ scheme. Our use of two different intermediate schemes
may provide some estimate of this term, but it is possible that the
effect in converting between the two intermediate schemes is smaller
than in converting to $\MSbar$. A smaller flavor-singlet
renormalization factor would make both $g_A^{u+d+s}$ and
$f_\eta^0/f_\eta^8$ more consistent with expectations. This highlights
the need for higher-order conversion factors. In the flavor-nonsinglet
case, these factors have been computed up to three-loop order for some
operators~\cite{Gracey:2003yr,Gracey:2006zr}. As lattice calculations
of disconnected diagrams have made great progress, there is now a need
for similar matching calculations in the flavor-singlet sector.

Since this work was performed using only one lattice ensemble, we
do not provide an estimate of systematic uncertainties due to the
heavier-than-physical pion mass or due to discretization effects. The
former have been investigated in many lattice calculations of the
isovector axial charge, where generally only modest effects have been
seen. Generalizing this, we don't expect large dependence on the pion
mass for $G_A^q(Q^2)$. On the other hand, the $G_P$ form factors ---
especially the isovector one --- will have a significant dependence on
light quark masses due to the presence of pseudoscalar
poles. Discretization effects for this lattice ensemble have been
studied in Ref.~\cite{Yoon:2016jzj}, where it is compared with another
ensemble with similar pion mass and smaller lattice spacing. The
isovector axial charge computed on the two ensembles is consistent
within one standard deviation, or about 3\%, which gives a rough
estimate of uncertainty due to finite lattice spacing. We expect that
these effects are of similar size for other nucleon matrix elements
involving the axial current.

We found that the statistical correlations between the values of a
form factor at different $Q^2$ behave differently for connected and
disconnected diagrams. In the latter case, data with different spatial
momentum transfers are nearly uncorrelated. This has the result of
better constraining fits to the form factors; using these fits, we
were able to obtain a precise value for the strange axial charge on
our ensemble, $g_A^s=-0.0240(21)(11)$, which is consistent with
previous lattice calculations.

For $G_A(Q^2)$, the disconnected diagrams are small compared with the
connected ones. For instance, $g_A^{u+d,\text{disc}}/g_A^{u+d}=-0.17$,
and the strange disconnected diagrams are about half as large as the
light ones. However, this is somewhat larger than we saw for the
electromagnetic form factors~\cite{Green:2015wqa}, where the
disconnected light magnetic moment, $\mu^{u+d,\text{disc}}\approx
0.11$, is about 4\% of the full experimental value
$\mu^{u+d}=3(\mu^p+\mu^n)\approx 2.6$, and the disconnected $G_E(Q^2)$
is even smaller relative to the full experimental form factor. This
may change closer to the physical pion mass, since the disconnected
light-quark matrix elements are expected to grow as the quark mass is
decreased.

For $G_P(Q^2)$, the situation is different, with disconnected diagrams
not nearly as suppressed. This can be understood from the dominant
influence of the pseudoscalar poles in these form factors, which leads
to a significant cancellation between the connected and disconnected
contributions to $G_P^{u+d}(Q^2)$. As the pion mass is decreased
toward the physical point, we expect that $G_P^{u+d}(Q^2)$ will vary
only mildly, but at low $Q^2$ the individual connected and
disconnected contributions will become much larger since the location
of the pion pole will approach $Q^2=0$. This growing cancellation may
make it difficult to obtain a good signal for the full form factor at
the physical pion mass and low $Q^2$.

\begin{acknowledgments}
  Computations for this work were carried out on
  facilities of the USQCD Collaboration, which are funded by the
  Office of Science of the U.S. Department of Energy, on facilities provided
  by XSEDE, funded by National Science Foundation grant \#ACI-1053575,
  and at Forschungszentrum Jülich.

  During this research several of us were supported in part by the
  U.S.\ Department of Energy Office of Nuclear Physics under grants
  \#DE--FG02--94ER40818 (JG, SM, JN, and AP), \#DE--SC--0011090 (JN),
  \#DE--FG02--96ER40965 (ME), \#DE--FC02--12ER41890 (JL),
  \#DE--FG02--04ER41302 (KO), \#DE--AC02--05HC11231 (SS), and
  \#DE--AC05--06OR23177 under which JSA operates the Thomas Jefferson
  National Accelerator Facility (KO). Support was also received from
  National Science Foundation grants \#CCF--121834 (JL) and
  \#PHY--1520996 (SM), the RIKEN Foreign Postdoctoral Researcher
  program (SS), the RHIC Physics Fellow Program of the RIKEN BNL
  Research Center (SM), Deutsche Forschungsgemeinschaft grant
  SFB--TRR~55 (SK), and the PRISMA Cluster of Excellence at the
  University of Mainz (JG).

  Calculations were performed with the Chroma software
  suite~\cite{Edwards:2004sx}, using QUDA~\cite{Clark:2009wm} with
  multi-GPU support~\cite{Babich:2011:SLQ:2063384.2063478}, as well as
  the Qlua software suite~\cite{Qlua}.
\end{acknowledgments}

\appendix*

\section{\label{app:fit_params}Form factor fit parameters}

In this appendix we give parameters for the form factor fits and the
estimated total uncertainty. Recall that we performed fits of the form
\begin{equation}
  G(Q^2) = \sum_{k=0}^5 a_k z(Q^2)^k,
\end{equation}
where $z(Q^2)$ is given in Eq.~\eqref{eq:z_Q2},
$t_\text{cut}=(3m_\pi)^2$, and we used the central value
$am_\pi=0.1833$. The parameters are given in
Tab.~\ref{tab:zexp_params}, where for each fit we have also given the
correlation matrix. For the $G_P$ form factors, we give parameters for
fits to $G(Q^2)=a^2(Q^2+m^2)G_P(Q^2)$, where $m$ is either $m_\pi$ or
$m_\eta$; we used the value $am_\eta=0.3342$. The fit curves and outer
error bands for the physical form factors shown in
Sec.~\ref{sec:formfacs} (i.e., excluding the individual connected and
disconnected parts) can be reproduced using the data in this table.

\begin{table}\
  \centering
  \begin{tabular}{r|D{.}{.}{3.10}|D{.}{.}{2.3}D{.}{.}{2.3}D{.}{.}{2.3}D{.}{.}{2.3}D{.}{.}{2.3}D{.}{.}{2.3}}
    \hline\hline
  \multicolumn{8}{c}{$G_A^{u-d}(Q^2)$}\\\hline
  k & \multicolumn{1}{c|}{$a_k$} & \multicolumn{6}{c}{Correlation matrix} \\\hline
  0 &  1.208(20)    & 1 & -0.771 &  0.365 & -0.101 &  0.067 &  0.058 \\
  1 & -3.985(332)   &   &  1     & -0.767 &  0.509 &  0.229 &  0.168 \\
  2 &  0.877(1.639) &   &        &  1     & -0.911 & -0.471 & -0.299 \\
  3 &  7.730(4.783) &   &        &        &  1     &  0.416 &  0.201 \\
  4 &  4.324(3.101) &   &        &        &        &  1     &  0.963 \\
  5 &  1.615(1.417) &   &        &        &        &        &  1     \\
    \hline\hline
  \multicolumn{8}{c}{$G_A^{u+d}(Q^2)$}\\\hline
  k & \multicolumn{1}{c|}{$a_k$} & \multicolumn{6}{c}{Correlation matrix} \\\hline
  0 &  0.517(18)    & 1 & -0.712 &  0.430 & -0.215 & -0.271 & -0.323 \\
  1 & -1.582(274)   &   &  1     & -0.821 &  0.427 &  0.459 &  0.467 \\
  2 &  0.947(1.975) &   &        &  1     & -0.822 & -0.826 & -0.803 \\
  3 & -0.853(5.519) &   &        &        &  1     &  0.990 &  0.965 \\
  4 & -0.534(2.451) &   &        &        &        &  1     &  0.991 \\
  5 & -0.214(745)   &   &        &        &        &        &  1     \\
    \hline\hline
  \multicolumn{8}{c}{$G_A^{s}(Q^2)$}\\\hline
  k & \multicolumn{1}{c|}{$a_k$} & \multicolumn{6}{c}{Correlation matrix} \\\hline
  0 & -0.0240(24)   & 1 & -0.678 &  0.478 &  0.314 &  0.177 &  0.094 \\
  1 &  0.0577(386)  &   &  1     & -0.943 & -0.802 & -0.656 & -0.575 \\
  2 &  0.0274(1445) &   &        &  1     &  0.845 &  0.687 &  0.598 \\
  3 & -0.0079(507 ) &   &        &        &  1     &  0.963 &  0.911 \\
  4 & -0.0049(156)  &   &        &        &        &  1     &  0.987 \\
  5 & -0.0017(43)   &   &        &        &        &        &  1     \\
    \hline\hline
  \multicolumn{8}{c}{$a^2(Q^2+m_\pi^2)G_P^{u-d}(Q^2)$}\\\hline
  k & \multicolumn{1}{c|}{$a_k$} & \multicolumn{6}{c}{Correlation matrix} \\\hline
  0 &  1.613(174)   & 1 & -0.882 &  0.518 &  0.052 &  0.056 &  0.107 \\
  1 & -3.997(1.583) &   &  1     & -0.838 &  0.086 &  0.118 &  0.078 \\
  2 & -3.946(5.729) &   &        &  1     & -0.365 & -0.423 & -0.403 \\
  3 &  4.078(5.339) &   &        &        &  1     &  0.991 &  0.974 \\
  4 &  1.825(2.345) &   &        &        &        &  1     &  0.994 \\
  5 &  0.506(675)   &   &        &        &        &        &  1     \\
    \hline\hline
  \multicolumn{8}{c}{$a^2(Q^2+m_\eta^2)G_P^{u+d}(Q^2)$}\\\hline
  k & \multicolumn{1}{c|}{$a_k$} & \multicolumn{6}{c}{Correlation matrix} \\\hline
  0 &  0.690(128)   & 1 & -0.948 &  0.815 &  0.305 &  0.074 & -0.022 \\
  1 & -2.805(1.929) &   &  1     & -0.941 & -0.194 &  0.071 &  0.174 \\
  2 &  3.790(6.441) &   &        &  1     &  0.200 & -0.083 & -0.192 \\
  3 &  0.829(3.428) &   &        &        &  1     &  0.957 &  0.909 \\
  4 &  0.045(1.336) &   &        &        &        &  1     &  0.990 \\
  5 & -0.031(360)   &   &        &        &        &        &  1     \\
    \hline\hline
  \multicolumn{8}{c}{$a^2(Q^2+m_\eta^2)G_P^{s}(Q^2)$}\\\hline
  k & \multicolumn{1}{c|}{$a_k$} & \multicolumn{6}{c}{Correlation matrix} \\\hline
  0 & -0.121(17)  & 1 & -0.868 &  0.664 &  0.719 &  0.714 &  0.689 \\
  1 &  0.256(223) &   &  1     & -0.935 & -0.909 & -0.838 & -0.765 \\
  2 &  0.538(766) &   &        &  1     &  0.949 &  0.858 &  0.771 \\
  3 &  0.191(293) &   &        &        &  1     &  0.976 &  0.931 \\
  4 &  0.049(83)  &   &        &        &        &  1     &  0.988 \\
  5 &  0.011(20)  &   &        &        &        &        &  1     \\
    \hline\hline
  \end{tabular}
  \caption{Parameters for the $z$-expansion fits, along with their correlation matrices. The latter are symmetric, and we have omitted redundant entries.}
  \label{tab:zexp_params}
\end{table}

\bibliography{disconnected_axial}
\bibliographystyle{utphys-noitalics}

\end{document}